\documentclass{nature_plusfigure}
\usepackage{graphicx}
\usepackage{subcaption}
\usepackage{aas_macros,amsmath,amssymb,xcolor,url}
\usepackage{multirow}
\usepackage{longtable,hhline}
\usepackage{upgreek}
\usepackage{hyperref}
\usepackage{graphicx}
\usepackage{rotating}
\usepackage[numbers]{natbib}
\usepackage{booktabs} 
\usepackage[none]{hyphenat}
\usepackage{xspace}
\newcommand{\msun}{M$_{\odot}$}

\usepackage[symbol]{footmisc}
\usepackage{cleveref}

\usepackage{multibib}
\newcites{meth}{Methods References}

\title{Decadal pre-explosion activity and circumstellar interaction in a supernova}

\author{Ting-Wan~Chen$^{1}$\footnote{twchen@astro.ncu.edu.tw}, Amar~Aryan$^{1}$\footnote{amar@astro.ncu.edu.tw}, Sheng~Yang$^{2}$\footnote{sheng.yang@hnas.ac.cn}, Stephen~J.~Smartt$^{3}$, Takashi~J.~Moriya$^{4,5,6}$, 
Se\'{a}n~J.~Brennan$^{8}$, 
Maximilian~D.~Stritzinger$^{7}$, Bailey~Martin$^{9}$, Matt~Nicholl$^{10}$, Albert~K.~H.~Kong$^{11,12}$, James~H.~Gillanders$^{3}$, Anirban~Dutta$^{1}$, Brian~P.~Schmidt$^{9}$, Yu-Chi~Cheng$^{1}$, Mark~E.~Huber$^{13}$, Cheng-Han~Lai$^{1}$, Chien-Hsiu~Lee$^{14}$, Yu-Hsing~Lee$^{1}$, Chow-Choong~Ngeow$^{1}$, Ken~W.~Smith$^{3,10}$, Christopher~Ashall$^{13}$, Katie~Auchettl$^{15}$, Chris~R.~Burns$^{16}$, Kenneth~C.~Chambers$^{13}$, Zhi-Yue~Chen$^{1}$, Thomas~de~Boer$^{13}$, Eric~Y.~Hsiao$^{17}$, Khoa~Ngo~Thanh~Ho$^{1}$, 
Willem~B.~Hoogendam$^{13,18}$, David~O.~Jones$^{19}$,
Erkki~Kankare$^{20}$, Tom~L.~Killestein$^{21}$, Hanindyo~Kuncarayakti$^{20,22}$,
Meng-Han~Lee$^{1}$, Chuan-Jui~Li$^{23}$, Chien-Cheng~Lin$^{13}$, Christopher~Lidman$^{9}$, Thomas~B.~Lowe$^{13}$, Eugene~A.~Magnier$^{13}$, Kyle~Medler$^{13}$, Anais~M\"oller$^{24}$, Thomas~Moore$^{25}$, 
Nidia~Morrell$^{26}$, Gregory~S.~H.~Paek$^{13}$, 
Cameron~M.~Pfeffer$^{13,18}$,
Da-Chun~Qiang$^{2}$, Liana~Rauf$^{9}$, Thomas~M.~Reynolds$^{20,27,28}$, Aiswarya~Sankar.K$^{1}$, 
Shubham~Srivastav$^{3}$, Jack~Tweddle$^{3}$, Richard~Wainscoat$^{13}$, Ze-Ning~Wang$^{2}$, Huangfei~Xiao$^{17}$, Zonghong~Zhu$^{29,30}$
}

\begin{document}

\maketitle

\begin{small}
\begin{affiliations}
\label{sec:affiliations}

\item Graduate Institute of Astronomy, National Central University, 300 Jhongda Road, 32001 Jhongli, Taiwan 
\item Institute for Gravitational Wave Astronomy, Henan Academy of Sciences, Zhengzhou 450046, Henan, China 
\item Astrophysics sub-Department, Department of Physics, University of Oxford, Keble Road, Oxford OX1 3RH, UK 
\item National Astronomical Observatory of Japan, National Institutes of Natural Sciences, 2-21-1 Osawa, Mitaka, Tokyo 181-8588, Japan 
\item Graduate Institute for Advanced Studies, SOKENDAI, 2-21-1 Osawa, Mitaka, Tokyo 181-8588, Japan 
\item School of Physics and Astronomy, Monash University, Clayton, VIC 3800, Australia 
\item Department of Physics and Astronomy, Aarhus University, Ny Munkegade 120, DK-8000 Aarhus C, Denmark 
\item Max Planck Institute for Extraterrestrial Physics, Max-Planck-Gesellschaft, Giessenbachstra\ss e 1, Garching, 85748 
\item The Research School of Astronomy and Astrophysics, The Australian National University, Canberra, ACT 2611, Australia  
\item Astrophysics Research Centre, School of Mathematics and Physics, Queen's University Belfast  
\item Institute of Astronomy, National Tsing Hua University, No. 101 Sect. 2 Kuang-Fu Road, 30013 Hsinchu, Taiwan  
\item Institute of Space Engineering, National Tsing Hua University, No. 101 Sect. 2 Kuang-Fu Road, 30013 Hsinchu, Taiwan 
\item  Institute for Astronomy, University of Hawai'i, 2680 Woodlawn Drive, Honolulu, HI 96822, USA  
\item  Citra Space Corporation, 4815 List Dr. Ste 102, Colorado Springs, CO 80919, USA 
\item  School of Physics, The University of Melbourne, VIC 3010, Australia  
\item The Observatories of the Carnegie Institution for Science, 813 Santa Barbara St., Pasadena, CA 91101, USA  
\item Department of Physics, Florida State University, 77 Chieftan Way, Tallahassee, FL 32306, USA  
\item National Science Foundation Graduate Research Fellowship  
\item Institute for Astronomy, University of Hawai'i, 640 N. A'ohoku Pl., Hilo, HI 96720, USA  
\item Tuorla observatory, Department of Physics and Astronomy, University of Turku, 20014 Turku, Finland  
\item  Department of Physics, University of Warwick, Gibbet Hill Road, Coventry CV4 7AL, UK 
\item Finnish Centre for Astronomy with ESO (FINCA), FI-20014 University of Turku, Finland  
\item Graduate Institute of Applied Physics, National Chengchi University, Taipei 116026, Taiwan  
\item Centre for Astrophysics and Supercomputing, Swinburne University of Technology, John St, Hawthorn, VIC 3122, Australia 
\item Space Telescope Science Institute, 3700 San Martin Drive, Baltimore, MD 21218, USA 
\item Las Campanas Observatory, Carnegie Observatories, Casilla 601, La Serena, Chile  
\item Niels Bohr Institute, University of Copenhagen, Jagtvej 128, DK-2200, Copenhagen N, Denmark  
\item Cosmic Dawn Center (DAWN)  
\item Department of Astronomy, Beijing Normal University, Beijing, China   
\item School of Physics and Technology, Wuhan University, Wuhan 430072, China   

\end{affiliations}
\end{small}

\clearpage

\begin{abstract}
{
When a massive star explodes as a supernova, crucial information about its immediate environment is lost within hours. Here we report rapid optical observations from Lulin Observatory of the broad-lined Type Ic supernova SN~2026gzf, beginning 1.25\,hours after Einstein Probe detected the X-ray transient EP260321a. Our data led to the discovery of the optical counterpart and showed a luminous blue first-day excess that cannot be reproduced by standard radioactive models. We find that interaction between the ejecta and $\approx 0.02$\,\msun\ of circumstellar material accounts for the early excess. Archival Panoramic Survey Telescope and Rapid Response System (Pan-STARRS) images show variability at the explosion site over the previous $\sim 12$\,years, with the source brightening by a factor of $\sim 1.5$ in the final $\sim 3$\,years before explosion, providing rare evidence for pre-explosion activity in a stripped-envelope progenitor system.
The precursor brightening suggests enhanced eruptive mass loss during late-stage oxygen burning before core collapse, while an additional silicon-burning episode shortly before explosion may have created the compact nearby material responsible for the X-ray shock-breakout signal. SN~2026gzf therefore offers the first view of how a stripped progenitor modifies its immediate environment shortly before death, linking long-term precursor variability, circumstellar interaction and the explosion itself.
}
\end{abstract}

\bigskip

The first photons emerge from a supernova (SN) explosion when the blast wave launched in the core reaches the stellar surface. 
For typical stellar radii ($0.1-1000$ solar radii), this results in a shock breakout transient lasting only seconds to hours, and a spectrum that evolves rapidly from X-ray to optical wavelengths as the shocked ejecta adiabatically expand. The emission encodes the pre-explosion structure of the star and its immediate environment, but detection is extremely challenging due to the rapid timescale, especially for `stripped-envelope' Type Ib and Ic SNe (explosions of relatively compact stars that have lost their hydrogen and sometimes helium layers). Only a handful of stripped-envelope SNe have ever been caught in X-rays at shock breakout, most famously SN~2006aj\cite{2006Natur.442.1008C,2006Natur.442.1011P,2006Natur.442.1018M} and SN~2008D\cite{2008Natur.453..469S,2008Sci...321.1185M}.

\begin{figure*}
    \centering
    
    \begin{subfigure}{\linewidth}
        \centering
        \includegraphics[width=\linewidth]{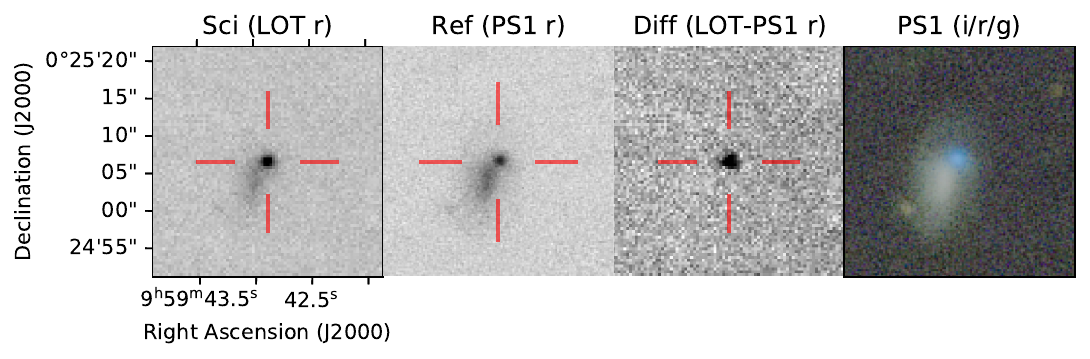}
    \end{subfigure}
    
    \vspace{0.5em}
    
    \begin{subfigure}{\linewidth}
        \centering
        \includegraphics[width=\linewidth]{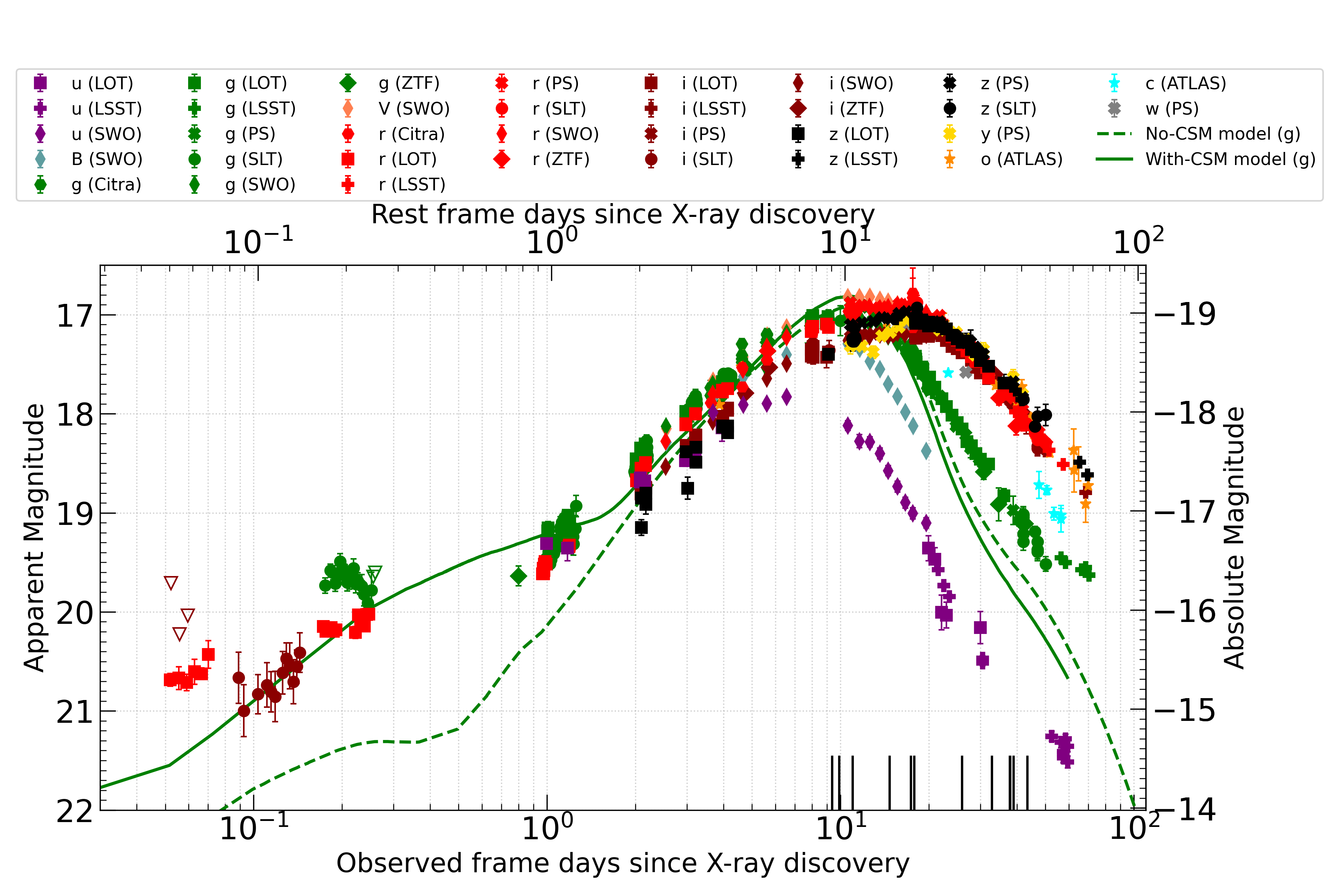}
    \end{subfigure}
    
    \caption{\textbf{Discovery and early optical evolution of SN~2026gzf.} 
    \textbf{Upper panel,} From left to right: the LOT $r$-band discovery image obtained by Kinder, the Pan-STARRS $r$-band reference image, the corresponding difference image, and a Pan-STARRS colour composite ($i$/$r$/$g$). The red cross marks the position of SN~2026gzf. The transient position is coincident with a blue compact source. \textbf{Lower panel,} Multi-band light curves of SN~2026gzf, including the earliest Lulin follow-up observations and later survey photometry. The logarithmic x-axis highlights the high-cadence first-day light curve and the rapid early optical rise following the X-ray trigger. All the magnitudes in this figure are corrected for galactic extinction assuming A$_{\rm V}$ = 0.067 mag. Upper limits in corresponding filters are shown by downward open triangles. {\bf The $g$-band hydrodynamic models with and without CSM are also indicated.} }
    \label{fig:discovery_lc}
\end{figure*}

On 21st March 2026 at 12:23 UTC (MJD 61120.516; hereafter T$_{0}$) the Einstein Probe (EP) satellite\cite{2022hxga.book...86Y} detected the fast X-ray transient EP260321a\cite{2026GCN.44068....1H}, which was rapidly identified as a luminous soft X-ray source consistent with SN shock breakout\cite{2026GCN.44075....1H} (W.~Yuan et al. in prep.; see Methods\,\ref{sec_med:SBO}). Within 1.25\,hours of the EP detection, we initiated rapid optical follow-up observations of the field with the Lulin observatory\cite{2026GCN.44070....1L}, carried out as part of the Kinder program\cite{2025ApJ...983...86C} (Methods\,\ref{sec_med:OT}). In our images we identified a blue, rising transient at $\alpha = [\mathrm{09{:}59{:}42.88}]$,
$\delta = [+\mathrm{00{:}25{:}06.11}]$ (Fig.\,\ref{fig:discovery_lc}, upper panel). The transient is spatially coincident with both a resolved galaxy at redshift\cite{2001MNRAS.328.1039C} $z=0.0345$ and a very blue point-like source
visible in archival images from the Panoramic Survey Telescope and Rapid Response System (Pan-STARRS)\cite{2016arXiv161205560C}. Our dense photometric monitoring over the course of the night showed that the source brightened by 0.66\,mag in the $r$ band within 4.68\,hours, and confirmed the transient as the optical counterpart to EP260321a\cite{2026GCN.44081....1A,2026GCN.44089....1S}. 
We continued observing over the subsequent weeks using a range of ground-based telescopes, capturing in detail the rise, peak and decline of the optical light curve (Fig.\,\ref{fig:discovery_lc}, lower panel). All photometry was performed after subtraction of host-galaxy template images to extract the true SN flux (see Methods\,\ref{sec_med:data_collection}).

\begin{figure}
    \centering
    \includegraphics[width=\linewidth]{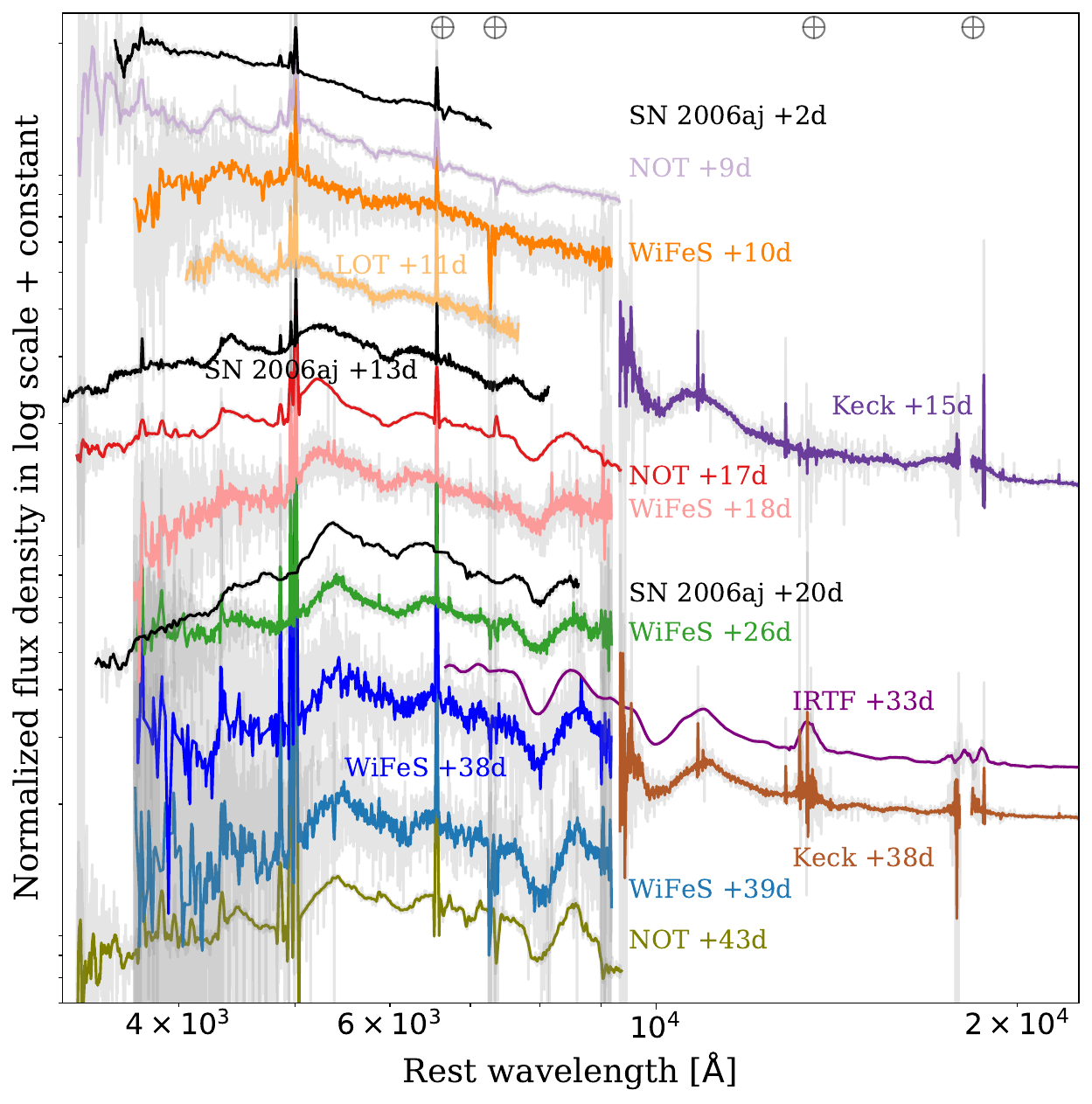}
    \caption{\textbf{Spectroscopic evolution of SN~2026gzf.} Rest-frame spectral sequence of SN~2026gzf from $+9$ to $+43$ days relative to the X-ray trigger ($T_{0}$ = MJD = 61120.516). Optical spectra from NOT, ANU 2.3-m/WiFeS and LOT are shown together with NIR spectra from Keck II/NIRES and IRTF/SpeX. All spectra are normalized and vertically offset for clarity. The coloured curves show the displayed binned spectra, while the grey curves indicate the original unbinned data. Earth symbols mark wavelength regions strongly affected by telluric absorption. Narrow host-galaxy H\,II-region emission lines are superposed on the SN features. Spectra of SN~2006aj at similar phases are overplotted for comparison.}
    \label{fig:spectra}
\end{figure}

Spectroscopic follow-up established that the transient was an emerging broad-lined Type\,Ic SN\cite{2026GCN.44092....1X, 2026GCN.44105....1C, 2026GCN.44107....1R} (A.~Martin-Carrillo et al. in prep.) at a redshift consistent with the nearby galaxy, linking the X-ray flash directly to the onset of a stripped-envelope SN. This was reported to the International Astronomical Union Transient Name Server and given the name SN~2026gzf\footnote{https://www.wis-tns.org/object/2026gzf}. We obtained optical spectra from the Nordic Optical Telescope (NOT)\cite{2010ASSP...14..211D}, the Australian National University (ANU) 2.3-m telescope\cite{2007Ap&SS.310..255D} and the LOT, and near-infrared (NIR) spectra from the W. M. Keck Observatory Telescope II and the 3.2-m NASA Infrared Telescope Facility (IRTF). The spectra of SN~2026gzf (Fig.\,\ref{fig:spectra}) show a close resemblance to broad-lined Type Ic SNe such as SN~2006aj\cite{2008AstBu..63..228S,2006ApJ...645L..21M,2006Natur.442.1011P}.
The spectroscopic modelling using \textsc{TARDIS} \cite{2014MNRAS.440..387K} indicated the contributions to spectra from Mg\,{\sc i}, Mg~{\sc ii}, Ca\,{\sc ii}, Fe\,{\sc ii}, and He\,{\sc i} (in the NIR), see Methods\,\ref{sec_med:spec_2026gzf}. 

While the redshift inferred from the spectrum of SN~2026gzf is consistent with the nearby galaxy, its coordinates are also consistent, within uncertainties, with an underlying blue compact source seen in Pan-STARRS (see Fig.\,\ref{fig:discovery_lc}, upper panel) as well as in images from the Sloan Digital Sky Survey\cite{2000AJ....120.1579Y} and the Dark Energy Spectroscopic Instrument Legacy Imaging Surveys (DESI-LS)\cite{2019AJ....157..168D} (see Methods\,\ref{sec_med:pre-burst activities}). 
Direct aperture photometry on archival images from Pan-STARRS, ZTF, Subaru and DESI-LS further shows that the site hosts a persistent baseline source with mutually consistent flux levels across surveys, and that this source is systematically brighter in the blue than in the redder optical bands.
The apparent point source could be a compact star-forming region or unresolved dwarf galaxy. A foreground star is disfavored as the source shows no measurable proper motion in Gaia Data Release 3\cite{gaia2023dr3}. 
Moreover, the probability of a chance alignment with an unrelated foreground-like point source of comparable brightness is low, being \(<10^{-4}\) in the \(r\) band for our fiducial calculation and remaining \(<1.4\times10^{-3}\) under substantially more conservative assumptions.
The archival blue compact source, the Pan-STARRS precursor residuals, and the later SN positions are all mutually consistent within the astrometric uncertainties, supporting a common physical site.

\begin{figure}
    \centering
    \includegraphics[width=\linewidth]{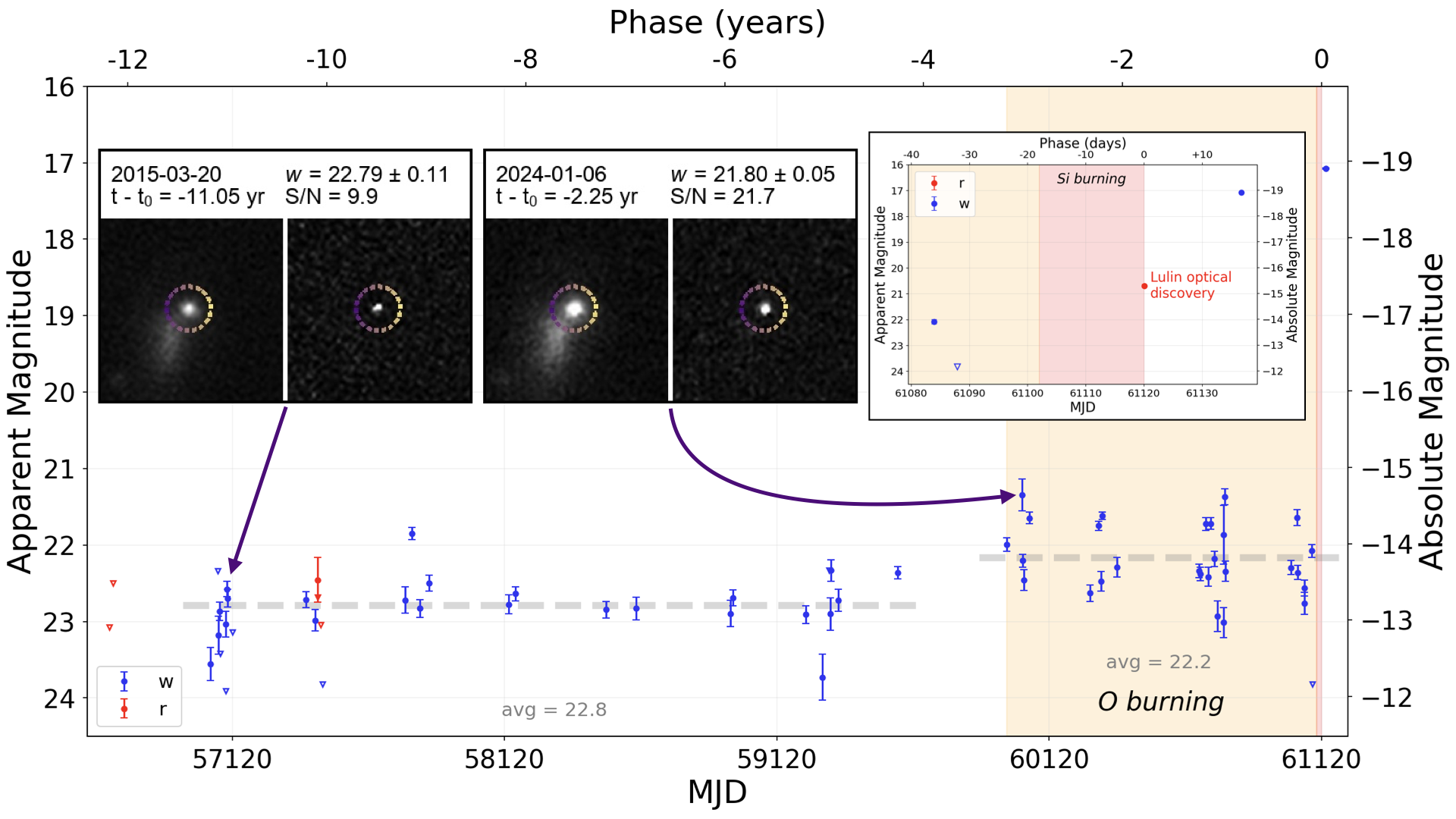}
    \caption{\textbf{Precursor activities of SN~2026gzf.} 
    Historical Pan-STARRS $w$-band data show variability at the progenitor site over the final $\sim12$\,years before the SN explosion. Owing to the lack of earlier w-band coverage, we cannot exclude the possibility that similar activity was already present at earlier times. 
    The precursor brightened from an average $w\approx22.8$\,mag ($M_w \approx -13.2$\,mag) during $-12$ to $-3$\,yr before explosion to $w\approx22.2$\,mag ($M_w \approx -13.8$) over the final $\sim3$\,yr. This is consistent with enhanced mass loss on a timescale comparable to late oxygen burning in a $\sim15$\,\msun\ progenitor. 
    An additional, more compact mass-loss episode during the final silicon-burning stage ($\sim18$\,d) may have produced the dense nearby circumstellar material responsible for the X-ray shock-breakout signal, although this period is not covered by our observations.   
     The upper inset panels show $0.28' \times 0.28'$ science and template-subtracted images at the site of SN~2026gzf for selected Pan-STARRS $w$-band data (north up, east to the left). }
    \label{fig:ps1_precursor}
\end{figure}

Remarkably, archival photometry shows that the blue compact source was variable even prior to the SN. We analyse 60 epochs of Pan-STARRS imaging in the broad optical $w$ filter obtained between 2015 and 2026, using a deep stack of earlier Pan-STARRS images as a reference (see Methods\,\ref{sec_med:pre-burst activities}). Image differencing with respect to the reference reveals variable excess flux at this position during the 12\,years before the EP trigger (confirmed using additional images in the $r$ filter). The excess is not constant, but rather exhibits significant pre-explosion variability (Fig.~\ref{fig:ps1_precursor}) with no evidence of periodicity. 
This further disfavors an ordinary foreground variable star as the origin of the blue compact source and instead points to pre-explosion activity from the SN progenitor embedded within a compact star-forming region. 

Most intriguingly, we find that the precursor activity became brighter toward the final years before explosion. The precursor had an average brightness of $w\approx22.8$\,mag (this implies an absolute magnitude of $M_w \approx -13.2$\,mag at $z=0.0345$) from $-12$ to $-3$\,yr before explosion, but then brightened significantly to $w\approx22.2$\,mag ($M_w \approx -13.8$) over the final $\sim3$\,yr, corresponding to an increase in luminosity by a factor of $\sim1.5$. This trend suggests that the progenitor underwent enhanced mass ejection and mass loss during the last few years before core collapse. This timescale is broadly consistent with the expected duration of oxygen burning in a $\sim15$\,\msun\
progenitor\cite{2002RvMP...74.1015W,2017MNRAS.467.3347K}. For a progenitor of similar mass, silicon burning lasts only $\sim18$\,d, raising the possibility of an even more violent episode of mass loss shortly before explosion, although we have no observations over that interval. Material expelled on such timescales could naturally account for the compact circumstellar material (CSM) responsible for the EP shock-breakout signal. Taking the X-ray peak timescale of $\sim700$\,s, the shock-breakout-producing CSM radius was estimated to be around $3\times10^{2}$ solar radii (W.~Yuan et al. in prep.), consistent with the idea that the dense CSM was located very close to the progenitor at the time of explosion.

Our earliest Lulin optical data were obtained during the transition from high-energy breakout emission to the initial optical rise of SN~2026gzf (Fig.~\ref{fig:discovery_lc}, bottom panel). 
While the main rise to peak is powered by the decay of radioactive $^{56}$Ni\cite{1982ApJ...253..785A}, the earliest emission can in principle also contain a contribution from shock-cooling of the outer ejecta, which is expected to be faint and short-lived for the compact progenitors of Type~Ic SNe\cite{2017hsn..book..967W}. 
Our data within the first few hours after explosion show that SN~2026gzf is more than 1\,mag brighter than expected compared to a hydrodynamic model (see Methods \ref{sec_med:early_excess}), which is powered only by $^{56}$Ni (Fig.\,\ref{fig:discovery_lc}, bottom panel). 
While the main peak is well reproduced with this model, this excess suggests a further powering source is required in the first few hours. 

We investigate a simple spherical hydrodynamic model in which the SN ejecta interact with dense CSM.
Based on the shock breakout signals observed by EP, we adopted the CSM radius of $3\times 10^{13}$\,cm (W.~Yuan et al. in prep.).
Our numerical light-curve modelling shows that the early luminosity excess observed in SN~2026gzf can be explained by ejecta-CSM interaction if the CSM mass is $\sim 0.02$\,\msun, assuming a wind-like CSM density profile (see Methods\,\ref{sec_med:LC_modeling}). 
For a typical wind velocity of Wolf-Rayet stars of 1000\,$\mathrm{km~s^{-1}}$, as appropriate for a commonly invoked progenitor channel of stripped-envelope SNe including some Type Ic events, this implies the mass-loss rate of the progenitor was as high as $2\,\mathrm{M_\odot~yr^{-1}}$ in the last 3.5\,days before the explosion. Such a high mass-loss rate cannot be achieved by the standard wind mass loss\cite{2004ApJ...616..525O}, and some dynamical processes likely drive the extreme mass loss\cite{2018MNRAS.476.1853F}.
Independently, a simple $^{56}$Ni-powered MOSFiT model reproduces the main multi-band light curve well, except for the luminous excess during the first day after explosion (Methods\,\ref{sec_med:mosfit}), supporting the view that the main peak is powered by radioactive heating while the earliest emission requires an additional interaction component.

This estimate assumes spherical symmetry, whereas many broad-line Type Ic SNe are accompanied by gamma-ray bursts (GRBs), known to drive collimated jets. The early emission (including the X-rays) from SN~2026gzf does not resemble a typical GRB afterglow. Dense CSM could also help to explain the lack of GRB signature, causing the jet to stall\cite{2025ApJ...986L...4H}, as the further analysis by A.~Martin-Carrillo, et al. (in prep.) finds no signature of a relativistic jet in SN~2026gzf. 
The X-ray shock breakout radiation exhibits a simple blackbody spectrum with the temperature around one million Kelvin indicating the presence of a CSM shell with a radius about 300 solar radii (W.~Yuan et al. in prep.).
In some cases, the energy that heats the CSM could be injected by the jet -- so-called cocoon emission -- which may also explain the early flux excess\cite{2025ApJ...985...21Z}. In either case, the existence of the dense CSM is required to explain the early flux excess. 

\begin{figure}
    \centering
    \includegraphics[width=\linewidth]{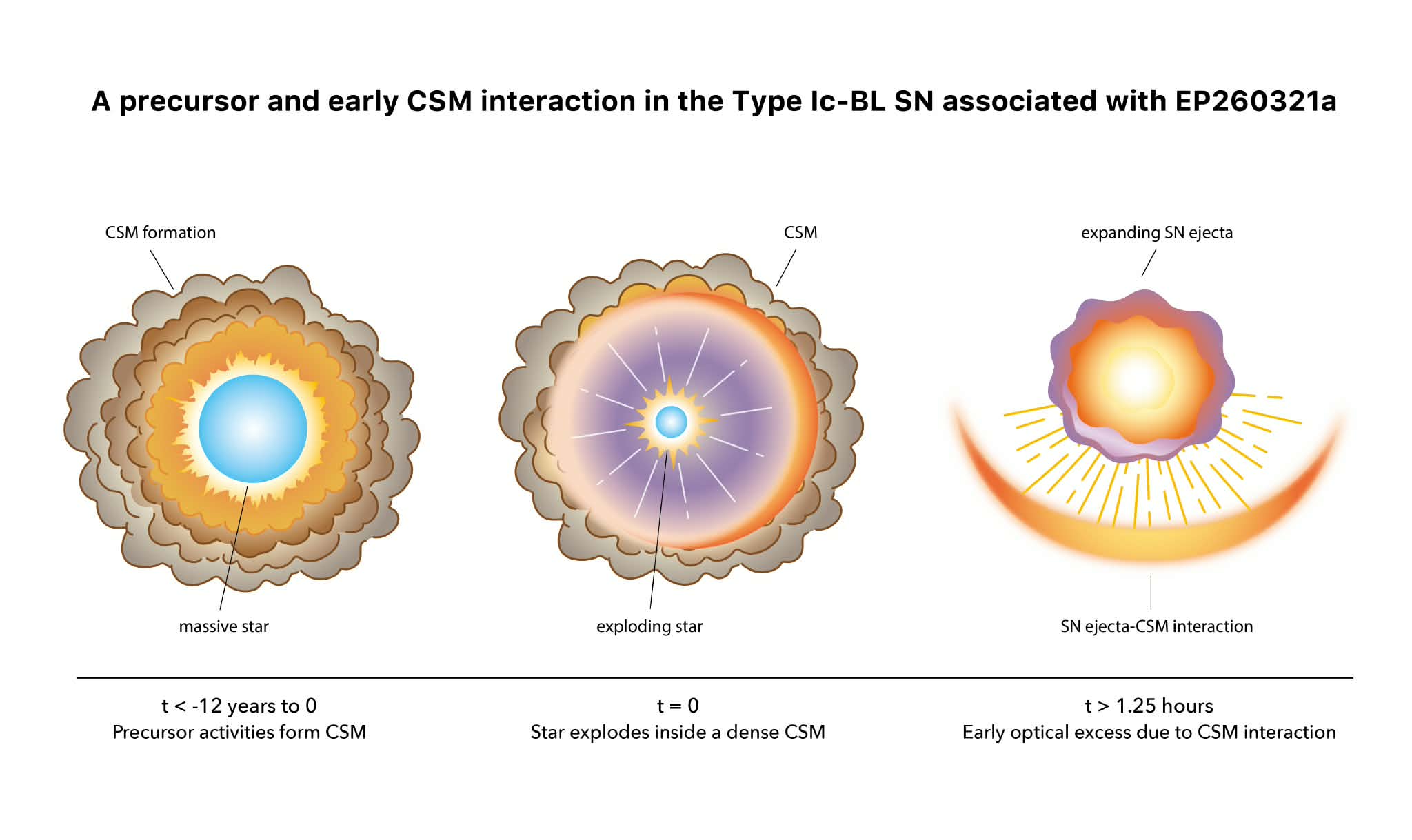}
    \caption{\textbf{Schematic scenario for SN~2026gzf.} \textbf{Left,} precursor activity over at least the $\sim12$\,years before explosion produces CSM around the progenitor system. \textbf{Middle,} at t=0, the star explodes inside this nearby dense CSM, giving rise to the soft X-ray flash detected by Einstein Probe. \textbf{Right,} by t$>1.25$\,h, interaction between the rapidly expanding SN ejecta and the nearby CSM contributes to the bright, blue first-day optical excess. The schematic therefore links the decade-long precursor variability, the EP X-ray trigger, the optical counterpart discovery at 1.25\,h, the blue first-day excess, and the subsequent rise into a normal Ic-BL SN.}
    \label{fig:scenario}  
\end{figure}

Our simple CSM interaction picture naturally explains the early excess in the optical wavelengths and provides an origin for this CSM through late-time variability of the progenitor. This interpretation also explains the unexpectedly long duration of the soft X-ray flash detected by EP, as the shock breakout could occur in extended dense CSM. Together, the optical modelling and the X-ray properties indicate that SN~2026gzf exploded in a compact and complex environment shaped by recent pre-supernova mass loss. Figure~\ref{fig:scenario} presents a schematic illustration of this scenario. 
Moreover, the weak radio detection at $+13.6$\,d\cite{2026GCN.44239....1O} may support the possibility of the presence of CSM shells formed by sporadic stronger precursor activities (see Methods\,\ref{sec_med:csm_shells}). 

This is not the first Type Ic SN to exhibit brighter-than-expected shock cooling signatures, and indeed the early emission from these events can be quite diverse (see Methods \ref{sec_med:comparison}). Some such as SN~2020lao\cite{2026A&A...708A.305S} exhibit no shock-cooling signatures, while others like SN~2018gep\cite{2019ApJ...887..169H} also shows very bright shock-cooling emission. SN~2018gep also exhibited precursor emission from the progenitor, though this was only detected in the weeks (rather than years) immediately before explosion. The light curve of SN~2026gzf around peak most closely resembles prototypical broad-lined Type Ic SNe such as SN~2006aj\cite{2014ApJS..213...19B}.
However, most of such objects are discovered either via GRB follow-up, in which case the GRB afterglow complicates any measurement of the early SN emission, or through wide-field optical surveys, which do not precisely measure the moment of explosion. 
The soft X-ray discovery and early optical follow-up of SN~2026gzf result in a uniquely well constrained explosion time and early light curve.

SN~2026gzf shows unusually luminous and pronounced pre-explosion variability. 
Its precursor reaches $\sim -14$\,mag, comparable only to the remarkably bright precursor of SN~1961V\cite{2011ApJ...737...76K}, while its short-timescale fluctuations resemble the strong variability seen in SN~2009ip\cite{2013ApJ...767....1P}, with variations of $>0.5$\,mag and up to $\sim3$\,mag (see Methods\,\ref{sec_med:comparison_precursor}). Although these comparison objects likely arose from different progenitor channels, they suggest that violent variability may also occur in stripped, Wolf–Rayet-like progenitor systems. Such decade-long precursor activity has not previously been detected in a stripped-envelope progenitor, 
besides they are intrinsically rare also largely because the source remains faint, at around (apparent) 23\,mag, below the depth routinely reached by most current wide-field surveys. 
SN~2026gzf may therefore represent the first BL-Ic SN with a decade-long pre-explosion activities reaching such an extreme luminosity, which offers a timely preview of the long-term precursor behaviour that future deep surveys such as Rubin LSST may uncover in much larger numbers.  

The local environment of SN~2026gzf provides further insight into its progenitor. We obtained WiFeS integral-field unit (IFU) spectroscopy to examine the integrated host galaxy and the compact star-forming region where the SN occurred. Modelling the spectral energy distributions with \textsc{prospector}\cite{2021ApJS..254...22J} (see Methods\,\ref{sec_med:host}) suggests stellar masses of $\log_{10}(M_\ast/{\rm M}_\odot)=9.04^{+0.09}_{-0.08}$ for the galaxy and $\log_{10}(M_\ast/{\rm M}_\odot)=8.09^{+0.23}_{-0.07}$ for the blue compact source. The global host properties are consistent with a low-metallicity star-forming system, with a dust-corrected star-formation rate of $0.37 \pm 0.08$\,\msun\,yr$^{-1}$ and a gas-phase metallicity of $12+\log(\mathrm{O/H})=8.00\pm0.03$. The spectrum extracted at the SN position, albeit contaminated by SN light, yields a similarly low metallicity of $12+\log(\mathrm{O/H})=8.05\pm0.04$ and a local star-formation rate of $0.17 \pm 0.04$\,\msun\,yr$^{-1}$.
The latter values are based on a 3x3-spaxel extraction and are therefore not directly comparable to the global measurements unless the same aperture definition is adopted. Moreover, the data do not allow a clean separation of the blue compact source from the surrounding host emission, and the local measurements should be regarded only as representative of the SN vicinity rather than the exact SN sightline. Overall, however, it is clear from the WiFeS data that SN~2026gzf exploded in a low-metallicity star-forming environment, with $\sim0.23\,{\rm Z}_\odot$ and a high specific star-formation rate of $\sim1.37$\,Gyr$^{-1}$, placing the SN site at the low-metallicity end of the broad-lined Type Ic SN host population\cite{2018A&A...617A.105J,2020ApJ...892..153M} (see also A.~Martin-Carrillo et al. in prep.).

Taken together, our rapid-response optical observations of SN~2026gzf, the EP shock-breakout detection, and deep archival imaging provide a rare view of a stripped-envelope SN progenitor from the final decade of its life to the moment of explosion. Rather than representing three disconnected observables, the X-ray flash, the blue first-day excess, and the decade-long precursor variability point to a single picture in which the progenitor system underwent enhanced activity before collapse and exploded into nearby material produced shortly beforehand.
The historical light curve demonstrates that SN~2026gzf experienced dramatic variability and enhanced mass loss in the years before core collapse, with a clear increase in precursor luminosity during the final $\sim3$ years.
The CSM produced by this late-phase instability shaped both the shock breakout signal and the earliest optical light curve.
SN~2026gzf thus provides direct evidence that pre-SN instability and early circumstellar interaction are physically linked in at least some broad-lined Type Ic explosions. Compared with the small number of similarly early events, SN~2026gzf strengthens the emerging view that broad-lined Type~Ic SNe are not homogeneous at the earliest times. Rapid optical follow-up of X-ray-selected explosions will therefore be critical for probing the diversity of pre-SN mass-loss histories and their implications for massive-star evolution\cite{2025ApJ...978L..21S}.

\clearpage
\noindent\textbf{Figures.} \\

\clearpage

\noindent {\bf References}
\bibliographystyle{naturemag}
\bibliography{astroph_references}

\begin{thebibliography}{10}
\expandafter\ifx\csname url\endcsname\relax
  \def\url#1{\texttt{#1}}\fi
\expandafter\ifx\csname urlprefix\endcsname\relax\def\urlprefix{URL }\fi
\providecommand{\bibinfo}[2]{#2}
\providecommand{\eprint}[2][]{\url{#2}}

\bibitem{2006Natur.442.1008C}
\bibinfo{author}{{Campana}, S.} \emph{et~al.}
\newblock \bibinfo{title}{{The association of GRB 060218 with a supernova and the evolution of the shock wave}}.
\newblock \emph{\bibinfo{journal}{\nat}} \textbf{\bibinfo{volume}{442}}, \bibinfo{pages}{1008--1010} (\bibinfo{year}{2006}).
\newblock \eprint{astro-ph/0603279}.

\bibitem{2006Natur.442.1011P}
\bibinfo{author}{{Pian}, E.} \emph{et~al.}
\newblock \bibinfo{title}{{An optical supernova associated with the X-ray flash XRF 060218}}.
\newblock \emph{\bibinfo{journal}{\nat}} \textbf{\bibinfo{volume}{442}}, \bibinfo{pages}{1011--1013} (\bibinfo{year}{2006}).
\newblock \eprint{astro-ph/0603530}.

\bibitem{2006Natur.442.1018M}
\bibinfo{author}{{Mazzali}, P.~A.} \emph{et~al.}
\newblock \bibinfo{title}{{A neutron-star-driven X-ray flash associated with supernova SN 2006aj}}.
\newblock \emph{\bibinfo{journal}{\nat}} \textbf{\bibinfo{volume}{442}}, \bibinfo{pages}{1018--1020} (\bibinfo{year}{2006}).
\newblock \eprint{astro-ph/0603567}.

\bibitem{2008Natur.453..469S}
\bibinfo{author}{{Soderberg}, A.~M.} \emph{et~al.}
\newblock \bibinfo{title}{{An extremely luminous X-ray outburst at the birth of a supernova}}.
\newblock \emph{\bibinfo{journal}{\nat}} \textbf{\bibinfo{volume}{453}}, \bibinfo{pages}{469--474} (\bibinfo{year}{2008}).
\newblock \eprint{0802.1712}.

\bibitem{2008Sci...321.1185M}
\bibinfo{author}{{Mazzali}, P.~A.} \emph{et~al.}
\newblock \bibinfo{title}{{The Metamorphosis of Supernova SN 2008D/XRF 080109: A Link Between Supernovae and GRBs/Hypernovae}}.
\newblock \emph{\bibinfo{journal}{Science}} \textbf{\bibinfo{volume}{321}}, \bibinfo{pages}{1185} (\bibinfo{year}{2008}).
\newblock \eprint{0807.1695}.

\bibitem{2022hxga.book...86Y}
\bibinfo{author}{{Yuan}, W.}, \bibinfo{author}{{Zhang}, C.}, \bibinfo{author}{{Chen}, Y.} \& \bibinfo{author}{{Ling}, Z.}
\newblock \bibinfo{title}{{The Einstein Probe Mission}}.
\newblock In \bibinfo{editor}{{Bambi}, C.} \& \bibinfo{editor}{{Sangangelo}, A.} (eds.) \emph{\bibinfo{booktitle}{Handbook of X-ray and Gamma-ray Astrophysics}}, \bibinfo{pages}{86} (\bibinfo{year}{2022}).

\bibitem{2026GCN.44068....1H}
\bibinfo{author}{{Huang}, Q.~J.} \emph{et~al.}
\newblock \bibinfo{title}{{EP260321a: Einstein Probe detection of an X-ray transient}}.
\newblock \emph{\bibinfo{journal}{GRB Coordinates Network}} \textbf{\bibinfo{volume}{44068}}, \bibinfo{pages}{1} (\bibinfo{year}{2026}).

\bibitem{2026GCN.44075....1H}
\bibinfo{author}{{Huang}, Q.~J.} \emph{et~al.}
\newblock \bibinfo{title}{{EP260321a: refined analysis of the EP-WXT and EP-FXT observations, implying a possible supernova shock breakout candidate}}.
\newblock \emph{\bibinfo{journal}{GRB Coordinates Network}} \textbf{\bibinfo{volume}{44075}}, \bibinfo{pages}{1} (\bibinfo{year}{2026}).

\bibitem{2026GCN.44070....1L}
\bibinfo{author}{{Lee}, M.-H.} \emph{et~al.}
\newblock \bibinfo{title}{{EP260321a: Kinder observations detect a blue variable star and set limits on a source from the z =0.034 galaxy within the error circle}}.
\newblock \emph{\bibinfo{journal}{GRB Coordinates Network}} \textbf{\bibinfo{volume}{44070}}, \bibinfo{pages}{1} (\bibinfo{year}{2026}).

\bibitem{2025ApJ...983...86C}
\bibinfo{author}{{Chen}, T.-W.} \emph{et~al.}
\newblock \bibinfo{title}{{Discovery and Extensive Follow-up of SN 2024ggi, a Nearby Type IIP Supernova in NGC 3621}}.
\newblock \emph{\bibinfo{journal}{\apj}} \textbf{\bibinfo{volume}{983}}, \bibinfo{pages}{86} (\bibinfo{year}{2025}).
\newblock \eprint{2406.09270}.

\bibitem{2001MNRAS.328.1039C}
\bibinfo{author}{{Colless}, M.} \emph{et~al.}
\newblock \bibinfo{title}{{The 2dF Galaxy Redshift Survey: spectra and redshifts}}.
\newblock \emph{\bibinfo{journal}{\mnras}} \textbf{\bibinfo{volume}{328}}, \bibinfo{pages}{1039--1063} (\bibinfo{year}{2001}).
\newblock \eprint{astro-ph/0106498}.

\bibitem{2016arXiv161205560C}
\bibinfo{author}{{Chambers}, K.~C.} \emph{et~al.}
\newblock \bibinfo{title}{{The Pan-STARRS1 Surveys}}.
\newblock \emph{\bibinfo{journal}{arXiv e-prints}} \bibinfo{pages}{arXiv:1612.05560} (\bibinfo{year}{2016}).
\newblock \eprint{1612.05560}.

\bibitem{2026GCN.44081....1A}
\bibinfo{author}{{Aryan}, A.} \emph{et~al.}
\newblock \bibinfo{title}{{EP260321a: further Kinder optical observations}}.
\newblock \emph{\bibinfo{journal}{GRB Coordinates Network}} \textbf{\bibinfo{volume}{44081}}, \bibinfo{pages}{1} (\bibinfo{year}{2026}).

\bibitem{2026GCN.44089....1S}
\bibinfo{author}{{Sankar. K}, A.} \emph{et~al.}
\newblock \bibinfo{title}{{EP260321a: Continued brightening from Kinder follow-up of the emerging supernova candidate}}.
\newblock \emph{\bibinfo{journal}{GRB Coordinates Network}} \textbf{\bibinfo{volume}{44089}}, \bibinfo{pages}{1} (\bibinfo{year}{2026}).

\bibitem{2026GCN.44092....1X}
\bibinfo{author}{{Xu}, D.} \emph{et~al.}
\newblock \bibinfo{title}{{EP260321a / AT2026gzf: VLT/X-shooter detection of supernova-like spectral features at z = 0.0344}}.
\newblock \emph{\bibinfo{journal}{GRB Coordinates Network}} \textbf{\bibinfo{volume}{44092}}, \bibinfo{pages}{1} (\bibinfo{year}{2026}).

\bibitem{2026GCN.44105....1C}
\bibinfo{author}{{Corcoran}, G.} \emph{et~al.}
\newblock \bibinfo{title}{{EP260321a: VLT/FORS2 spectroscopy confirmation of an associated type Ic-BL supernova SN 2026gzf}}.
\newblock \emph{\bibinfo{journal}{GRB Coordinates Network}} \textbf{\bibinfo{volume}{44105}}, \bibinfo{pages}{1} (\bibinfo{year}{2026}).

\bibitem{2026GCN.44107....1R}
\bibinfo{author}{{Rastinejad}, J.}, \bibinfo{author}{{Srinivasaragavan}, G.} \& \bibinfo{author}{{Ahumada}, T.}
\newblock \bibinfo{title}{{EP260321a: Gemini-South confirmation of a supernova}}.
\newblock \emph{\bibinfo{journal}{GRB Coordinates Network}} \textbf{\bibinfo{volume}{44107}}, \bibinfo{pages}{1} (\bibinfo{year}{2026}).

\bibitem{2010ASSP...14..211D}
\bibinfo{author}{{Djupvik}, A.~A.} \& \bibinfo{author}{{Andersen}, J.}
\newblock \bibinfo{title}{{The Nordic Optical Telescope}}.
\newblock In \bibinfo{editor}{{Diego}, J.~M.}, \bibinfo{editor}{{Goicoechea}, L.~J.}, \bibinfo{editor}{{Gonz{\'a}lez-Serrano}, J.~I.} \& \bibinfo{editor}{{Gorgas}, J.} (eds.) \emph{\bibinfo{booktitle}{Highlights of Spanish Astrophysics V}}, vol.~\bibinfo{volume}{14} of \emph{\bibinfo{series}{Astrophysics and Space Science Proceedings}}, \bibinfo{pages}{211} (\bibinfo{year}{2010}).
\newblock \eprint{0901.4015}.

\bibitem{2007Ap&SS.310..255D}
\bibinfo{author}{{Dopita}, M.} \emph{et~al.}
\newblock \bibinfo{title}{{The Wide Field Spectrograph (WiFeS)}}.
\newblock \emph{\bibinfo{journal}{\apss}} \textbf{\bibinfo{volume}{310}}, \bibinfo{pages}{255--268} (\bibinfo{year}{2007}).
\newblock \eprint{0705.0287}.

\bibitem{2008AstBu..63..228S}
\bibinfo{author}{{Sonbas}, E.} \emph{et~al.}
\newblock \bibinfo{title}{{Stellar-wind envelope around the massive supernova progenitor XRF/GRB 060218/SN 2006aj}}.
\newblock \emph{\bibinfo{journal}{Astrophysical Bulletin}} \textbf{\bibinfo{volume}{63}}, \bibinfo{pages}{228--243} (\bibinfo{year}{2008}).
\newblock \eprint{0805.2657}.

\bibitem{2006ApJ...645L..21M}
\bibinfo{author}{{Modjaz}, M.} \emph{et~al.}
\newblock \bibinfo{title}{{Early-Time Photometry and Spectroscopy of the Fast Evolving SN 2006aj Associated with GRB 060218}}.
\newblock \emph{\bibinfo{journal}{\apjl}} \textbf{\bibinfo{volume}{645}}, \bibinfo{pages}{L21--L24} (\bibinfo{year}{2006}).
\newblock \eprint{astro-ph/0603377}.

\bibitem{2014MNRAS.440..387K}
\bibinfo{author}{{Kerzendorf}, W.~E.} \& \bibinfo{author}{{Sim}, S.~A.}
\newblock \bibinfo{title}{{A spectral synthesis code for rapid modelling of supernovae}}.
\newblock \emph{\bibinfo{journal}{\mnras}} \textbf{\bibinfo{volume}{440}}, \bibinfo{pages}{387--404} (\bibinfo{year}{2014}).
\newblock \eprint{1401.5469}.

\bibitem{2000AJ....120.1579Y}
\bibinfo{author}{{York}, D.~G.} \emph{et~al.}
\newblock \bibinfo{title}{{The Sloan Digital Sky Survey: Technical Summary}}.
\newblock \emph{\bibinfo{journal}{\aj}} \textbf{\bibinfo{volume}{120}}, \bibinfo{pages}{1579--1587} (\bibinfo{year}{2000}).
\newblock \eprint{astro-ph/0006396}.

\bibitem{2019AJ....157..168D}
\bibinfo{author}{{Dey}, A.} \emph{et~al.}
\newblock \bibinfo{title}{{Overview of the DESI Legacy Imaging Surveys}}.
\newblock \emph{\bibinfo{journal}{\aj}} \textbf{\bibinfo{volume}{157}}, \bibinfo{pages}{168} (\bibinfo{year}{2019}).
\newblock \eprint{1804.08657}.

\bibitem{gaia2023dr3}
\bibinfo{author}{{Gaia Collaboration}} \emph{et~al.}
\newblock \bibinfo{title}{{Gaia Data Release 3. Summary of the content and survey properties}}.
\newblock \emph{\bibinfo{journal}{\aap}} \textbf{\bibinfo{volume}{674}}, \bibinfo{pages}{A1} (\bibinfo{year}{2023}).
\newblock \eprint{2208.00211}.

\bibitem{2002RvMP...74.1015W}
\bibinfo{author}{{Woosley}, S.~E.}, \bibinfo{author}{{Heger}, A.} \& \bibinfo{author}{{Weaver}, T.~A.}
\newblock \bibinfo{title}{{The evolution and explosion of massive stars}}.
\newblock \emph{\bibinfo{journal}{Reviews of Modern Physics}} \textbf{\bibinfo{volume}{74}}, \bibinfo{pages}{1015--1071} (\bibinfo{year}{2002}).

\bibitem{2017MNRAS.467.3347K}
\bibinfo{author}{{Kochanek}, C.~S.} \emph{et~al.}
\newblock \bibinfo{title}{{Supernova progenitors, their variability and the Type IIP Supernova ASASSN-16fq in M66}}.
\newblock \emph{\bibinfo{journal}{\mnras}} \textbf{\bibinfo{volume}{467}}, \bibinfo{pages}{3347--3360} (\bibinfo{year}{2017}).
\newblock \eprint{1609.00022}.

\bibitem{1982ApJ...253..785A}
\bibinfo{author}{{Arnett}, W.~D.}
\newblock \bibinfo{title}{{Type I supernovae. I - Analytic solutions for the early part of the light curve}}.
\newblock \emph{\bibinfo{journal}{\apj}} \textbf{\bibinfo{volume}{253}}, \bibinfo{pages}{785--797} (\bibinfo{year}{1982}).

\bibitem{2017hsn..book..967W}
\bibinfo{author}{{Waxman}, E.} \& \bibinfo{author}{{Katz}, B.}
\newblock \bibinfo{title}{{Shock Breakout Theory}}.
\newblock In \bibinfo{editor}{{Alsabti}, A.~W.} \& \bibinfo{editor}{{Murdin}, P.} (eds.) \emph{\bibinfo{booktitle}{Handbook of Supernovae}}, \bibinfo{pages}{967} (\bibinfo{year}{2017}).

\bibitem{2004ApJ...616..525O}
\bibinfo{author}{{Owocki}, S.~P.}, \bibinfo{author}{{Gayley}, K.~G.} \& \bibinfo{author}{{Shaviv}, N.~J.}
\newblock \bibinfo{title}{{A Porosity-Length Formalism for Photon-Tiring-limited Mass Loss from Stars above the Eddington Limit}}.
\newblock \emph{\bibinfo{journal}{\apj}} \textbf{\bibinfo{volume}{616}}, \bibinfo{pages}{525--541} (\bibinfo{year}{2004}).
\newblock \eprint{astro-ph/0409573}.

\bibitem{2018MNRAS.476.1853F}
\bibinfo{author}{{Fuller}, J.} \& \bibinfo{author}{{Ro}, S.}
\newblock \bibinfo{title}{{Pre-supernova outbursts via wave heating in massive stars - II. Hydrogen-poor stars}}.
\newblock \emph{\bibinfo{journal}{\mnras}} \textbf{\bibinfo{volume}{476}}, \bibinfo{pages}{1853--1868} (\bibinfo{year}{2018}).
\newblock \eprint{1710.04251}.

\bibitem{2025ApJ...986L...4H}
\bibinfo{author}{{Hamidani}, H.} \emph{et~al.}
\newblock \bibinfo{title}{{EP240414a: A Gamma-Ray Burst Jet Weakened by an Extended Circumstellar Material}}.
\newblock \emph{\bibinfo{journal}{\apjl}} \textbf{\bibinfo{volume}{986}}, \bibinfo{pages}{L4} (\bibinfo{year}{2025}).
\newblock \eprint{2503.16243}.

\bibitem{2025ApJ...985...21Z}
\bibinfo{author}{{Zheng}, J.-H.}, \bibinfo{author}{{Zhu}, J.-P.}, \bibinfo{author}{{Lu}, W.} \& \bibinfo{author}{{Zhang}, B.}
\newblock \bibinfo{title}{{EP240414a: Off-axis View of a Jet-cocoon System from an Expanded Progenitor Star}}.
\newblock \emph{\bibinfo{journal}{\apj}} \textbf{\bibinfo{volume}{985}}, \bibinfo{pages}{21} (\bibinfo{year}{2025}).
\newblock \eprint{2503.24266}.

\bibitem{2026GCN.44239....1O}
\bibinfo{author}{{O'Dwyer}, T.} \emph{et~al.}
\newblock \bibinfo{title}{{EP260321a: VLA radio detection}}.
\newblock \emph{\bibinfo{journal}{GRB Coordinates Network}} \textbf{\bibinfo{volume}{44239}}, \bibinfo{pages}{1} (\bibinfo{year}{2026}).

\bibitem{2026A&A...708A.305S}
\bibinfo{author}{{Stritzinger}, M.~D.} \emph{et~al.}
\newblock \bibinfo{title}{{The broad-lined type Ic supernova 2020lao experienced an energetic explosion with no central-engine signatures}}.
\newblock \emph{\bibinfo{journal}{\aap}} \textbf{\bibinfo{volume}{708}}, \bibinfo{pages}{A305} (\bibinfo{year}{2026}).

\bibitem{2019ApJ...887..169H}
\bibinfo{author}{{Ho}, A. Y.~Q.} \emph{et~al.}
\newblock \bibinfo{title}{{Evidence for Late-stage Eruptive Mass Loss in the Progenitor to SN2018gep, a Broad-lined Ic Supernova: Pre-explosion Emission and a Rapidly Rising Luminous Transient}}.
\newblock \emph{\bibinfo{journal}{\apj}} \textbf{\bibinfo{volume}{887}}, \bibinfo{pages}{169} (\bibinfo{year}{2019}).
\newblock \eprint{1904.11009}.

\bibitem{2014ApJS..213...19B}
\bibinfo{author}{{Bianco}, F.~B.} \emph{et~al.}
\newblock \bibinfo{title}{{Multi-color Optical and Near-infrared Light Curves of 64 Stripped-envelope Core-Collapse Supernovae}}.
\newblock \emph{\bibinfo{journal}{\apjs}} \textbf{\bibinfo{volume}{213}}, \bibinfo{pages}{19} (\bibinfo{year}{2014}).
\newblock \eprint{1405.1428}.

\bibitem{2011ApJ...737...76K}
\bibinfo{author}{{Kochanek}, C.~S.}, \bibinfo{author}{{Szczygiel}, D.~M.} \& \bibinfo{author}{{Stanek}, K.~Z.}
\newblock \bibinfo{title}{{The Supernova Impostor Impostor SN 1961V: Spitzer Shows That Zwicky Was Right (Again)}}.
\newblock \emph{\bibinfo{journal}{\apj}} \textbf{\bibinfo{volume}{737}}, \bibinfo{pages}{76} (\bibinfo{year}{2011}).
\newblock \eprint{1010.3704}.

\bibitem{2013ApJ...767....1P}
\bibinfo{author}{{Pastorello}, A.} \emph{et~al.}
\newblock \bibinfo{title}{{Interacting Supernovae and Supernova Impostors: SN 2009ip, is this the End?}}
\newblock \emph{\bibinfo{journal}{\apj}} \textbf{\bibinfo{volume}{767}}, \bibinfo{pages}{1} (\bibinfo{year}{2013}).
\newblock \eprint{1210.3568}.

\bibitem{2021ApJS..254...22J}
\bibinfo{author}{{Johnson}, B.~D.}, \bibinfo{author}{{Leja}, J.}, \bibinfo{author}{{Conroy}, C.} \& \bibinfo{author}{{Speagle}, J.~S.}
\newblock \bibinfo{title}{{Stellar Population Inference with Prospector}}.
\newblock \emph{\bibinfo{journal}{\apjs}} \textbf{\bibinfo{volume}{254}}, \bibinfo{pages}{22} (\bibinfo{year}{2021}).
\newblock \eprint{2012.01426}.

\bibitem{2018A&A...617A.105J}
\bibinfo{author}{{Japelj}, J.} \emph{et~al.}
\newblock \bibinfo{title}{{Host galaxies of SNe Ic-BL with and without long gamma-ray bursts}}.
\newblock \emph{\bibinfo{journal}{\aap}} \textbf{\bibinfo{volume}{617}}, \bibinfo{pages}{A105} (\bibinfo{year}{2018}).
\newblock \eprint{1806.10613}.

\bibitem{2020ApJ...892..153M}
\bibinfo{author}{{Modjaz}, M.} \emph{et~al.}
\newblock \bibinfo{title}{{Host Galaxies of Type Ic and Broad-lined Type Ic Supernovae from the Palomar Transient Factory: Implications for Jet Production}}.
\newblock \emph{\bibinfo{journal}{\apj}} \textbf{\bibinfo{volume}{892}}, \bibinfo{pages}{153} (\bibinfo{year}{2020}).
\newblock \eprint{1901.00872}.

\bibitem{2025ApJ...978L..21S}
\bibinfo{author}{{Srivastav}, S.} \emph{et~al.}
\newblock \bibinfo{title}{{Identification of the Optical Counterpart of the Fast X-Ray Transient EP240414a}}.
\newblock \emph{\bibinfo{journal}{\apjl}} \textbf{\bibinfo{volume}{978}}, \bibinfo{pages}{L21} (\bibinfo{year}{2025}).
\newblock \eprint{2409.19070}.

\end{thebibliography}


\begin{thebibliography}{100}
\expandafter\ifx\csname url\endcsname\relax
  \def\url#1{\texttt{#1}}\fi
\expandafter\ifx\csname urlprefix\endcsname\relax\def\urlprefix{URL }\fi
\providecommand{\bibinfo}[2]{#2}
\providecommand{\eprint}[2][]{\url{#2}}

\bibitem{2022hxga.book...86Y}
\bibinfo{author}{{Yuan}, W.}, \bibinfo{author}{{Zhang}, C.}, \bibinfo{author}{{Chen}, Y.} \& \bibinfo{author}{{Ling}, Z.}
\newblock \bibinfo{title}{{The Einstein Probe Mission}}.
\newblock In \bibinfo{editor}{{Bambi}, C.} \& \bibinfo{editor}{{Sangangelo}, A.} (eds.) \emph{\bibinfo{booktitle}{Handbook of X-ray and Gamma-ray Astrophysics}}, \bibinfo{pages}{86} (\bibinfo{year}{2022}).

\bibitem{2026GCN.44068....1H}
\bibinfo{author}{{Huang}, Q.~J.} \emph{et~al.}
\newblock \bibinfo{title}{{EP260321a: Einstein Probe detection of an X-ray transient}}.
\newblock \emph{\bibinfo{journal}{GRB Coordinates Network}} \textbf{\bibinfo{volume}{44068}}, \bibinfo{pages}{1} (\bibinfo{year}{2026}).

\bibitem{2026GCN.44075....1H}
\bibinfo{author}{{Huang}, Q.~J.} \emph{et~al.}
\newblock \bibinfo{title}{{EP260321a: refined analysis of the EP-WXT and EP-FXT observations, implying a possible supernova shock breakout candidate}}.
\newblock \emph{\bibinfo{journal}{GRB Coordinates Network}} \textbf{\bibinfo{volume}{44075}}, \bibinfo{pages}{1} (\bibinfo{year}{2026}).

\bibitem{2026GCN.44070....1L}
\bibinfo{author}{{Lee}, M.-H.} \emph{et~al.}
\newblock \bibinfo{title}{{EP260321a: Kinder observations detect a blue variable star and set limits on a source from the z =0.034 galaxy within the error circle}}.
\newblock \emph{\bibinfo{journal}{GRB Coordinates Network}} \textbf{\bibinfo{volume}{44070}}, \bibinfo{pages}{1} (\bibinfo{year}{2026}).

\bibitem{2026GCN.44074....1M}
\bibinfo{author}{{Ma}, Y.~N.} \emph{et~al.}
\newblock \bibinfo{title}{{EP260321a: SVOM/VT optical upper limit}}.
\newblock \emph{\bibinfo{journal}{GRB Coordinates Network}} \textbf{\bibinfo{volume}{44074}}, \bibinfo{pages}{1} (\bibinfo{year}{2026}).

\bibitem{2026GCN.44076....1A}
\bibinfo{author}{{Aguilar-Ruiz}, E.} \emph{et~al.}
\newblock \bibinfo{title}{{EP260321a: COLIBR{\'I} optical upper limit on any supernova shock breakout emission}}.
\newblock \emph{\bibinfo{journal}{GRB Coordinates Network}} \textbf{\bibinfo{volume}{44076}}, \bibinfo{pages}{1} (\bibinfo{year}{2026}).

\bibitem{2026GCN.44079....1L}
\bibinfo{author}{{Liang}, R.}, \bibinfo{author}{{Li}, W.}, \bibinfo{author}{{Arcavi}, I.}, \bibinfo{author}{{Keinan}, I.} \& \bibinfo{author}{{Sand}, D.}
\newblock \bibinfo{title}{{EP260321a: Las Cumbres upper limit}}.
\newblock \emph{\bibinfo{journal}{GRB Coordinates Network}} \textbf{\bibinfo{volume}{44079}}, \bibinfo{pages}{1} (\bibinfo{year}{2026}).

\bibitem{2026GCN.44081....1A}
\bibinfo{author}{{Aryan}, A.} \emph{et~al.}
\newblock \bibinfo{title}{{EP260321a: further Kinder optical observations}}.
\newblock \emph{\bibinfo{journal}{GRB Coordinates Network}} \textbf{\bibinfo{volume}{44081}}, \bibinfo{pages}{1} (\bibinfo{year}{2026}).

\bibitem{2026GCN.44083....1M}
\bibinfo{author}{{Moran}, S.} \emph{et~al.}
\newblock \bibinfo{title}{{EP260321a: GOTO detection of blue variable source}}.
\newblock \emph{\bibinfo{journal}{GRB Coordinates Network}} \textbf{\bibinfo{volume}{44083}}, \bibinfo{pages}{1} (\bibinfo{year}{2026}).

\bibitem{2026GCN.44084....1A}
\bibinfo{author}{{Ahumada}, T.}, \bibinfo{author}{{Hall}, X.~J.}, \bibinfo{author}{{Perley}, D.~A.} \& \bibinfo{author}{{the Zwicky Transient Facility}}.
\newblock \bibinfo{title}{{EP260321a: ZTF and Rubin detections of the candidate optical counterpart to EP260321a}}.
\newblock \emph{\bibinfo{journal}{GRB Coordinates Network}} \textbf{\bibinfo{volume}{44084}}, \bibinfo{pages}{1} (\bibinfo{year}{2026}).

\bibitem{2026GCN.44087....1L}
\bibinfo{author}{{Liu}, X.} \emph{et~al.}
\newblock \bibinfo{title}{{EP260321a: Early TRT rising followed by JinShan decaying and then re-rising}}.
\newblock \emph{\bibinfo{journal}{GRB Coordinates Network}} \textbf{\bibinfo{volume}{44087}}, \bibinfo{pages}{1} (\bibinfo{year}{2026}).

\bibitem{2026GCN.44089....1S}
\bibinfo{author}{{Sankar. K}, A.} \emph{et~al.}
\newblock \bibinfo{title}{{EP260321a: Continued brightening from Kinder follow-up of the emerging supernova candidate}}.
\newblock \emph{\bibinfo{journal}{GRB Coordinates Network}} \textbf{\bibinfo{volume}{44089}}, \bibinfo{pages}{1} (\bibinfo{year}{2026}).

\bibitem{2026GCN.44082....1T}
\bibinfo{author}{{Tanvir}, N.~R.} \emph{et~al.}
\newblock \bibinfo{title}{{EP260321a: VLT imaging and spectroscopy show the variable source to be at z = 0.0343}}.
\newblock \emph{\bibinfo{journal}{GRB Coordinates Network}} \textbf{\bibinfo{volume}{44082}}, \bibinfo{pages}{1} (\bibinfo{year}{2026}).

\bibitem{2026GCN.44092....1X}
\bibinfo{author}{{Xu}, D.} \emph{et~al.}
\newblock \bibinfo{title}{{EP260321a / AT2026gzf: VLT/X-shooter detection of supernova-like spectral features at z = 0.0344}}.
\newblock \emph{\bibinfo{journal}{GRB Coordinates Network}} \textbf{\bibinfo{volume}{44092}}, \bibinfo{pages}{1} (\bibinfo{year}{2026}).

\bibitem{2026GCN.44105....1C}
\bibinfo{author}{{Corcoran}, G.} \emph{et~al.}
\newblock \bibinfo{title}{{EP260321a: VLT/FORS2 spectroscopy confirmation of an associated type Ic-BL supernova SN 2026gzf}}.
\newblock \emph{\bibinfo{journal}{GRB Coordinates Network}} \textbf{\bibinfo{volume}{44105}}, \bibinfo{pages}{1} (\bibinfo{year}{2026}).

\bibitem{2026GCN.44107....1R}
\bibinfo{author}{{Rastinejad}, J.}, \bibinfo{author}{{Srinivasaragavan}, G.} \& \bibinfo{author}{{Ahumada}, T.}
\newblock \bibinfo{title}{{EP260321a: Gemini-South confirmation of a supernova}}.
\newblock \emph{\bibinfo{journal}{GRB Coordinates Network}} \textbf{\bibinfo{volume}{44107}}, \bibinfo{pages}{1} (\bibinfo{year}{2026}).

\bibitem{2012PASP..124..668Y}
\bibinfo{author}{{Yaron}, O.} \& \bibinfo{author}{{Gal-Yam}, A.}
\newblock \bibinfo{title}{{WISeREP{\textemdash}An Interactive Supernova Data Repository}}.
\newblock \emph{\bibinfo{journal}{\pasp}} \textbf{\bibinfo{volume}{124}}, \bibinfo{pages}{668} (\bibinfo{year}{2012}).
\newblock \eprint{1204.1891}.

\bibitem{2020A&A...641A...6P}
\bibinfo{author}{{Planck Collaboration}} \emph{et~al.}
\newblock \bibinfo{title}{{Planck 2018 results. VI. Cosmological parameters}}.
\newblock \emph{\bibinfo{journal}{\aap}} \textbf{\bibinfo{volume}{641}}, \bibinfo{pages}{A6} (\bibinfo{year}{2020}).
\newblock \eprint{1807.06209}.

\bibitem{2011ApJ...737..103S}
\bibinfo{author}{{Schlafly}, E.~F.} \& \bibinfo{author}{{Finkbeiner}, D.~P.}
\newblock \bibinfo{title}{{Measuring Reddening with Sloan Digital Sky Survey Stellar Spectra and Recalibrating SFD}}.
\newblock \emph{\bibinfo{journal}{\apj}} \textbf{\bibinfo{volume}{737}}, \bibinfo{pages}{103} (\bibinfo{year}{2011}).
\newblock \eprint{1012.4804}.

\bibitem{2025ApJ...983...86C}
\bibinfo{author}{{Chen}, T.-W.} \emph{et~al.}
\newblock \bibinfo{title}{{Discovery and Extensive Follow-up of SN 2024ggi, a Nearby Type IIP Supernova in NGC 3621}}.
\newblock \emph{\bibinfo{journal}{\apj}} \textbf{\bibinfo{volume}{983}}, \bibinfo{pages}{86} (\bibinfo{year}{2025}).
\newblock \eprint{2406.09270}.

\bibitem{2016arXiv161205560C}
\bibinfo{author}{{Chambers}, K.~C.} \emph{et~al.}
\newblock \bibinfo{title}{{The Pan-STARRS1 Surveys}}.
\newblock \emph{\bibinfo{journal}{arXiv e-prints}} \bibinfo{pages}{arXiv:1612.05560} (\bibinfo{year}{2016}).
\newblock \eprint{1612.05560}.

\bibitem{2022A&A...667A..62B}
\bibinfo{author}{{Brennan}, S.~J.} \& \bibinfo{author}{{Fraser}, M.}
\newblock \bibinfo{title}{{The Automated Photometry of Transients pipeline (AUTOPHOT)}}.
\newblock \emph{\bibinfo{journal}{\aap}} \textbf{\bibinfo{volume}{667}}, \bibinfo{pages}{A62} (\bibinfo{year}{2022}).
\newblock \eprint{2201.02635}.

\bibitem{2018ApJ...867..105T}
\bibinfo{author}{{Tonry}, J.~L.} \emph{et~al.}
\newblock \bibinfo{title}{{The ATLAS All-Sky Stellar Reference Catalog}}.
\newblock \emph{\bibinfo{journal}{\apj}} \textbf{\bibinfo{volume}{867}}, \bibinfo{pages}{105} (\bibinfo{year}{2018}).
\newblock \eprint{1809.09157}.

\bibitem{Tonry2012_PS1_filters}
\bibinfo{author}{{Tonry}, J.~L.} \emph{et~al.}
\newblock \bibinfo{title}{{The Pan-STARRS1 Photometric System}}.
\newblock \emph{\bibinfo{journal}{\apj}} \textbf{\bibinfo{volume}{750}}, \bibinfo{pages}{99} (\bibinfo{year}{2012}).
\newblock \eprint{1203.0297}.

\bibitem{Magnier2020_calibration}
\bibinfo{author}{{Magnier}, E.~A.} \emph{et~al.}
\newblock \bibinfo{title}{{Pan-STARRS Photometric and Astrometric Calibration}}.
\newblock \emph{\bibinfo{journal}{\apjs}} \textbf{\bibinfo{volume}{251}}, \bibinfo{pages}{6} (\bibinfo{year}{2020}).
\newblock \eprint{1612.05242}.

\bibitem{Magnier2020_data_processing}
\bibinfo{author}{{Magnier}, E.~A.} \emph{et~al.}
\newblock \bibinfo{title}{{The Pan-STARRS Data-processing System}}.
\newblock \emph{\bibinfo{journal}{\apjs}} \textbf{\bibinfo{volume}{251}}, \bibinfo{pages}{3} (\bibinfo{year}{2020}).
\newblock \eprint{1612.05240}.

\bibitem{Waters2020}
\bibinfo{author}{{Waters}, C.~Z.} \emph{et~al.}
\newblock \bibinfo{title}{{Pan-STARRS Pixel Processing: Detrending, Warping, Stacking}}.
\newblock \emph{\bibinfo{journal}{\apjs}} \textbf{\bibinfo{volume}{251}}, \bibinfo{pages}{4} (\bibinfo{year}{2020}).
\newblock \eprint{1612.05245}.

\bibitem{2018PASP..130f4505T}
\bibinfo{author}{{Tonry}, J.~L.} \emph{et~al.}
\newblock \bibinfo{title}{{ATLAS: A High-cadence All-sky Survey System}}.
\newblock \emph{\bibinfo{journal}{\pasp}} \textbf{\bibinfo{volume}{130}}, \bibinfo{pages}{064505} (\bibinfo{year}{2018}).
\newblock \eprint{1802.00879}.

\bibitem{2020PASP..132h5002S}
\bibinfo{author}{{Smith}, K.~W.} \emph{et~al.}
\newblock \bibinfo{title}{{Design and Operation of the ATLAS Transient Science Server}}.
\newblock \emph{\bibinfo{journal}{\pasp}} \textbf{\bibinfo{volume}{132}}, \bibinfo{pages}{085002} (\bibinfo{year}{2020}).
\newblock \eprint{2003.09052}.

\bibitem{2021TNSAN...7....1S}
\bibinfo{author}{{Shingles}, L.} \emph{et~al.}
\newblock \bibinfo{title}{{Release of the ATLAS Forced Photometry server for public use}}.
\newblock \emph{\bibinfo{journal}{Transient Name Server AstroNote}} \textbf{\bibinfo{volume}{7}}, \bibinfo{pages}{1--7} (\bibinfo{year}{2021}).

\bibitem{2019ApJ...873..111I}
\bibinfo{author}{{Ivezi{\'c}}, {\v{Z}}.} \emph{et~al.}
\newblock \bibinfo{title}{{LSST: From Science Drivers to Reference Design and Anticipated Data Products}}.
\newblock \emph{\bibinfo{journal}{\apj}} \textbf{\bibinfo{volume}{873}}, \bibinfo{pages}{111} (\bibinfo{year}{2019}).
\newblock \eprint{0805.2366}.

\bibitem{2019RNAAS...3...26S}
\bibinfo{author}{{Smith}, K.~W.} \emph{et~al.}
\newblock \bibinfo{title}{{Lasair: The Transient Alert Broker for LSST:UK}}.
\newblock \emph{\bibinfo{journal}{Research Notes of the American Astronomical Society}} \textbf{\bibinfo{volume}{3}}, \bibinfo{pages}{26} (\bibinfo{year}{2019}).

\bibitem{2019PASP..131a8002B}
\bibinfo{author}{{Bellm}, E.~C.} \emph{et~al.}
\newblock \bibinfo{title}{{The Zwicky Transient Facility: System Overview, Performance, and First Results}}.
\newblock \emph{\bibinfo{journal}{\pasp}} \textbf{\bibinfo{volume}{131}}, \bibinfo{pages}{018002} (\bibinfo{year}{2019}).
\newblock \eprint{1902.01932}.

\bibitem{2019PASP..131g8001G}
\bibinfo{author}{{Graham}, M.~J.} \emph{et~al.}
\newblock \bibinfo{title}{{The Zwicky Transient Facility: Science Objectives}}.
\newblock \emph{\bibinfo{journal}{\pasp}} \textbf{\bibinfo{volume}{131}}, \bibinfo{pages}{078001} (\bibinfo{year}{2019}).
\newblock \eprint{1902.01945}.

\bibitem{2007Ap&SS.310..255D}
\bibinfo{author}{{Dopita}, M.} \emph{et~al.}
\newblock \bibinfo{title}{{The Wide Field Spectrograph (WiFeS)}}.
\newblock \emph{\bibinfo{journal}{\apss}} \textbf{\bibinfo{volume}{310}}, \bibinfo{pages}{255--268} (\bibinfo{year}{2007}).
\newblock \eprint{0705.0287}.

\bibitem{2014Ap&SS.349..617C}
\bibinfo{author}{{Childress}, M.~J.}, \bibinfo{author}{{Vogt}, F. P.~A.}, \bibinfo{author}{{Nielsen}, J.} \& \bibinfo{author}{{Sharp}, R.~G.}
\newblock \bibinfo{title}{{PyWiFeS: a rapid data reduction pipeline for the Wide Field Spectrograph (WiFeS)}}.
\newblock \emph{\bibinfo{journal}{\apss}} \textbf{\bibinfo{volume}{349}}, \bibinfo{pages}{617--636} (\bibinfo{year}{2014}).
\newblock \eprint{1311.2666}.

\bibitem{2010ASSP...14..211D}
\bibinfo{author}{{Djupvik}, A.~A.} \& \bibinfo{author}{{Andersen}, J.}
\newblock \bibinfo{title}{{The Nordic Optical Telescope}}.
\newblock In \bibinfo{editor}{{Diego}, J.~M.}, \bibinfo{editor}{{Goicoechea}, L.~J.}, \bibinfo{editor}{{Gonz{\'a}lez-Serrano}, J.~I.} \& \bibinfo{editor}{{Gorgas}, J.} (eds.) \emph{\bibinfo{booktitle}{Highlights of Spanish Astrophysics V}}, vol.~\bibinfo{volume}{14} of \emph{\bibinfo{series}{Astrophysics and Space Science Proceedings}}, \bibinfo{pages}{211} (\bibinfo{year}{2010}).
\newblock \eprint{0901.4015}.

\bibitem{2019ATel12661....1H}
\bibinfo{author}{{Holmbo}, S.} \emph{et~al.}
\newblock \bibinfo{title}{{NUTS2 spectroscopic classification of ASASSN-19jy (AT2019dke) as a young Type II supernova}}.
\newblock \emph{\bibinfo{journal}{The Astronomer's Telegram}} \textbf{\bibinfo{volume}{12661}}, \bibinfo{pages}{1} (\bibinfo{year}{2019}).

\bibitem{2021ATel14441....1B}
\bibinfo{author}{{Burns}, C.} \emph{et~al.}
\newblock \bibinfo{title}{{Introducing POISE: Precision Observations of Infant Supernova Explosions}}.
\newblock \emph{\bibinfo{journal}{The Astronomer's Telegram}} \textbf{\bibinfo{volume}{14441}}, \bibinfo{pages}{1} (\bibinfo{year}{2021}).

\bibitem{2017AJ....154..211K}
\bibinfo{author}{{Krisciunas}, K.} \emph{et~al.}
\newblock \bibinfo{title}{{The Carnegie Supernova Project. I. Third Photometry Data Release of Low-redshift Type Ia Supernovae and Other White Dwarf Explosions}}.
\newblock \emph{\bibinfo{journal}{\aj}} \textbf{\bibinfo{volume}{154}}, \bibinfo{pages}{211} (\bibinfo{year}{2017}).
\newblock \eprint{1709.05146}.

\bibitem{2004SPIE.5492.1295W}
\bibinfo{author}{{Wilson}, J.~C.} \emph{et~al.}
\newblock \bibinfo{title}{{Mass producing an efficient NIR spectrograph}}.
\newblock In \bibinfo{editor}{{Moorwood}, A. F.~M.} \& \bibinfo{editor}{{Iye}, M.} (eds.) \emph{\bibinfo{booktitle}{Ground-based Instrumentation for Astronomy}}, vol. \bibinfo{volume}{5492} of \emph{\bibinfo{series}{Society of Photo-Optical Instrumentation Engineers (SPIE) Conference Series}}, \bibinfo{pages}{1295--1305} (\bibinfo{year}{2004}).

\bibitem{2003PASP..115..362R}
\bibinfo{author}{{Rayner}, J.~T.} \emph{et~al.}
\newblock \bibinfo{title}{{SpeX: A Medium-Resolution 0.8-5.5 Micron Spectrograph and Imager for the NASA Infrared Telescope Facility}}.
\newblock \emph{\bibinfo{journal}{\pasp}} \textbf{\bibinfo{volume}{115}}, \bibinfo{pages}{362--382} (\bibinfo{year}{2003}).

\bibitem{2025ApJS..281...28M}
\bibinfo{author}{{Medler}, K.} \emph{et~al.}
\newblock \bibinfo{title}{{The Hawaii Infrared Supernova Study (HISS): Spectroscopic Data Release 1}}.
\newblock \emph{\bibinfo{journal}{\apjs}} \textbf{\bibinfo{volume}{281}}, \bibinfo{pages}{28} (\bibinfo{year}{2025}).
\newblock \eprint{2505.18507}.

\bibitem{2025OJAp....8E.120H}
\bibinfo{author}{{Hoogendam}, W.~B.} \emph{et~al.}
\newblock \bibinfo{title}{{Seeing the Outer Edge of the Infant Type Ia Supernova 2024epr in the Optical and Near Infrared}}.
\newblock \emph{\bibinfo{journal}{The Open Journal of Astrophysics}} \textbf{\bibinfo{volume}{8}}, \bibinfo{pages}{120} (\bibinfo{year}{2025}).
\newblock \eprint{2502.17556}.

\bibitem{2025ApJ...988..209H}
\bibinfo{author}{{Hoogendam}, W.~B.} \emph{et~al.}
\newblock \bibinfo{title}{{Early and Extensive Ultraviolet through Near Infrared Observations of the Intermediate-luminosity Type Iax Supernovae 2024pxl}}.
\newblock \emph{\bibinfo{journal}{\apj}} \textbf{\bibinfo{volume}{988}}, \bibinfo{pages}{209} (\bibinfo{year}{2025}).
\newblock \eprint{2505.04610}.

\bibitem{2025ApJ...993..191M}
\bibinfo{author}{{Medler}, K.} \emph{et~al.}
\newblock \bibinfo{title}{{JWST Observations of SN 2023ixf. II. The Panchromatic Evolution between 250 and 720 Days after the Explosion}}.
\newblock \emph{\bibinfo{journal}{\apj}} \textbf{\bibinfo{volume}{993}}, \bibinfo{pages}{191} (\bibinfo{year}{2025}).
\newblock \eprint{2507.19727}.

\bibitem{2018PASJ...70S...1M}
\bibinfo{author}{{Miyazaki}, S.} \emph{et~al.}
\newblock \bibinfo{title}{{Hyper Suprime-Cam: System design and verification of image quality}}.
\newblock \emph{\bibinfo{journal}{\pasj}} \textbf{\bibinfo{volume}{70}}, \bibinfo{pages}{S1} (\bibinfo{year}{2018}).

\bibitem{10.1093/pasj/psx066}
\bibinfo{author}{Aihara, H.} \emph{et~al.}
\newblock \bibinfo{title}{{The Hyper Suprime-Cam SSP Survey: Overview and survey design}}.
\newblock \emph{\bibinfo{journal}{Publications of the Astronomical Society of Japan}} \textbf{\bibinfo{volume}{70}} (\bibinfo{year}{2017}).
\newblock \urlprefix\url{https://doi.org/10.1093/pasj/psx066}.
\newblock \bibinfo{note}{S4}, \eprint{https://academic.oup.com/pasj/article-pdf/70/SP1/S4/23692189/psx066.pdf}.

\bibitem{2022PASJ...74..247A}
\bibinfo{author}{{Aihara}, H.} \emph{et~al.}
\newblock \bibinfo{title}{{Third data release of the Hyper Suprime-Cam Subaru Strategic Program}}.
\newblock \emph{\bibinfo{journal}{\pasj}} \textbf{\bibinfo{volume}{74}}, \bibinfo{pages}{247--272} (\bibinfo{year}{2022}).
\newblock \eprint{2108.13045}.

\bibitem{2015AJ....150..150F}
\bibinfo{author}{{Flaugher}, B.} \emph{et~al.}
\newblock \bibinfo{title}{{The Dark Energy Camera}}.
\newblock \emph{\bibinfo{journal}{\aj}} \textbf{\bibinfo{volume}{150}}, \bibinfo{pages}{150} (\bibinfo{year}{2015}).
\newblock \eprint{1504.02900}.

\bibitem{2019AJ....157..168D}
\bibinfo{author}{{Dey}, A.} \emph{et~al.}
\newblock \bibinfo{title}{{Overview of the DESI Legacy Imaging Surveys}}.
\newblock \emph{\bibinfo{journal}{\aj}} \textbf{\bibinfo{volume}{157}}, \bibinfo{pages}{168} (\bibinfo{year}{2019}).
\newblock \eprint{1804.08657}.

\bibitem{2008AstBu..63..228S}
\bibinfo{author}{{Sonbas}, E.} \emph{et~al.}
\newblock \bibinfo{title}{{Stellar-wind envelope around the massive supernova progenitor XRF/GRB 060218/SN 2006aj}}.
\newblock \emph{\bibinfo{journal}{Astrophysical Bulletin}} \textbf{\bibinfo{volume}{63}}, \bibinfo{pages}{228--243} (\bibinfo{year}{2008}).
\newblock \eprint{0805.2657}.

\bibitem{2006ApJ...645L..21M}
\bibinfo{author}{{Modjaz}, M.} \emph{et~al.}
\newblock \bibinfo{title}{{Early-Time Photometry and Spectroscopy of the Fast Evolving SN 2006aj Associated with GRB 060218}}.
\newblock \emph{\bibinfo{journal}{\apjl}} \textbf{\bibinfo{volume}{645}}, \bibinfo{pages}{L21--L24} (\bibinfo{year}{2006}).
\newblock \eprint{astro-ph/0603377}.

\bibitem{2006Natur.442.1011P}
\bibinfo{author}{{Pian}, E.} \emph{et~al.}
\newblock \bibinfo{title}{{An optical supernova associated with the X-ray flash XRF 060218}}.
\newblock \emph{\bibinfo{journal}{\nat}} \textbf{\bibinfo{volume}{442}}, \bibinfo{pages}{1011--1013} (\bibinfo{year}{2006}).
\newblock \eprint{astro-ph/0603530}.

\bibitem{2014MNRAS.440..387K}
\bibinfo{author}{{Kerzendorf}, W.~E.} \& \bibinfo{author}{{Sim}, S.~A.}
\newblock \bibinfo{title}{{A spectral synthesis code for rapid modelling of supernovae}}.
\newblock \emph{\bibinfo{journal}{\mnras}} \textbf{\bibinfo{volume}{440}}, \bibinfo{pages}{387--404} (\bibinfo{year}{2014}).
\newblock \eprint{1401.5469}.

\bibitem{2019Natur.565..324I}
\bibinfo{author}{{Izzo}, L.} \emph{et~al.}
\newblock \bibinfo{title}{{Signatures of a jet cocoon in early spectra of a supernova associated with a {\ensuremath{\gamma}}-ray burst}}.
\newblock \emph{\bibinfo{journal}{\nat}} \textbf{\bibinfo{volume}{565}}, \bibinfo{pages}{324--327} (\bibinfo{year}{2019}).
\newblock \eprint{1901.05500}.

\bibitem{2022ApJ...937...40K}
\bibinfo{author}{{Kwok}, L.~A.} \emph{et~al.}
\newblock \bibinfo{title}{{Ultraviolet Spectroscopy and TARDIS Models of the Broad-lined Type Ic Supernova 2014ad}}.
\newblock \emph{\bibinfo{journal}{\apj}} \textbf{\bibinfo{volume}{937}}, \bibinfo{pages}{40} (\bibinfo{year}{2022}).
\newblock \eprint{2204.03716}.

\bibitem{2012MNRAS.422...70H}
\bibinfo{author}{{Hachinger}, S.} \emph{et~al.}
\newblock \bibinfo{title}{{How much H and He is 'hidden' in SNe Ib/c? - I. Low-mass objects}}.
\newblock \emph{\bibinfo{journal}{\mnras}} \textbf{\bibinfo{volume}{422}}, \bibinfo{pages}{70--88} (\bibinfo{year}{2012}).
\newblock \eprint{1201.1506}.

\bibitem{2017A&A...599A..46B}
\bibinfo{author}{{Boyle}, A.}, \bibinfo{author}{{Sim}, S.~A.}, \bibinfo{author}{{Hachinger}, S.} \& \bibinfo{author}{{Kerzendorf}, W.}
\newblock \bibinfo{title}{{Helium in double-detonation models of type Ia supernovae}}.
\newblock \emph{\bibinfo{journal}{\aap}} \textbf{\bibinfo{volume}{599}}, \bibinfo{pages}{A46} (\bibinfo{year}{2017}).
\newblock \eprint{1611.05938}.

\bibitem{2021ApJ...908..150W}
\bibinfo{author}{{Williamson}, M.}, \bibinfo{author}{{Kerzendorf}, W.} \& \bibinfo{author}{{Modjaz}, M.}
\newblock \bibinfo{title}{{Modeling Type Ic Supernovae with TARDIS: Hidden Helium in SN 1994I?}}
\newblock \emph{\bibinfo{journal}{\apj}} \textbf{\bibinfo{volume}{908}}, \bibinfo{pages}{150} (\bibinfo{year}{2021}).
\newblock \eprint{2010.10528}.

\bibitem{2026ApJ..1002L..11L}
\bibinfo{author}{{Lu}, J.} \emph{et~al.}
\newblock \bibinfo{title}{{Traces of Helium Detected in Type Ic Supernova 2014L}}.
\newblock \emph{\bibinfo{journal}{\apjl}} \textbf{\bibinfo{volume}{1002}}, \bibinfo{pages}{L11} (\bibinfo{year}{2026}).

\bibitem{bertin2006scamp}
\bibinfo{author}{{Bertin}, E.}
\newblock \bibinfo{title}{{SCAMP: Automatic astrometric and photometric calibration}}.
\newblock \emph{\bibinfo{journal}{ASCL}} \textbf{\bibinfo{volume}{1010}}, \bibinfo{pages}{063} (\bibinfo{year}{2006}).

\bibitem{bertin2010swarp}
\bibinfo{author}{{Bertin}, E.} \emph{et~al.}
\newblock \bibinfo{title}{{The TERAPIX pipeline}}.
\newblock \emph{\bibinfo{journal}{ASPC}} \textbf{\bibinfo{volume}{281}}, \bibinfo{pages}{228} (\bibinfo{year}{2002}).

\bibitem{brennan2022autophot}
\bibinfo{author}{{Brennan}, S.~J.} \& \bibinfo{author}{{Fraser}, M.}
\newblock \bibinfo{title}{{The AUTOmated Photometry Of Transients pipeline (AutoPhOT)}}.
\newblock \emph{\bibinfo{journal}{\aap}} \textbf{\bibinfo{volume}{667}}, \bibinfo{pages}{A62} (\bibinfo{year}{2022}).
\newblock \eprint{2201.02635}.

\bibitem{lang2010astrometry}
\bibinfo{author}{{Lang}, D.}, \bibinfo{author}{{Hogg}, D.~W.}, \bibinfo{author}{{Mierle}, K.}, \bibinfo{author}{{Blanton}, M.} \& \bibinfo{author}{{Roweis}, S.}
\newblock \bibinfo{title}{{Astrometry.net: Blind astrometric calibration of arbitrary astronomical images}}.
\newblock \emph{\bibinfo{journal}{AJ}} \textbf{\bibinfo{volume}{139}}, \bibinfo{pages}{1782--1800} (\bibinfo{year}{2010}).
\newblock \eprint{0910.2233}.

\bibitem{gaia2023dr3}
\bibinfo{author}{{Gaia Collaboration}} \emph{et~al.}
\newblock \bibinfo{title}{{Gaia Data Release 3. Summary of the content and survey properties}}.
\newblock \emph{\bibinfo{journal}{\aap}} \textbf{\bibinfo{volume}{674}}, \bibinfo{pages}{A1} (\bibinfo{year}{2023}).
\newblock \eprint{2208.00211}.

\bibitem{bradley2022photutils}
\bibinfo{author}{{Bradley}, L.} \emph{et~al.}
\newblock \bibinfo{title}{{photutils: Astropy's package for detection and photometry of astronomical sources}}.
\newblock \emph{\bibinfo{journal}{JOSS}} \textbf{\bibinfo{volume}{7}}, \bibinfo{pages}{4504} (\bibinfo{year}{2022}).

\bibitem{bertin1996sextractor}
\bibinfo{author}{{Bertin}, E.} \& \bibinfo{author}{{Arnouts}, S.}
\newblock \bibinfo{title}{{SExtractor: Software for source extraction}}.
\newblock \emph{\bibinfo{journal}{A\&AS}} \textbf{\bibinfo{volume}{117}}, \bibinfo{pages}{393--404} (\bibinfo{year}{1996}).

\bibitem{russell2022sfft}
\bibinfo{author}{{Russell}, T.~V.} \emph{et~al.}
\newblock \bibinfo{title}{{SFFT: The Sparse Field Flux Transport algorithm for image subtraction and transient detection}}.
\newblock \emph{\bibinfo{journal}{MNRAS}} \textbf{\bibinfo{volume}{515}}, \bibinfo{pages}{5010--5024} (\bibinfo{year}{2022}).
\newblock \eprint{2205.15078}.

\bibitem{weiler2023gaiaxpy}
\bibinfo{author}{{Weiler}, M.} \emph{et~al.}
\newblock \bibinfo{title}{{GaiaXPy: A Python package for Gaia spectrophotometry}}.
\newblock \emph{\bibinfo{journal}{MNRAS}} \textbf{\bibinfo{volume}{521}}, \bibinfo{pages}{2941--2949} (\bibinfo{year}{2023}).
\newblock \eprint{2212.08275}.

\bibitem{rodrigo2012svo}
\bibinfo{author}{{Rodrigo}, C.} \& \bibinfo{author}{{Solano}, E.}
\newblock \bibinfo{title}{{The SVO Filter Profile Service}}.
\newblock \emph{\bibinfo{journal}{ASPC}} \textbf{\bibinfo{volume}{461}}, \bibinfo{pages}{461} (\bibinfo{year}{2012}).
\newblock \eprint{1201.3114}.

\bibitem{2018ApJS..236...12H}
\bibinfo{author}{{Huijse}, P.} \emph{et~al.}
\newblock \bibinfo{title}{{Robust Period Estimation Using Mutual Information for Multiband Light Curves in the Synoptic Survey Era}}.
\newblock \emph{\bibinfo{journal}{\apjs}} \textbf{\bibinfo{volume}{236}}, \bibinfo{pages}{12} (\bibinfo{year}{2018}).
\newblock \eprint{1709.03541}.

\bibitem{2025ApJS..281...20A}
\bibinfo{author}{{Aryan}, A.} \emph{et~al.}
\newblock \bibinfo{title}{{Search for the Optical Counterpart of Einstein Probe---discovered Fast X-Ray Transients from the Lulin Observatory}}.
\newblock \emph{\bibinfo{journal}{\apjs}} \textbf{\bibinfo{volume}{281}}, \bibinfo{pages}{20} (\bibinfo{year}{2025}).
\newblock \eprint{2504.21096}.

\bibitem{blinnikov1999}
\bibinfo{author}{{Blinnikov}, S.~I.}
\newblock \bibinfo{title}{{Modeling of the light curve and parameters of the supernova 1987A}}.
\newblock \emph{\bibinfo{journal}{Astronomy Letters}} \textbf{\bibinfo{volume}{25}}, \bibinfo{pages}{359--368} (\bibinfo{year}{1999}).

\bibitem{2006Natur.442.1018M}
\bibinfo{author}{{Mazzali}, P.~A.} \emph{et~al.}
\newblock \bibinfo{title}{{A neutron-star-driven X-ray flash associated with supernova SN 2006aj}}.
\newblock \emph{\bibinfo{journal}{\nat}} \textbf{\bibinfo{volume}{442}}, \bibinfo{pages}{1018--1020} (\bibinfo{year}{2006}).
\newblock \eprint{astro-ph/0603567}.

\bibitem{2020MNRAS.497.1619M}
\bibinfo{author}{{Moriya}, T.~J.}, \bibinfo{author}{{Suzuki}, A.}, \bibinfo{author}{{Takiwaki}, T.}, \bibinfo{author}{{Pan}, Y.-C.} \& \bibinfo{author}{{Blinnikov}, S.~I.}
\newblock \bibinfo{title}{{Systematic investigation of the effect of $^{56}$Ni mixing in the early photospheric velocity evolution of stripped-envelope supernovae}}.
\newblock \emph{\bibinfo{journal}{\mnras}} \textbf{\bibinfo{volume}{497}}, \bibinfo{pages}{1619--1626} (\bibinfo{year}{2020}).
\newblock \eprint{2007.04438}.

\bibitem{2018ApJS..236....6G}
\bibinfo{author}{{Guillochon}, J.} \emph{et~al.}
\newblock \bibinfo{title}{{MOSFiT: Modular Open Source Fitter for Transients}}.
\newblock \emph{\bibinfo{journal}{\apjs}} \textbf{\bibinfo{volume}{236}}, \bibinfo{pages}{6} (\bibinfo{year}{2018}).
\newblock \eprint{1710.02145}.

\bibitem{1982ApJ...253..785A}
\bibinfo{author}{{Arnett}, W.~D.}
\newblock \bibinfo{title}{{Type I supernovae. I - Analytic solutions for the early part of the light curve}}.
\newblock \emph{\bibinfo{journal}{\apj}} \textbf{\bibinfo{volume}{253}}, \bibinfo{pages}{785--797} (\bibinfo{year}{1982}).

\bibitem{2017ApJ...850...55N}
\bibinfo{author}{{Nicholl}, M.}, \bibinfo{author}{{Guillochon}, J.} \& \bibinfo{author}{{Berger}, E.}
\newblock \bibinfo{title}{{The Magnetar Model for Type I Superluminous Supernovae. I. Bayesian Analysis of the Full Multicolor Light-curve Sample with MOSFiT}}.
\newblock \emph{\bibinfo{journal}{\apj}} \textbf{\bibinfo{volume}{850}}, \bibinfo{pages}{55} (\bibinfo{year}{2017}).
\newblock \eprint{1706.00825}.

\bibitem{2022ApJ...941..107G}
\bibinfo{author}{{Gomez}, S.}, \bibinfo{author}{{Berger}, E.}, \bibinfo{author}{{Nicholl}, M.}, \bibinfo{author}{{Blanchard}, P.~K.} \& \bibinfo{author}{{Hosseinzadeh}, G.}
\newblock \bibinfo{title}{{Luminous Supernovae: Unveiling a Population between Superluminous and Normal Core-collapse Supernovae}}.
\newblock \emph{\bibinfo{journal}{\apj}} \textbf{\bibinfo{volume}{941}}, \bibinfo{pages}{107} (\bibinfo{year}{2022}).
\newblock \eprint{2204.08486}.

\bibitem{2025ApJ...980L..44M}
\bibinfo{author}{{Moore}, T.} \emph{et~al.}
\newblock \bibinfo{title}{{SN 2023zaw: The Low-energy Explosion of an Ultrastripped Star}}.
\newblock \emph{\bibinfo{journal}{\apjl}} \textbf{\bibinfo{volume}{980}}, \bibinfo{pages}{L44} (\bibinfo{year}{2025}).
\newblock \eprint{2405.13596}.

\bibitem{2019ApJ...887..169H}
\bibinfo{author}{{Ho}, A. Y.~Q.} \emph{et~al.}
\newblock \bibinfo{title}{{Evidence for Late-stage Eruptive Mass Loss in the Progenitor to SN2018gep, a Broad-lined Ic Supernova: Pre-explosion Emission and a Rapidly Rising Luminous Transient}}.
\newblock \emph{\bibinfo{journal}{\apj}} \textbf{\bibinfo{volume}{887}}, \bibinfo{pages}{169} (\bibinfo{year}{2019}).
\newblock \eprint{1904.11009}.

\bibitem{2025A&A...700A.200F}
\bibinfo{author}{{Finneran}, G.}, \bibinfo{author}{{Cotter}, L.} \& \bibinfo{author}{{Martin-Carrillo}, A.}
\newblock \bibinfo{title}{{Velocity evolution of broad-lined type-Ic supernovae with and without gamma-ray bursts}}.
\newblock \emph{\bibinfo{journal}{\aap}} \textbf{\bibinfo{volume}{700}}, \bibinfo{pages}{A200} (\bibinfo{year}{2025}).
\newblock \eprint{2411.11503}.

\bibitem{2026GCN.44239....1O}
\bibinfo{author}{{O'Dwyer}, T.} \emph{et~al.}
\newblock \bibinfo{title}{{EP260321a: VLA radio detection}}.
\newblock \emph{\bibinfo{journal}{GRB Coordinates Network}} \textbf{\bibinfo{volume}{44239}}, \bibinfo{pages}{1} (\bibinfo{year}{2026}).

\bibitem{2026GCN.44403....1J}
\bibinfo{author}{{Nayana A.}, J.}, \bibinfo{author}{{Wiston}, E.}, \bibinfo{author}{{Margutti}, R.}, \bibinfo{author}{{Sfaradi}, I.} \& \bibinfo{author}{{Chornock}, R.}
\newblock \bibinfo{title}{{EP260321a: Upper limits from ATCA radio observations}}.
\newblock \emph{\bibinfo{journal}{GRB Coordinates Network}} \textbf{\bibinfo{volume}{44403}}, \bibinfo{pages}{1} (\bibinfo{year}{2026}).

\bibitem{2026GCN.44357....1O}
\bibinfo{author}{{O'Dwyer}, T.} \emph{et~al.}
\newblock \bibinfo{title}{{EP260321a: VLA radio follow-up}}.
\newblock \emph{\bibinfo{journal}{GRB Coordinates Network}} \textbf{\bibinfo{volume}{44357}}, \bibinfo{pages}{1} (\bibinfo{year}{2026}).

\bibitem{2011AJ....141..163C}
\bibinfo{author}{{Clocchiatti}, A.}, \bibinfo{author}{{Suntzeff}, N.~B.}, \bibinfo{author}{{Covarrubias}, R.} \& \bibinfo{author}{{Candia}, P.}
\newblock \bibinfo{title}{{The Ultimate Light Curve of SN 1998bw/GRB 980425}}.
\newblock \emph{\bibinfo{journal}{\aj}} \textbf{\bibinfo{volume}{141}}, \bibinfo{pages}{163} (\bibinfo{year}{2011}).
\newblock \eprint{1106.1695}.

\bibitem{2014ApJS..213...19B}
\bibinfo{author}{{Bianco}, F.~B.} \emph{et~al.}
\newblock \bibinfo{title}{{Multi-color Optical and Near-infrared Light Curves of 64 Stripped-envelope Core-Collapse Supernovae}}.
\newblock \emph{\bibinfo{journal}{\apjs}} \textbf{\bibinfo{volume}{213}}, \bibinfo{pages}{19} (\bibinfo{year}{2014}).
\newblock \eprint{1405.1428}.

\bibitem{2020ApJ...902...86H}
\bibinfo{author}{{Ho}, A. Y.~Q.} \emph{et~al.}
\newblock \bibinfo{title}{{SN 2020bvc: A Broad-line Type Ic Supernova with a Double-peaked Optical Light Curve and a Luminous X-Ray and Radio Counterpart}}.
\newblock \emph{\bibinfo{journal}{\apj}} \textbf{\bibinfo{volume}{902}}, \bibinfo{pages}{86} (\bibinfo{year}{2020}).
\newblock \eprint{2004.10406}.

\bibitem{2026A&A...708A.305S}
\bibinfo{author}{{Stritzinger}, M.~D.} \emph{et~al.}
\newblock \bibinfo{title}{{The broad-lined type Ic supernova 2020lao experienced an energetic explosion with no central-engine signatures}}.
\newblock \emph{\bibinfo{journal}{\aap}} \textbf{\bibinfo{volume}{708}}, \bibinfo{pages}{A305} (\bibinfo{year}{2026}).

\bibitem{2025ApJ...982L..47V}
\bibinfo{author}{{van Dalen}, J. N.~D.} \emph{et~al.}
\newblock \bibinfo{title}{{The Einstein Probe Transient EP240414a: Linking Fast X-Ray Transients, Gamma-Ray Bursts, and Luminous Fast Blue Optical Transients}}.
\newblock \emph{\bibinfo{journal}{\apjl}} \textbf{\bibinfo{volume}{982}}, \bibinfo{pages}{L47} (\bibinfo{year}{2025}).
\newblock \eprint{2409.19056}.

\bibitem{2025ApJ...988L..13R}
\bibinfo{author}{{Rastinejad}, J.~C.} \emph{et~al.}
\newblock \bibinfo{title}{{EP 250108a/SN 2025kg: Observations of the Most Nearby Broad-line Type Ic Supernova Following an Einstein Probe Fast X-Ray Transient}}.
\newblock \emph{\bibinfo{journal}{\apjl}} \textbf{\bibinfo{volume}{988}}, \bibinfo{pages}{L13} (\bibinfo{year}{2025}).
\newblock \eprint{2504.08889}.

\bibitem{2025arXiv251210239S}
\bibinfo{author}{{Srinivasaragavan}, G.~P.} \emph{et~al.}
\newblock \bibinfo{title}{{EP250827b/SN 2025wkm: An X-ray Flash-Supernova Powered by a Central Engine and Circumstellar Interaction}}.
\newblock \emph{\bibinfo{journal}{arXiv e-prints}} \bibinfo{pages}{arXiv:2512.10239} (\bibinfo{year}{2025}).
\newblock \eprint{2512.10239}.

\bibitem{2008Natur.453..469S}
\bibinfo{author}{{Soderberg}, A.~M.} \emph{et~al.}
\newblock \bibinfo{title}{{An extremely luminous X-ray outburst at the birth of a supernova}}.
\newblock \emph{\bibinfo{journal}{\nat}} \textbf{\bibinfo{volume}{453}}, \bibinfo{pages}{469--474} (\bibinfo{year}{2008}).
\newblock \eprint{0802.1712}.

\bibitem{2008Sci...321.1185M}
\bibinfo{author}{{Mazzali}, P.~A.} \emph{et~al.}
\newblock \bibinfo{title}{{The Metamorphosis of Supernova SN 2008D/XRF 080109: A Link Between Supernovae and GRBs/Hypernovae}}.
\newblock \emph{\bibinfo{journal}{Science}} \textbf{\bibinfo{volume}{321}}, \bibinfo{pages}{1185} (\bibinfo{year}{2008}).
\newblock \eprint{0807.1695}.

\bibitem{2011ApJ...737...76K}
\bibinfo{author}{{Kochanek}, C.~S.}, \bibinfo{author}{{Szczygiel}, D.~M.} \& \bibinfo{author}{{Stanek}, K.~Z.}
\newblock \bibinfo{title}{{The Supernova Impostor Impostor SN 1961V: Spitzer Shows That Zwicky Was Right (Again)}}.
\newblock \emph{\bibinfo{journal}{\apj}} \textbf{\bibinfo{volume}{737}}, \bibinfo{pages}{76} (\bibinfo{year}{2011}).
\newblock \eprint{1010.3704}.

\bibitem{2007Natur.447..829P}
\bibinfo{author}{{Pastorello}, A.} \emph{et~al.}
\newblock \bibinfo{title}{{A giant outburst two years before the core-collapse of a massive star}}.
\newblock \emph{\bibinfo{journal}{\nat}} \textbf{\bibinfo{volume}{447}}, \bibinfo{pages}{829--832} (\bibinfo{year}{2007}).
\newblock \eprint{astro-ph/0703663}.

\bibitem{2013ApJ...767....1P}
\bibinfo{author}{{Pastorello}, A.} \emph{et~al.}
\newblock \bibinfo{title}{{Interacting Supernovae and Supernova Impostors: SN 2009ip, is this the End?}}
\newblock \emph{\bibinfo{journal}{\apj}} \textbf{\bibinfo{volume}{767}}, \bibinfo{pages}{1} (\bibinfo{year}{2013}).
\newblock \eprint{1210.3568}.

\bibitem{2024A&A...684L..18B}
\bibinfo{author}{{Brennan}, S.~J.} \emph{et~al.}
\newblock \bibinfo{title}{{Spectroscopic observations of progenitor activity 100 days before a Type Ibn supernova}}.
\newblock \emph{\bibinfo{journal}{\aap}} \textbf{\bibinfo{volume}{684}}, \bibinfo{pages}{L18} (\bibinfo{year}{2024}).
\newblock \eprint{2401.15148}.

\bibitem{2010AJ....139.1451S}
\bibinfo{author}{{Smith}, N.} \emph{et~al.}
\newblock \bibinfo{title}{{Discovery of Precursor Luminous Blue Variable Outbursts in Two Recent Optical Transients: The Fitfully Variable Missing Links UGC 2773-OT and SN 2009ip}}.
\newblock \emph{\bibinfo{journal}{\aj}} \textbf{\bibinfo{volume}{139}}, \bibinfo{pages}{1451--1467} (\bibinfo{year}{2010}).
\newblock \eprint{0909.4792}.

\bibitem{2022ApJ...938...57W}
\bibinfo{author}{{Woosley}, S.~E.} \& \bibinfo{author}{{Smith}, N.}
\newblock \bibinfo{title}{{SN 1961V: A Pulsational Pair-instability Supernova}}.
\newblock \emph{\bibinfo{journal}{\apj}} \textbf{\bibinfo{volume}{938}}, \bibinfo{pages}{57} (\bibinfo{year}{2022}).
\newblock \eprint{2205.06386}.

\bibitem{2015MNRAS.452.1567C}
\bibinfo{author}{{Chen}, T.-W.} \emph{et~al.}
\newblock \bibinfo{title}{{The host galaxy and late-time evolution of the superluminous supernova PTF12dam}}.
\newblock \emph{\bibinfo{journal}{\mnras}} \textbf{\bibinfo{volume}{452}}, \bibinfo{pages}{1567--1586} (\bibinfo{year}{2015}).
\newblock \eprint{1409.7728}.

\bibitem{2014AJ....147...60S}
\bibinfo{author}{{Smith}, B.~J.} \emph{et~al.}
\newblock \bibinfo{title}{{Extra-nuclear Starbursts: Young Luminous Hinge Clumps in Interacting Galaxies}}.
\newblock \emph{\bibinfo{journal}{\aj}} \textbf{\bibinfo{volume}{147}}, \bibinfo{pages}{60} (\bibinfo{year}{2014}).
\newblock \eprint{1401.0338}.

\bibitem{2013A&A...559A.114M}
\bibinfo{author}{{Marino}, R.~A.} \emph{et~al.}
\newblock \bibinfo{title}{{The O3N2 and N2 abundance indicators revisited: improved calibrations based on CALIFA and T$_{e}$-based literature data}}.
\newblock \emph{\bibinfo{journal}{\aap}} \textbf{\bibinfo{volume}{559}}, \bibinfo{pages}{A114} (\bibinfo{year}{2013}).
\newblock \eprint{1307.5316}.

\bibitem{2022JOSS....7.4508M}
\bibinfo{author}{{M{\"u}ller-Bravo}, T.} \& \bibinfo{author}{{Galbany}, L.}
\newblock \bibinfo{title}{{HostPhot: global and local photometry of galaxies hosting supernovae or other transients}}.
\newblock \emph{\bibinfo{journal}{The Journal of Open Source Software}} \textbf{\bibinfo{volume}{7}}, \bibinfo{pages}{4508} (\bibinfo{year}{2022}).
\newblock \eprint{2208.08117}.

\bibitem{2021ApJS..254...22J}
\bibinfo{author}{{Johnson}, B.~D.}, \bibinfo{author}{{Leja}, J.}, \bibinfo{author}{{Conroy}, C.} \& \bibinfo{author}{{Speagle}, J.~S.}
\newblock \bibinfo{title}{{Stellar Population Inference with Prospector}}.
\newblock \emph{\bibinfo{journal}{\apjs}} \textbf{\bibinfo{volume}{254}}, \bibinfo{pages}{22} (\bibinfo{year}{2021}).
\newblock \eprint{2012.01426}.

\bibitem{2007ApJ...657...810}
\bibinfo{author}{Draine, B.~T.} \& \bibinfo{author}{Li, A.}
\newblock \bibinfo{title}{Infrared emission from interstellar dust. iv. the silicate-graphite-pah model in the post-spitzer era}.
\newblock \emph{\bibinfo{journal}{The Astrophysical Journal}} \textbf{\bibinfo{volume}{657}}, \bibinfo{pages}{810--837} (\bibinfo{year}{2007}).

\end{thebibliography}

\newpage

\begin{methods}

\section{X-ray shock breakout}
\label{sec_med:SBO}
Einstein Probe (EP)\citemeth{2022hxga.book...86Y} is a space X-ray mission designed to monitor the soft X-ray sky for transient phenomena and to perform rapid onboard follow-up of newly discovered events. By combining the Wide-field X-ray Telescope (WXT) with the more sensitive Follow-up X-ray Telescope (FXT), EP is optimized for the discovery and early characterization of short lived high energy transients.
EP260321a was discovered by EP-WXT on 21 March 2026, 12:23:07 UTC (MJD 61120.516; T$_{0}$) as a fast, soft X-ray transient\citemeth{2026GCN.44068....1H}. The initial WXT exposure lasted only 
$\sim432$\,s before being interrupted by the spacecraft's autonomous follow-up sequence triggered by the transient itself, and thus represents only a lower limit on the true duration of the X-ray emission. Follow-up observations with the onboard EP-FXT started 12\,min later and revealed a previously uncatalogued, rapidly fading X-ray source within the WXT localization region. The X-ray spectra are well described by absorbed blackbody models with temperatures of $164^{+40}_{-29}$\,eV and $121^{+3}_{-3}$\,eV in the WXT and FXT data, respectively. A galaxy at $z=0.0345$ lies within the refined X-ray position; if associated, the transient reached a luminosity of $\sim2.2\times10^{44}$\,erg\,s$^{-1}$. These properties make EP260321a a compelling candidate for the shock breakout phase of a SN\citemeth{2026GCN.44075....1H}. 
Further analysis and results from the X-ray data are presented in W.~Yuan et al. in prep.

\section{Discovery of the optical counterpart}
\label{sec_med:OT}
The search for the optical counterpart to EP260321a began immediately after the EP trigger. 
Our Kinder/Lulin campaign obtained the first optical discovery of the counterpart to EP260321a, starting only 1.25\,hr after the X-ray detection\citemeth{2026GCN.44070....1L}. In these earliest Lulin observations, we identified a blue source at $\alpha,\delta = 09^{\rm h}59^{\rm m}42^{\rm s}88,\ +00^\circ25'06''3$ (J2000), projected near a galaxy (SDSS J095942.99+002504.0) within the EP localisation region (Fig.~\ref{fig:discovery_lc}). We noted that
this SDSS galaxy has a redshift of $z=0.034$, as independently measured in the 2dF and DESI Legacy Survey DR1 spectroscopic catalogues.
At first, this source was plausibly interpreted as an unrelated variable star, as no unambiguous new optical transient had yet emerged and several contemporaneous observations reported only upper limits\citemeth{2026GCN.44070....1L,2026GCN.44074....1M, 2026GCN.44076....1A, 2026GCN.44079....1L}.
However, the picture changed over the following day. Our continued optical monitoring showed that the source brightened\citemeth{2026GCN.44081....1A}, while independent detections from other facilities established a consistent counterpart at the same position\citemeth{2026GCN.44083....1M, 2026GCN.44084....1A}. Together with the early optical evolution\citemeth{2026GCN.44087....1L}, these data suggested that the initial emission was not dominated by a persistent variable star, but instead traced a rapidly evolving transient associated with EP260321a\citemeth{2026GCN.44089....1S}. 
Subsequent spectroscopy confirmed this interpretation\citemeth{2026GCN.44082....1T}. Broad SN-like features were identified at the redshift of the nearby galaxy, excluding a Galactic origin and demonstrating that the optical counterpart was an emerging broad-lined Type Ic SN\citemeth{2026GCN.44092....1X, 2026GCN.44105....1C, 2026GCN.44107....1R}. The combined X-ray and optical behaviour therefore supports a picture in which SN~2026gzf was discovered at an exceptionally early phase, when the SN emission was only beginning to emerge.

We note that a variable blue source at the same position as SN~2026gzf was previously identified in Pan-STARRS1 imaging (internally named PS18csa) on MJD 58465.55 (2018 December 13) with a $w$-band magnitude, $21.66\pm0.15$\,mag. This detection was subsequently reported to the International Astronomical Union Transient Name Server as AT~2018mtl. This detection corresponds to one of the episodes of precursor variability of SN~2026gzf (see Fig.~\ref{fig:ps1_precursor}).

\section{Photometric and Spectroscopic data}
\label{sec_med:data_collection}
Our analysis is built by the high cadence optical light curves obtained with the Lulin Observatory telescopes (LOT and SLT), which provide the core dataset of this work and capture the earliest evolution of SN~2026gzf. To improve the temporal coverage and mitigate weather-related gaps, we supplemented the Lulin data with additional imaging from Pan-STARRS, ATLAS, LSST, and public ZTF observations. At the earliest phases, we further included data from the Citra telescope and the Swope telescope obtained through POISE, which help bridge gaps in coverage caused by longitudinal separation.

In addition to the imaging data, we obtained optical spectra of SN~2026gzf near maximum light with the WiFeS, the NOT, and the LOT spectrographs. The WiFeS integral-field unit (IFU) spectroscopy also provides spatially resolved information on the host galaxy environment, enabling investigation of the nearby H\,II region and the local environment of the SN. We further include two NIR spectrum obtained with Keck~II/NIRES and IRTF/SpeX.

A log of the photometric and the spectroscopic observations are provided in Tables\,\ref{tab:spec_log} and \,\ref{tab:phot_log}, and the extinction-corrected spectra are presented in Fig.~\ref{fig:spectra}. All spectra reported here have been uploaded to WISeREP\citemeth{2012PASP..124..668Y}. We use AB magnitudes in this work. We adopt a flat $\Lambda$CDM cosmology with parameters $H_0 = 67.4\,\mathrm{km}\,\mathrm{s}^{-1}\,\mathrm{Mpc}^{-1}$, $\Omega_{\mathrm{m}} = 0.315$, and $\Omega_{\Lambda} = 0.685$, from the \textit{Planck} 2018 collaboration \citemeth{2020A&A...641A...6P}. Thus the distant module $\mu$ = 35.98, luminosity distance = 156.54\,Mpc. The Milky Way extinction is $A_{V}=0.067$\,mag\citemeth{2011ApJ...737..103S}.

The details of the instrument configuration are as follows:

\subsection{Kinder}

We initiated follow-up observations of EP260321a as part of the Kinder collaboration\citemeth{2025ApJ...983...86C} 
using the Lulin One-meter Telescope (LOT), equipped with the SOPHIA 2048BX camera, and the 40-cm Seisi Lulin Telescope (SLT)\footnote{``Seisi'' is a word from the Tsou, an Indigenous people of Taiwan, referring to a minivet. The telescope is named in recognition of the longstanding connection between Lulin Observatory and its Tsou resident assistants.}, equipped with an Andor EMCCD camera. 
The data were processed with a custom-built pipeline\footnote{https://hdl.handle.net/11296/98q6x4} that performs standard bias, dark, and flat-field corrections. 
Unless explicitly stated otherwise, all photometric measurements were performed using a homogeneous approach, including template subtraction, in which we used Pan-STARRS stacked images as templates. 
Whenever the observed filter matched the Pan-STARRS system, we used the corresponding Pan-STARRS image\citemeth{2016arXiv161205560C} as the subtraction template. 
The SN brightness was then measured with point spread function (PSF) photometry on the subtracted images using \textsc{AutoPhOT}\citemeth{2022A&A...667A..62B}, and the resulting magnitudes were calibrated using field stars from the ATLAS Refcat2 catalogue\citemeth{2018ApJ...867..105T}. 
To preserve the highest possible temporal resolution, we performed photometry directly on the individual exposure images. However, the observations obtained near peak ($\sim$MJD 61129 and MJD 61131) were affected by moonlight, so the individual images were combined to improve the signal-to-noise ratio (see details in the observing log).
For $u$-band measurements, the template and calibration were from SDSS. 

Follow-up optical spectra were obtained with the LOT using the Shelyak UVEX spectrograph. The instrumental setup, consisting of a 300 line mm$^{-1}$ grating, a 50 $\mu$m slit, and the ASI533MM Pro CMOS camera under bin-2 readout mode, provides a resolving power of $\sim300$, a wavelength coverage of 4200--7950\,\AA, and a dispersion of 2.49\,\AA\ pixel$^{-1}$. Standard reductions were applied, including bias and dark subtraction, flat-field correction with tungsten-illuminated dome flats, wavelength calibration using the internal ArNe lamp, and flux calibration with spectrophotometric standards GD 108 observed at similar airmass to the target. We adopted the A-B dithering pattern during the observation of the spectra, with an offset of 5'' along the R.A. direction, for a better background subtraction. 

\subsection{Citra}

We obtained images with the 0.5-m Citra Space Corporation's PlaneWave telescope in New Mexico, which is equipped with a CMOS camera and a Sloan filter set.
The CMOS camera Moravian C5A-150M has a $1.2\times1.9$ degrees field of view, with a pixel scale of 1''. Observations were carried out in $g$ and $r$ filters, with individual exposure time of 60 seconds. The raw data were processed following standard calibration procedures, including bias and dark subtraction as well as flat-field corrections.

\subsection{Pan-STARRS}

We undertook a multi-band ($grizyw$) observational follow-up campaign of SN~2026gzf with Pan-STARRS, commencing on MJD~61131.3. The PS telescope system consists of two 1.8-m telescopes (Pan-STARRS1 and Pan-STARRS2) located at the summit of Haleakala on the Hawaiian island of Maui\citemeth{2016arXiv161205560C}, and employs an SDSS-like filter system denoted as $grizy$, as well as a broad $w$ filter, which is a composite of the $gri$ filters\citemeth{Tonry2012_PS1_filters}. The imaging system provides a pixel scale of 0.26'' and a field of view of 7~deg$^{2}$\citemeth{2016arXiv161205560C}. All target images of SN~2026gzf were obtained with Pan-STARRS2, and were processed with the standard Pan-STARRS pipeline. Target images were astrometrically and photometrically calibrated, before having a historical Pan-STARRS1 $3 \pi$ reference image subtracted to produce a final difference image, from which we measure PSF photometry\citemeth{Magnier2020_calibration, Magnier2020_data_processing, Waters2020}.

\subsection{ATLAS}

Asteroid Terrestrial-impact Last Alert System (ATLAS) is a four-unit, 0.5-m wide-field survey system that monitors the entire sky\citemeth{2018PASP..130f4505T}. The transient is listed internally as ATLAS26dgs. The available ATLAS data were obtained in the cyan ($c$) and orange ($o$) bands, which roughly corresponds to the combined Pan-STARRS/SDSS $g+r$ and $r+i$ bands, respectively. We obtained the ATLAS points\citemeth{2020PASP..132h5002S} from the forced photometry service\citemeth{2021TNSAN...7....1S}.

\subsection{Rubin/LSST}

The Vera C. Rubin Observatory's Legacy Survey of Space and Time (LSST) is an upcoming wide-field, 
deep-optical synoptic survey that will scan the southern sky using an 8.4-meter telescope and a 
3.2-gigapixel camera, driving unprecedented discoveries in time-domain astronomy \citemeth{2019ApJ...873..111I}. To construct the optical light curves, we retrieved the photometric data for LSST object ID 
\texttt{314003014107006318} utilizing the Lasair broker \citemeth{2019RNAAS...3...26S}. 

\subsection{ZTF}

Zwicky Transient Facility (ZTF) is a wide-field optical synoptic survey carried out with the Palomar 48-inch telescope, using a 47 deg$^{2}$ camera and $g$,$r$, and $i$ filters for time-domain discovery and monitoring\citemeth{2019PASP..131a8002B, 2019PASP..131g8001G}. We supplemented our follow-up photometry with public forced ZTF measurements retrieved from the Lasair broker\citemeth{2019RNAAS...3...26S}. The transient is internally designated ZTF26aaonmha, and the available public light curve consists of $g$- and $r$-band observations. 

\subsection{WiFeS}

The Wide-Field Spectrograph (WiFeS) is an Integral Field Unit mounted on the Australian National University (ANU) 2.3m Telescope\citemeth{2007Ap&SS.310..255D}, located at Siding Spring Observatory in Coonabarabran, Australia. WiFeS spectroscopy of SN~2026gzf was originally obtained on the 31st of March 2026 near maximum light, using the Nod and Shuffle mode for three 1800s exposures, each made up of 900s on source and 900s on sky. We utilised the B3000 and R3000 gratings, along with the RT560 dichroic to produce R$\sim$3000 spectra from 3300--9200\,\AA. The data were reduced using the \textsc{PyWiFeS} pipeline\citemeth{2014Ap&SS.349..617C}. We further acquired an additional 4 WiFeS observations under darker sky conditions, and stacked the resulting datacubes to use in the spectral analysis of host galaxy properties.

\subsection{PRESTO and NUTS2}

Three optical spectra of SN~2026gzf were obtained with the 2.56-m  Nordic Optical Telescope (NOT)\citemeth{2010ASSP...14..211D} at the Roque de los Muchachos Observatory on La Palma, Spain, one as part of the PRESTO project and two by the NOT Unbiased Transient Survey 2 (NUTS2)\citemeth{2019ATel12661....1H}. The observations were carried out with the Alhambra Faint Object Spectrograph and Camera (ALFOSC) and the data were reduced following standard procedures and flux calibrated using observations of spectrophotometric standard stars obtained on the same nights.

\subsection{POISE}

Optical $ugriBV$ imaging was obtained as part of the Precision Observations of Infant SupernovaE (POISE) programme \citemeth{2021ATel14441....1B} using the Swope 1-m telescope at Las Campanas Observatory and reduced using the Carnegie Supernova Project (CSP) photometric pipeline \citemeth{2017AJ....154..211K}. The reductions included bias subtraction, flat-field correction, astrometric registration, and PSF photometry measured relative to a local sequence of field stars. The local sequence was calibrated using stars from the Refcat2 catalogue \citemeth{2018PASP..130f4505T} transformed to the natural system of the Swope telescope. Host-galaxy contamination was removed through image subtraction using Pan-STARRS1 (PS1) reference images. Specifically, PS1 $g$-band images were used for subtraction of the $B$, $V$, and $g$-band data, while PS1 $r$- and $i$-band images were used for subtraction of the corresponding $r$- and $i$-band observations, respectively. Image subtractions were performed with the ImageMatch subtraction software\footnote{\url{https://code.obs.carnegiescience.edu/Algorithms/imagematch}}. No template subtraction was applied to the $u$-band imaging.

\subsection{HISS}

Three NIR spectra were obtained on the 10-m W. M. Keck Observatory Telescope II (Keck II) with the Near-Infrared Echellette Spectrometer (NIRES)\citemeth{2004SPIE.5492.1295W} and the 3.2-m NASA Infrared Telescope Facility (IRTF) with the SpeX Spectrograph\citemeth{2003PASP..115..362R} at Maunakea, Hawaii.
 The observations are as part of the Hawaii Infrared Supernova Study (HISS)\citemeth{2025ApJS..281...28M,2025OJAp....8E.120H,2025ApJ...988..209H,2025ApJ...993..191M}. The data were reduced using the methods outlined in \citemeth{2025ApJS..281...28M}.

\subsection{HSC-SSP}
The Hyper-Suprime Cam (HSC)\citemeth{2018PASJ...70S...1M} is a powerful wide-field mosaic camera mounted on the prime focus of the Subaru 8.2-m telescope, which consists of 104 2k x 4k scientific CCDs with a mean pixel scale of 0.168". It has an effective field of view of 1.5 degrees in diameter, with a modified sloan \textit{griz} filter system. We collected the survey data under the framework of the Hyper-Suprime Cam Subaru Strategic Program (HSC-SSP)\citemeth{10.1093/pasj/psx066}, which is a dedicated survey to answer the cosmological problem. In the latest HSC-SSP Public Data Release version 3 (HSC-SSP PDR3)\citemeth{2022PASJ...74..247A}, we retrieved 43 well-calibrated 'CORR' level HSC snapshot products, separate in 14 nights. 

\subsection{DESI Legacy Imaging Surveys}
Historical pre-explosion imaging of the SN~2026gzf site was obtained with the Dark Energy Camera (DECam)\citemeth{2015AJ....150..150F} mounted on the V\'{\i}ctor M. Blanco 4-m telescope at the Cerro Tololo Inter-American Observatory. We retrieved archival pipeline-reduced science images in the $g$, $r$, $i$, and $z$ bands through the NOIRLab Astro Data Archive. These public archival data primarily consist of observations from the DESI Legacy Imaging Surveys\citemeth{2019AJ....157..168D} and earlier DECam programmes. The dataset comprises a total of 55 individual exposures, spanning from April 2013 to February 2020. These long-term multi-band observations provide critical data for assessing the precursor activity at the explosion site prior to the Einstein Probe trigger. 

\section{Spectroscopy of the broad-lined Type Ic SN~2026gzf and TARDIS modelling}
\label{sec_med:spec_2026gzf}

Figure\,\ref{fig:spectra} presents the rest-frame spectral sequence of SN~2026gzf from $+9$ (near maximum light) to $+43$\,days relative to the EP X-ray trigger time (MJD = 61120.516; $T_{0}$), based on optical observations obtained with NOT/ALFOSC, ANU 2.3-m/WiFeS and LOT/UVEX, together with NIR spectra from Keck II/NIRES and IRTF/SpeX. For visual comparison, the spectra are normalized and vertically offset by arbitrary constants. The optical sequence shows the broad absorption and emission features characteristic of a broad-lined Type~Ic SN. Superposed on these are strong narrow host-galaxy H\,II-region emission lines, including prominent H$\alpha$, H$\beta$, and [O~III] $\lambda\lambda4959,5007$. The NIR spectra extend the wavelength coverage to $\sim2.4\,\mu$m and provide additional constraints on the spectral energy distribution and line identifications at later phases. For context, we also overplot spectra of SN~2006aj at comparable epochs\citemeth{2008AstBu..63..228S,2006ApJ...645L..21M,2006Natur.442.1011P}, which illustrate the close overall resemblance in spectral evolution. This spectral sequence forms the basis for the line identification and radiative-transfer modelling discussed below.

We employ \textsc{TARDIS} \citemeth{2014MNRAS.440..387K} to model the observed optical-NIR spectra. For modelling of the NOT optical spectrum at $+17$\, day and Keck NIR spectrum at $+15$\, day, we used the same model generated with \textsc{TARDIS} at $+16$\,days since the explosion. The other two models are generated at $+33$\,days and $+38$\,days since explosion respectively.

The optical spectra have been scaled with $r$-band and the NIR spectra with $J$-band photometry.
In the \textsc{TARDIS} version (v2025.8.17) used in this work the code assumes homologous expansion of the ejecta, and all the features contributing to the spectrum for a given epoch to be above an inner velocity ($v_{inner}$). The following density profile has been used for the simulations - 
\begin{equation}
    \rho = \rho_{0} (\frac{v}{v_{0}})^{-n} (\frac{t_{exp}}{t_{0}})^{-3}
\end{equation}
with $\rho_{0}$ $=$ 4.0 $\times$ 10$^{-13}$ g~cm$^{-3}$ $v_{0}$ $=$ 13000 km~s$^{-1}$ at $t_{0}$ $=$ 10.0\,days.
We used a single power law with an index $n =$ 6 as is used in other studies \citemeth{2019Natur.565..324I, 2022ApJ...937...40K}. In our simulation we have used a uniform composition of elements above $v_{inner}$ with mass fractions X(He) = 0.0153, X(C) = 0.076, X(O) = 0.763, X(Na) = 0.015, X(Mg) = 0.076, X(Si) = 0.015, X(S) = 0.008, X(Ca) = 7.6 $\times$ $10^{-6}$, X(Ti) = 1.5 $\times$ $10^{-4}$, X(Cr) = 1.5 $\times$ $10^{-4}$, X(Fe) = 0.015, X($^{56}$Ni) = 0.015. The $^{56}$Ni mass fraction is defined at 10.0\,days. The inner velocity for our first epoch is placed at 16000\,km\,s$^{-1}$ and the outer velocity at 40000\,km\,s$^{-1}$. \textsc{TARDIS} assumes a variety of approximations to calculate the ionization and excitation state of the ejecta\citemeth{2014MNRAS.440..387K}. For this study we used the \texttt{nebular} mode for ionization and \texttt{dilute-lte} for excitation for all the elements except He. The \texttt{macroatom} mode was used for line interaction type.  We used an analytical approximation for He called \texttt{recomb-nlte} which calculates the He ion/level population relative to the He\,II ground state assuming He\,I ground-state population to be negligible \citemeth{2012MNRAS.422...70H, 2017A&A...599A..46B}. 

The effect is more prominently seen in the later epochs as the temperature drops. The strongest feature due to He\,I at 10380\,\AA ~is present in the spectra at $+$33 days and $+$38 days as is seen in other studies\citemeth{2017A&A...599A..46B, 2021ApJ...908..150W, 2026ApJ..1002L..11L}. The other He lines at $\sim$ 5877 Å, and $\sim$ 20587 Å have much lower Sobolev optical depth.
We find from our simulations that the major features contributing to the optical spectrum are Fe\,{\sc ii}, Si\,{\sc ii}, O\,{\sc i}, Na \,{\sc i} and in the NIR by Mg\,{\sc ii}, Ca\,{\sc ii}, Mg\,{\sc i}, He \,{\sc i}. We find that the Ca II feature around 8000 Å is absent in our model at +16 days (marked blue in the first panel of Figure \ref{fig:spectroscopic_modeling}), while using the same Ca mass fraction for other epochs, a pronounced Ca II NIR feature appears around 8000 \AA. 
In our modelling, we get a total ejecta mass of 1.15 $M_{\odot}$ above 13000\,km\,s$^{-1}$,
and a total He mass of 0.018 $M_{\odot}$ based on the composition and density profile used. 

\begin{figure}
    \centering
    \includegraphics[width=0.8\linewidth]{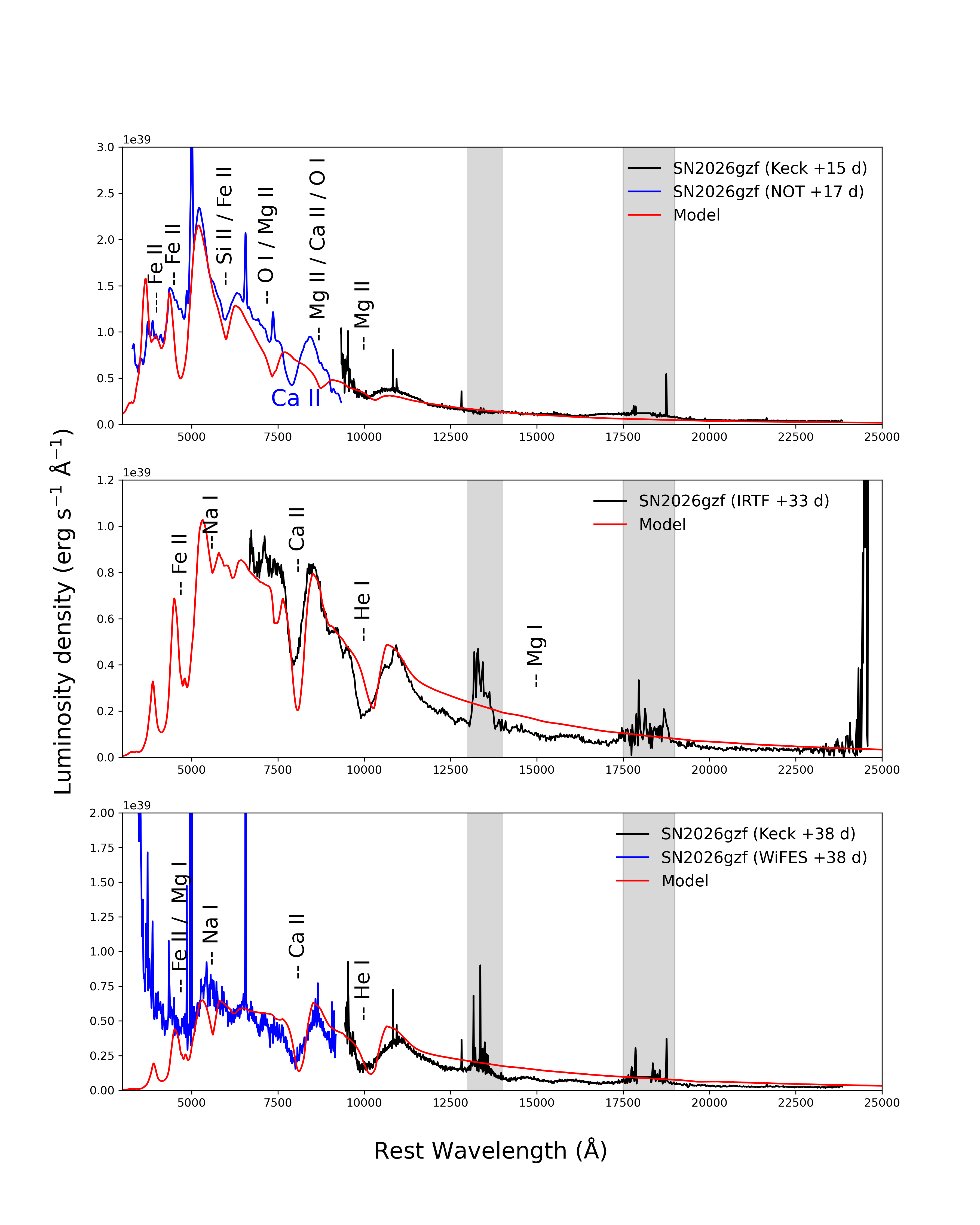}
    \caption{
        Model spectra of SN2026gzf at three different epochs. The observed spectra have been dereddened with $R_{V}$=3.1 and an extinction of $A_{V}$=0.067\,mag and deredshifted with $z$ = 0.03445. The flux spectra have been converted to luminosity using a $\mu$ = 35.99 mag. The data have been smoothed with a Gaussian filter. The telluric bands in the NIR spectra are marked with grey vertical regions. The features contributing to the spectrum are identified with \textsc{TARDIS} spectral element decomposition. 
            }
    \label{fig:spectroscopic_modeling}
\end{figure}

\section{Pre-burst activities}
\label{sec_med:pre-burst activities}

\begin{figure}
    \centering
    \includegraphics[width=\linewidth]{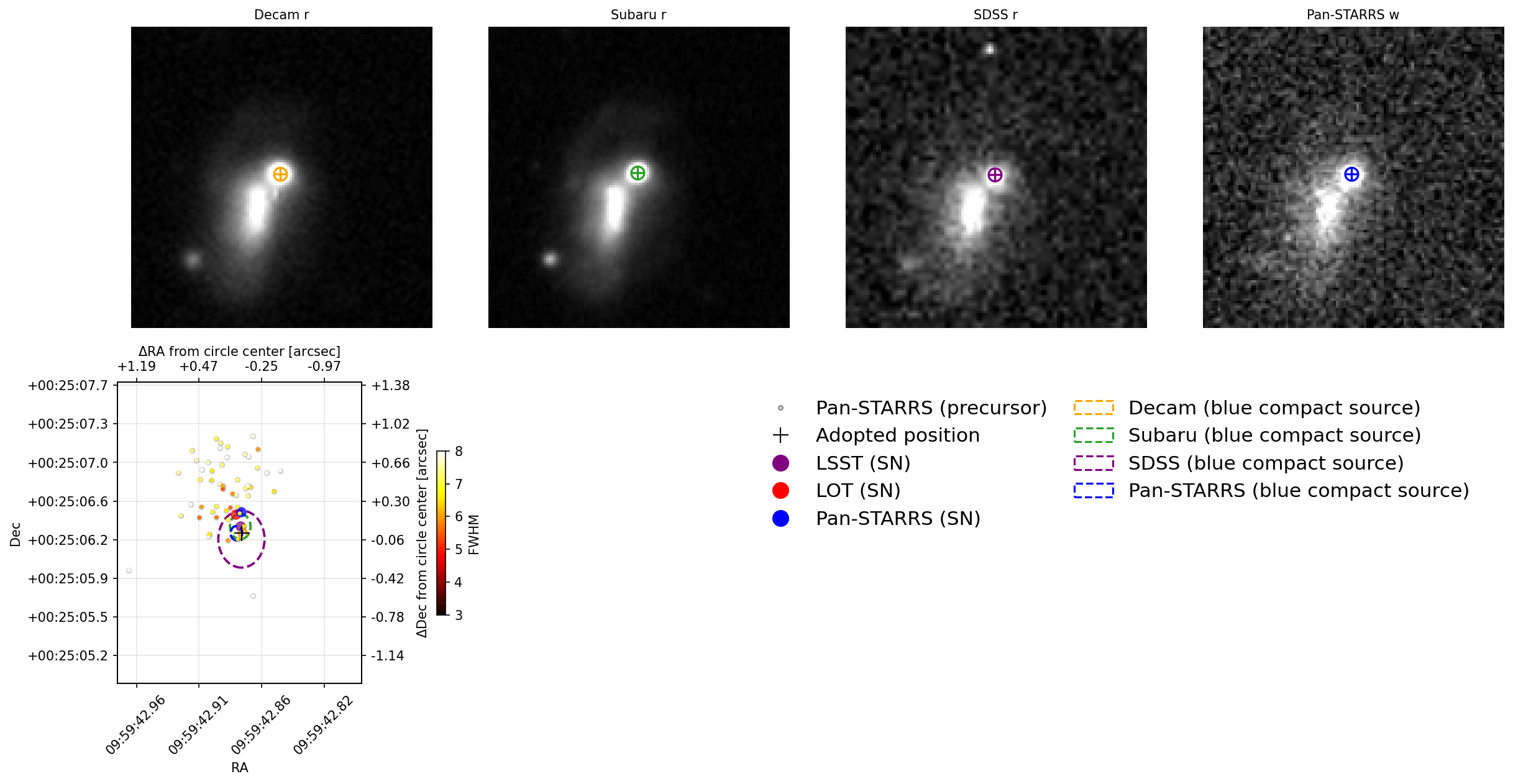}
    \caption{
        \textbf{Astrometric association between the historical blue compact source, the Pan-STARRS precursor residuals, and SN~2026gzf.}
        Top, archival cutouts of the explosion site from DECam $r$, Subaru $r$, SDSS $r$, and Pan-STARRS $w$, registered to a common astrometric frame. The coloured crosshairs mark the independently measured centroid of the blue compact source in each dataset. Bottom, sky coordinate comparison of the measured source centroids. The small open circles show the individual Pan-STARRS precursor positions measured from $w$-band difference image residuals, with the colour indicating the image seeing at each epoch. Filled coloured circles mark representative later SN positions from optical imaging. The dashed ellipses indicate the astrometric uncertainty of the blue compact source centroid in each archival dataset. The best-seeing precursor residuals lie very close to the SN and blue compact source positions, while small offsets are expected because the centroid measured from a difference image residual depends to some extent on the adopted reference frame. Overall, the pre-explosion source, the precursor residuals, and the SN positions are consistent within the astrometric uncertainties, supporting a common origin.
    }
    \label{fig:Astrometry}
\end{figure}

\subsection{Astrometric registration and source localisation}

We quantified the positional association between SN~2026gzf and the historical blue compact source, by which we mean the persistent blue source visible in archival direct images prior to explosion, rather than the later Pan-STARRS difference image residuals. In the present analysis we used archival direct images from DECam, Subaru/Hyper Suprime-Cam, SDSS, and Pan-STARRS, and compared these with representative post-explosion SN positions shown in Fig.~\ref{fig:Astrometry} and listed in Table~\ref{tab:astrometry}. Each archival image was tied to the \textit{Gaia} DR3 astrometric frame using nearby reference stars retrieved through VizieR (\texttt{I/355/gaiadr3}), retaining stars brighter than $G=20$~mag and excluding objects close to the image boundaries. For each image, the astrometric solution was refined in two stages. We first estimated a bulk translational offset by searching for the nearest local peaks around the WCS-predicted \textit{Gaia} positions within a coarse search radius of 30 pixels, using a centroid box with a half-size of 7 pixels. The resulting shifts were sigma-clipped at the $3\sigma$ level to derive a robust global offset for each frame. We then carried out a finer local search within 10 pixels and fitted an affine transformation between the predicted and observed reference star positions. Only matches within 4 pixels of the shifted prediction were retained in the affine fit, and a residual clip of 2.5 pixels was applied. The astrometric uncertainty for each image includes the rms residual of this local affine solution.

We measured the centroid of the historical blue compact source in each archival direct image using \texttt{photutils}, extracting a small cutout around the target position, estimating the local background from the border pixels of the cutout, subtracting this background, setting negative pixel values to zero, and computing the centroid from the remaining flux distribution. The centroid uncertainty was estimated from the weighted second moments of the source light distribution and combined in quadrature with the local astrometric rms. For archival images in which a stable local centroid could not be recovered, we retained the Gaia-registered prior-projected position rather than forcing an unstable centroid solution. The adopted pre-explosion position was defined from the archival direct image centroids of the blue compact source alone, and not from the Pan-STARRS precursor residuals or the later SN positions. The latter are shown separately in Fig.~\ref{fig:Astrometry} and Table~\ref{tab:astrometry} as consistency checks. The Pan-STARRS precursor positions were measured from $w$-band difference image residuals and are therefore expected to show small centroid shifts that depend to some extent on the adopted reference frame and subtraction template. Likewise, the later SN positions are not used to define the pre-explosion centroid, but instead to test whether the explosion occurred at the same site.

Using the weighted mean of the archival direct image centroids listed in Table~\ref{tab:astrometry}, we adopt a pre-explosion astrometric reference position of
$\alpha = [\mathrm{09{:}59{:}42.88}]$,
$\delta = [+\mathrm{00{:}25{:}06.3}]$
(J2000). The corresponding uncertainties are $\sigma_{\rm RA,sky}=0.0268$~arcsec and $\sigma_{\rm Dec}=0.0232$~arcsec. All archival direct image centroids of the blue compact source are mutually consistent with this adopted position within the quoted astrometric uncertainties, and the later SN positions are consistent with the same site to within the overall astrometric error budget. Unless otherwise noted, we use this adopted astrometric reference position throughout the paper as the fiducial coordinate of the explosion site.

\begin{figure}
    \centering
    \includegraphics[width=\linewidth]{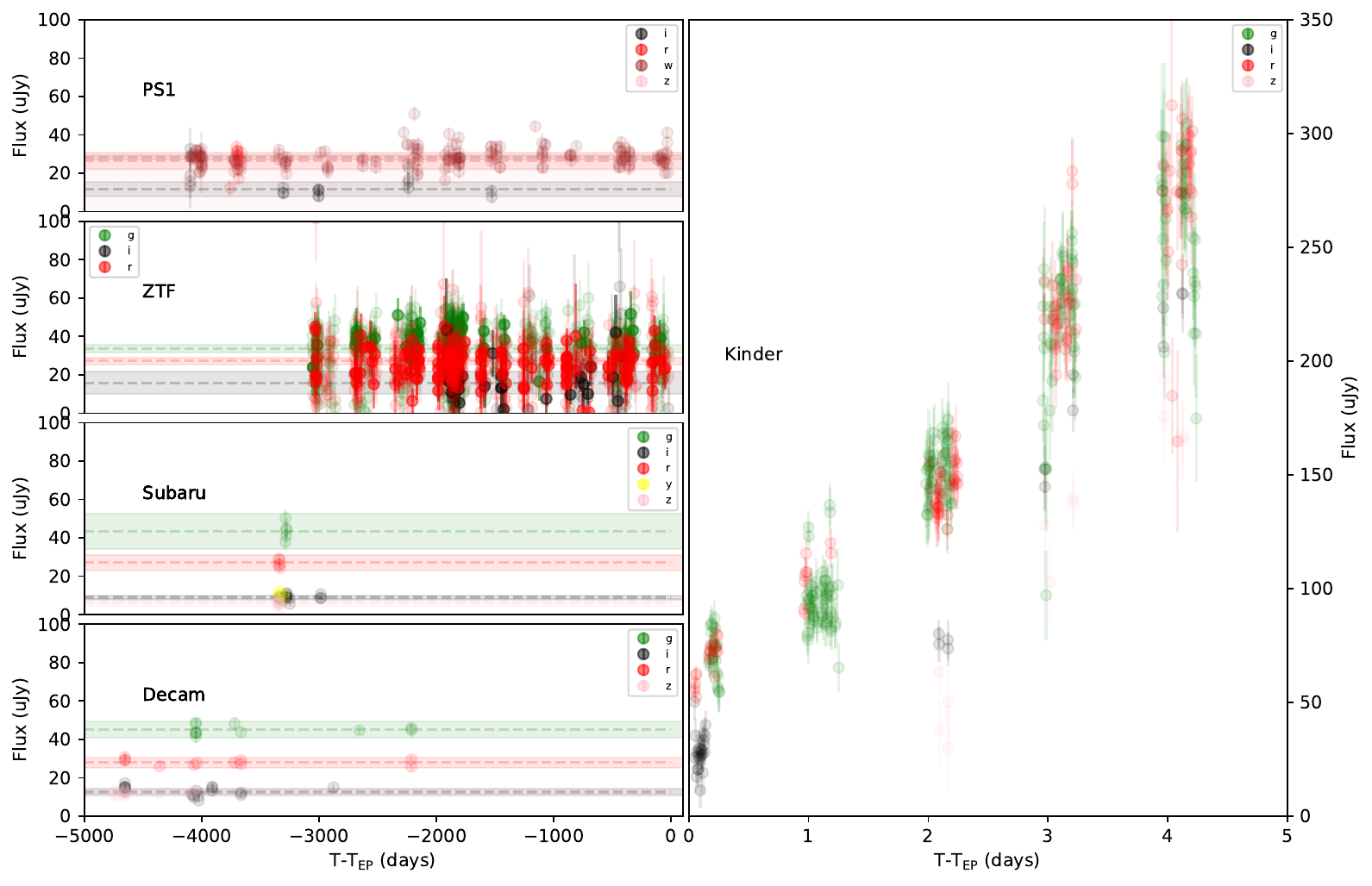}
    \caption{
        \textbf{Long-term pre-explosion variability and the early rise of SN~2026gzf.}
        Left, historical photometry measured at the explosion site from Pan-STARRS, ZTF, Subaru, and DECam. The pre-explosion measurements are based on direct aperture photometry at a fixed position and without template subtraction, in order to track the total flux at the site of the blue compact source in a survey consistent way. For each epoch, the source aperture radius was set to $0.8\times{\rm FWHM}$ and the local background was estimated from a concentric annulus with inner and outer radii of $1.0\times{\rm FWHM}$ and $1.5\times{\rm FWHM}$, respectively. The dashed lines and shaded bands indicate the mean pre-explosion flux level and its uncertainty in each band. Right, the rapid optical rise measured by the Kinder/Lulin follow-up during the first days after EP260321a. The strongest pre-explosion activity is seen in the blue bands, indicating that the explosion site was not fully quiescent before the final outburst.
    }
    \label{fig:preburst}
\end{figure}

\subsection{Archival aperture photometry and flux homogenisation}

To characterise the long-term behaviour of the explosion site, we measured circular aperture fluxes at the fixed adopted source position on archival images from Pan-STARRS1, ZTF, Subaru/Hyper Suprime-Cam, and DECam. All images were analysed with the same aperture photometry procedure using \texttt{photutils.aperture\_photometry}, in order to minimise survey dependent systematics. ZTF images were retrieved from IRSA, while the Pan-STARRS1, Subaru, and DECam images were taken from the corresponding survey or archival image products used throughout this work. 

For each frame, the source aperture radius was scaled to the image quality and set to $0.8\times{\rm FWHM}$, while the local background was estimated from a concentric annulus with inner and outer radii of $1.0\times{\rm FWHM}$ and $1.5\times{\rm FWHM}$, respectively. We adopted this relatively local background estimate because the source is projected on a structured host environment, so larger annuli risk incorporating unrelated galaxy substructure. The background level was estimated from sigma-clipped annular pixels, and the aperture-sum uncertainty was combined with the local background rms to obtain the final photometric uncertainty.
To minimise calibration offsets, all measurements were expressed in flux-density units (\(\mu\)Jy) and calibrated against the same local Pan-STARRS1 field-star sequence, using nearby unsaturated stars and the closest available Pan-STARRS1 bandpass for each image filter. This common calibration strategy was applied consistently to Pan-STARRS1, ZTF, Subaru and DECam images, so that the measured differences primarily reflect source variability, bandpass mismatch and host-background structure rather than differences in the adopted photometric reference system. We did not apply explicit S-corrections across surveys; any remaining inter-survey offsets are therefore treated conservatively as arising from a combination of bandpass mismatch, seeing differences and structured host-background contamination.
For the pre-explosion analysis, we used direct aperture photometry rather than PSF photometry on template subtracted images. Because a blue compact source is already present at the explosion site before outburst, the residual flux in a difference image depends sensitively on the adopted reference frame. In practice, different reference images can produce different residual amplitudes, making small long-term variations difficult to interpret robustly. Direct aperture photometry at a fixed position therefore provides a more stable and survey consistent way to assess the baseline flux level and flux variance of the unresolved pre-explosion source. 

The long-lived activity of the source site is also independently suggested by the fact that the PS1 team reported the source in 2018 as PS18csa (later TNS name AT~2018mtl\footnote{https://www.wis-tns.org/object/2018mtl}), discovered on 2018 December 13 at a $w$-band magnitude of 21.66. This motivated our closer examination of the pre-explosion variability in the Pan-STARRS1 data. However, because the direct aperture photometry used here measures the total flux at the site without template subtraction, we do not use it alone to quantify the amplitude of the precursor variability. Instead, in the following subsection we analyse the Pan-STARRS1 pre-explosion images with template subtraction and PSF-based photometry to confirm and characterise the variability more directly.

Independent aperture photometry from Pan-STARRS1, ZTF, Subaru, and DECam yields broadly consistent baseline fluxes at the source position (Table~\ref{tab:preflux}). In particular, the mean $r$-band fluxes measured from the four surveys are mutually consistent within the expected survey-to-survey scatter, while the source is systematically brighter in the blue than in the redder optical bands. Table~\ref{tab:preflux} also lists the corresponding AB magnitudes, computed from \(m_{\rm AB}=23.9-2.5\log_{10}(f_{\mu{\rm Jy}})\). In Methods 5.3, we use the band-specific \emph{Average} AB magnitudes listed in the final row of each filter group to define the magnitude ranges of the comparison samples in the chance-alignment calculation.

\subsection{Chance-alignment probability of an unrelated foreground source}

To quantify the probability that the pre-explosion blue compact source is an unrelated source projected by chance onto the position of SN~2026gzf, we estimated the local surface density of catalogue sources with comparable apparent brightness in a control field centred on the explosion site. Our primary calculation used the Pan-STARRS1 DR2 mean-source catalogue within a radius of 5 arcmin around the adopted astrometric reference position, excluding the immediate host region with a circular mask of radius 5 arcsec. For each optical band, the fiducial apparent magnitude of the blue compact source was taken from the band-specific \emph{Average} AB magnitudes listed in Table~\ref{tab:preflux}. Comparison samples were then selected within a magnitude interval of \(\pm0.5\) mag around the corresponding value for the fiducial calculation.

For a given comparison sample, the chance-alignment probability was computed analytically as
\[
P = 1 - \exp(-\pi r_{\rm match}^{2}\Sigma),
\]
where \(r_{\rm match}\) is the adopted matching radius and \(\Sigma\) is the local surface density of comparison sources in units of arcsec\(^{-2}\). Because the formal astrometric uncertainty of the adopted pre-explosion position is much smaller than the scatter among the pre-explosion and post-explosion positional measurements, we adopted a deliberately conservative matching radius of \(r_{\rm match}=0.3\) arcsec for the fiducial calculation. We report two related probabilities: \(P_{\rm any}\), using all catalogue sources of comparable apparent brightness, and \(P_{\rm point}\), restricting the comparison sample to unresolved point-like sources selected from the Pan-STARRS1 catalogue using PSF--Kron compactness criteria.

For the fiducial case (Run A in Table~\ref{tab:chancealign}), the inferred probabilities are low in all optical bands, with the smallest values obtained in the blue. Representative values are \(P_{\rm any}=3.70\times10^{-5}\) in \(g\) and \(2.07\times10^{-4}\) in \(r\), while restricting the comparison sample to point-like sources yields \(P_{\rm point}=1.50\times10^{-5}\) in \(g\) and \(8.90\times10^{-5}\) in \(r\). The corresponding probabilities in the redder bands are somewhat larger because the sky density of comparable sources is higher.
As robustness tests, we repeated the same calculation with a substantially larger matching radius of \(r_{\rm match}=0.5\) arcsec (Run B), and then with both \(r_{\rm match}=0.5\) arcsec and a wider magnitude interval of \(\Delta m=\pm1.0\) mag (Run C). As expected, the inferred probabilities increase under these deliberately more conservative assumptions, but remain low and do not alter the qualitative conclusion (Table~\ref{tab:chancealign}).
As an independent cross check, we also estimated the same probability by Monte Carlo sampling. We drew \(10^{5}\) random sky positions uniformly within the same 5-arcmin control field, rejected points falling inside the excluded host mask, and measured the fraction whose nearest comparison object lay within \(r_{\rm match}\). The Monte Carlo results are consistent with the analytic surface density estimates, supporting the robustness of the calculation.

These single band probabilities are themselves conservative, because they condition only on positional coincidence and comparable brightness in one filter at a time. The pre-explosion source is also unusually blue, and requiring a comparison object to match both the position and the observed colour of the blue compact source would reduce the chance-alignment probability further. We therefore regard the quoted single band values as conservative upper limits on the probability that the pre-explosion blue compact source is an unrelated foreground-like object.

\subsection{Photometry on Pan-STARRS pre-explosion images}

All Pan-Starrs pre-explosion images were stacked on a nightly basis using SCAMP\citemeth{bertin2006scamp} for astrometric refinement and SWarp\citemeth{bertin2010swarp}. Stacked science and template images were then processed through the AutoPhOT pipeline\citemeth{brennan2022autophot}. Astrometric calibration was performed using \texttt{astrometry.net}\citemeth{lang2010astrometry} to align the World Coordinate System to the Gaia DR3 reference frame\citemeth{gaia2023dr3}. Template images were aligned to the science frame using SCAMP\citemeth{bertin2006scamp} for astrometric refinement and SWarp\citemeth{bertin2010swarp} for resampling onto the science pixel grid, with distortion corrections. Alignment quality was verified through the rms scatter of matched sources. The PSFs were constructed empirically using PhotUtils\citemeth{bradley2022photutils} from isolated, high signal-to-noise stellar sources identified by SExtractor\citemeth{bertin1996sextractor}. Image subtraction was performed using the Sparse Field Flux Transport algorithm\citemeth{russell2022sfft}, which determines an optimal convolution kernel to match the template PSF to the science image. The template was convolved to match the science PSF such that transient sources retain the science frame characteristics in the difference image. SFFT operates in either sparse-field or crowded-field mode depending on the stellar density. Photometric calibration was propagated through the subtraction process, and difference image quality was assessed via background statistics and photometric consistency of non-variable sources.

Photometric calibration was performed using synthetic photometry from Gaia DR3 spectrophotometry. Reference star magnitudes in the target filter system were derived using GaiaXPy\citemeth{weiler2023gaiaxpy} to convolve the Gaia low-resolution spectra with filter transmission curves obtained from the Spanish Virtual Observatory (SVO) Filter Profile Service\citemeth{rodrigo2012svo}. This approach provides accurate synthetic magnitudes for calibration stars across a wide range of colours and spectral types, enabling consistent zero-point determination for each image. However, the magnitudes of the precursors represent a lower limit, as the contribution of the progenitor to the total light from the blue compact source at the time of the reference stack is unknown.

Limiting magnitudes were determined using a photometric injection and recovery framework. Artificial point sources with magnitudes spanning the expected detection threshold were generated using the empirical PSF and added at random positions across the difference image. The $3\sigma$ limiting magnitude was defined as the faintest injected magnitude at which sources were recovered with signal-to-noise ratio $\geq 3$ and positional accuracy within $0.5$ pixels. 

Figure\,\ref{fig:ps1_precursor} shows a subset of $w$-band differences images. For the majority of $w$-band images, a clear detection is seen at the position of SN~2026gzf

\subsection{Search for periodicity in the archival variability}

To test whether the historical variability could be explained by an ordinary periodic variable star, we searched for coherent periodicity in the pre-explosion Pan-STARRS1 light curve using the P4J periodogram package\citemeth{2018ApJS..236...12H}. We restricted the analysis to pre-trigger measurements and removed entries with non-finite times, fluxes, or flux uncertainties, as well as measurements with non-positive uncertainties. The periodogram was evaluated over trial periods between 1 and 1000\,days.

The resulting periodogram shows no significant peak over this range. 
In particular, no peak exceeds the adopted significance threshold, indicating that the archival PS light curve does not show convincing evidence for coherent periodic modulation. We therefore disfavor an ordinary periodic stellar variable as the origin of the pre-explosion variability.

\section{Early luminosity excess of SN~2026gzf}
\label{sec_med:early_excess}

Optical follow-up of EP260321a started at Lulin Observatory 1.25\,h after the EP-WXT trigger, using the 1-m LOT and the 40-cm SLT, as part of the Kinder (Kilonova Finder) programme\citemeth{2025ApJ...983...86C, 2025ApJS..281...20A}. The first observations began on 2026 March 21 at 13:38:02 UT (MJD 61120.568079), with SLT following 23\,s later. In this first epoch, LOT obtained six 300\,s $r$-band exposures, while SLT obtained 24 separate 300\,s $i$-band exposures. 
LOT later secured two additional $r$-band observing blocks. The second LOT block started at 4.15\,h after T$_{0}$) and comprised six 300\,s $r$-band exposures. After a $\sim44$\,min gap, LOT obtained a third block of seven 300\,s $r$-band exposures. During this later epoch, SLT obtained 24 frames of 300\,s $g$-band exposures. The resulting high-cadence light curve during the first hours is shown in the zoomed-in panel of Fig.,\ref{fig:discovery_lc}.
The observing log is given in Table~\ref{tab:phot_log}.
In particular, the $r$-band light curve brightened from 
$\sim20.7$\,mag to $\sim20.2$\,mag within five\,hours, corresponding to an increase of about 0.5\,mag. The 
$g$-band data also shows a rising trend, while the 
$i$-band measurements are broadly consistent with early brightening despite the sparser sampling. During these early stages, the SN exhibited very high temperatures, indicated by ($g-r$) colours of $\sim -0.5$\,mag. These data demonstrate that the source was still emerging and brightening rapidly during the first five hours of optical follow-up, and show that SN~2026gzf was caught during the transition from the high-energy breakout phase to the earliest optical emergence of the SN. Over the following days, the source continued to brighten and evolved to redder colours. 

\section{Hydrodynamic light curve modelling with STELLA}
\label{sec_med:LC_modeling}

We computed synthetic light curves to estimate explosion and CSM properties of SN~2026gzf by using the one-dimensional multi-frequency radiation hydrodynamic code \texttt{STELLA}\citemeth{blinnikov1999}. Motivated by the photometric and spectroscopic similarities to SN~2006aj, we assumed the same ejecta mass ($2~\mathrm{M_\odot}$) 
and explosion energy ($2\times 10^{51}~\mathrm{erg}$)\citemeth{2006Natur.442.1018M}. We adopted the $3.85~\mathrm{M_\odot}$ C+O star progenitor model of \citemeth{2020MNRAS.497.1619M} and set the mass cut at $1.85~\mathrm{M_\odot}$ to obtain the assumed ejecta mass. 
The wind-like dense CSM 
(
$\rho_\mathrm{CSM}= 10^{-3} (r/10^{10}\,\mathrm{cm})^{-2}\,\mathrm{g\,cm^{-3}}$ is attached on top of the progenitor. The progenitor radius is $9\times 10^9~\mathrm{cm}$ and the CSM density near the progenitor surface is about $10^{-3}~\mathrm{g~cm^{-3}}$. 
The radius of the dense CSM is assumed to be $3\times 10^{13}~\mathrm{cm}$ given the CSM radius estimate based on the X-ray flash (W.~Yuan et al. in prep.). 
The CSM mass is 0.02~M$_{\odot}$ and the required mass-loss rate to form the dense CSM is 2~M$_{\odot}$\,yr$^{-1}$ assuming the wind velocity of $1000\,\mathrm{km~s^{-1}}$.

The light-curve models with and without CSM are shown in Fig.~\ref{fig:discovery_lc}.  In this scenario, the CSM is sufficiently dense to produce shock breakout at its outer surface, and the duration of the X-ray flash corresponds to the light-travel time across the CSM. A $^{56}$Ni mass of $0.5~\mathrm{M_\odot}$ is required to match the peak brightness, and significant outward mixing of $^{56}$Ni in the ejecta is needed to reproduce the observed rise time of SN~2026gzf.

\section{Semi-analytic light-curve modelling with MOSFiT}
\label{sec_med:mosfit}

To independently estimate the explosion parameters that explain the main light curve peak, we fit the multi-band optical photometry with the Modular Open Source Fitter for Transients (MOSFiT)\citemeth{2018ApJS..236....6G}. We use a basic $^{56}$Ni-heated model following the Arnett formalism\citemeth{1982ApJ...253..785A}, and assume a blackbody spectral energy distribution with absorption at blue wavelengths to mimic the effects of line blanketing\citemeth{2017ApJ...850...55N,2022ApJ...941..107G}: $L_{\lambda}(\lambda) = L_{\lambda,{\rm BB}} (\lambda/\lambda_{\rm cut})^\alpha)$, where $\alpha$ and $\lambda_{\rm cut}$ are free parameters. The other physical parameters of the model are the ejecta mass $M_{\rm ej}$ and velocity $v_{\rm ej}$, the $^{56}$Ni fraction $f_{\rm Ni}$, the temperature at late times $T_{\rm min}$, and the gamma-ray trapping opacity $\kappa_\gamma$ (we fix the photon diffusion opacity at $\kappa=0.1$\,cm$^2$\,g$^{-1}$). Since our model does not include shock cooling, we do not fit to any data points prior to MJD 61123 (i.e. within the first $\approx3$ days of explosion. Given this simplification and the approximate nature of the model, we do not fix the explosion date to exactly match the EP detection, but instead make this a free parameter within a small window of 5 days. Finally, we include a white noise parameter $\sigma$ in the likelihood function (see Ref of \citemeth{2018ApJS..236....6G}). We use flat or log-flat priors for all parameters.

The fitting results are shown in Figs.\,\ref{fig:lc_modfit} and Y. We obtain an excellent fit to the light curve in all bands, with broad metal line absorption significantly affecting the spectrum only below $\approx 4500$\,\AA ($u$ and $g$ bands). The estimated explosion date is within 2 days of the true date from the EP detection. We find $M_{\rm ej}\approx 1$\,M$_\odot$, $v_{\rm ej}\approx11,000$\,km\,s$^{-1}$, and $M_{\rm Ni}\approx 0.4$\,M$_\odot$. Although the statistical errors on these parameters are $\sim10\%$, given the exquisite data, this is a highly simplified model and the systematic errors are likely much larger (see e.g.~discussion in Ref of \citemeth{2025ApJ...980L..44M}) and we suggest that given the degeneracies visible between the main physical parameters, these numbers are likely reliable only to within a factor $\sim2$. The best-fit velocity is consistent with the observed spectra, and the masses of the ejecta and $^{56}$Ni are consistent with those in the hydrodynamic models based on the similar event SN~2006aj\citemeth{2006Natur.442.1018M}, further motivating their use to study the early light curve.

\begin{figure}
    \centering
    \includegraphics[width=.8\linewidth]{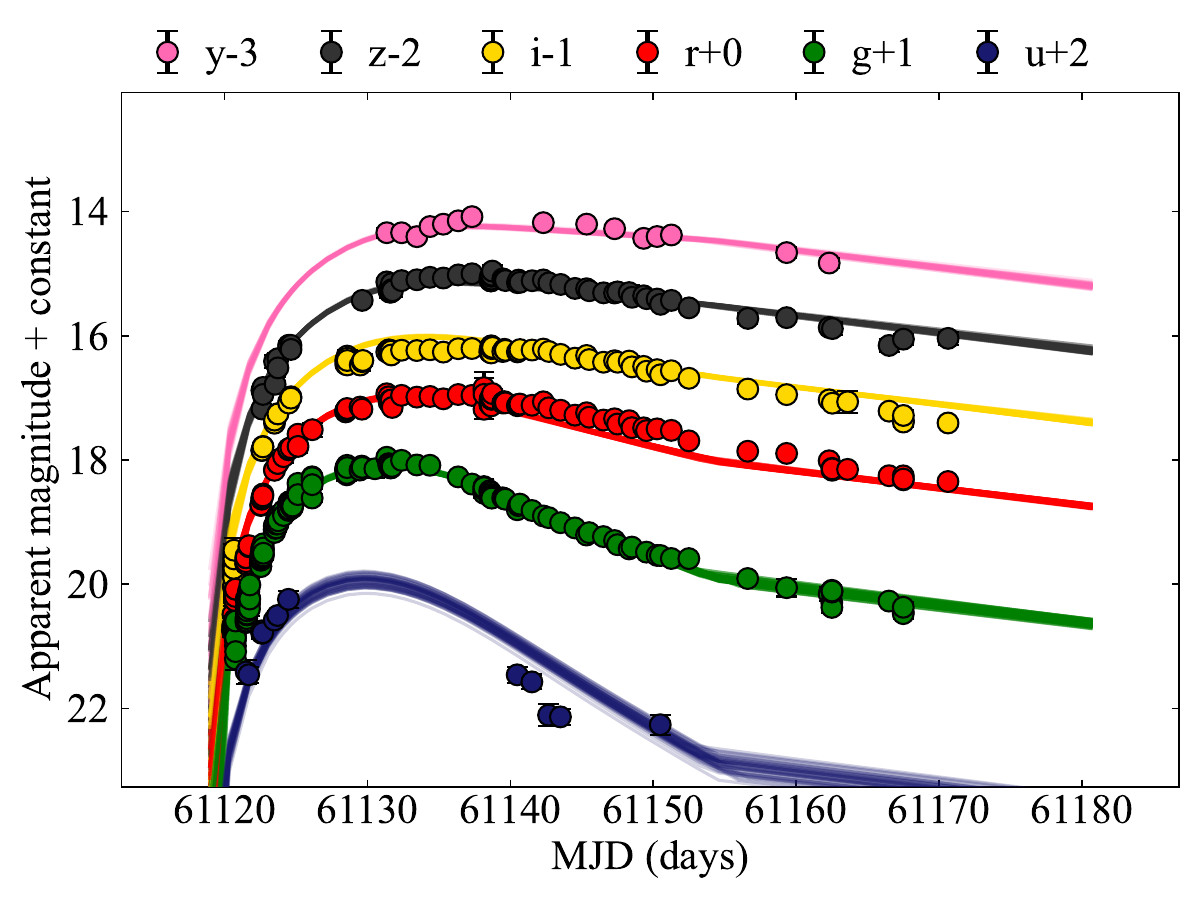}
    \caption{
        \textbf{MOSFiT fit to the multi-band optical light curve of SN~2026gzf.}
The points show the observed photometry, while the solid curves show the best-fit $^{56}$Ni-powered model based on the Arnett formalism. Data earlier than MJD~61123 were excluded from the fit because the model does not include shock-cooling or circumstellar-interaction emission. The model reproduces the main peak well in all bands and yields an explosion date within \(\sim 2\) days of the EP trigger, with representative parameters \(M_{\rm ej}\approx1\,{\rm M}_{\odot}\), \(v_{\rm ej}\approx 11{,}000\)~km~s\(^{-1}\), and \(M_{\rm Ni}\approx0.4\,{\rm M}_{\odot}\).} 
    \label{fig:lc_modfit}
\end{figure}

\section{On the presence of the CSM shells and the epochs of CSM interaction}
\label{sec_med:csm_shells}

Our hydrodynamic modelling suggests that the early excess in the light curves can be attributed to the presence of dense CSM.  The formation of CSM is supported by the precursor activities that started $\sim12$ years before the X-ray detection (Fig.\,\ref{fig:preburst}). As a robust example of Type Ic-BL SNe with precursor activity, SN~2018gep showed detection about 15 days before its explosion \citemeth{2019ApJ...887..169H}. The pre-SN detections of SN~2018gep were close to $-14$\,mag and were rather attributed to eruptive mass losses than a quiescent progenitor. Given the position of EP260321a in its host galaxy, a quiescent progenitor can be less favored. 
In Figure~\ref{fig:ps1_precursor}, we see six $w$-band kinks at MJDs 57779.5, 60021.3, 60315.5, 60696.5, 60761.3, and 61053.5. These kinks possibly mark more rigorous activities following stronger eruptions which eventually would have formed denser CSM shells. Assuming a wind velocity of 1000\,km\,s$^{-1}$ for Wolf-Ray stars, we used equation 5 of \citemeth{2019ApJ...887..169H} to estimate the radii of corresponding CSM shells. Corresponding to the above-mentioned six major eruptions, there would be CSM shells with radii $\sim5.8\times10^{14}$\,cm, $3.1\times10^{15}$\,cm, $3.7\times10^{15}$\,cm, $6.9\times10^{15}$\,cm, $9.5\times10^{15}$\,cm, and $2.9\times10^{16}$\,cm.

Next, we adopted a broken power-law distribution to the velocity evolution (of Si II) for Type Ic-BL SNe of the form presented in \citemeth{2025A&A...700A.200F} and estimated the tentative epochs of CSM interactions corresponding to the presence of CSM shells with the above-mentioned radii. The tentative epochs of CSM interaction for the last five eruptions (starting from the one at MJD 61053.5, only $\sim$\,67 days before the X-ray detection) would be $\sim$ +1d, +14d, +18d, +93d, and +2350d (MJD 60021.3 eruption). For the eruption at MJD 57779.5, the CSM shell radius is $2.9\times10^{16}$\,cm, thus, the SN-ejecta would eventually be too slow to reach there. With the tentative epochs of CSM-interactions estimated, the day-1 early excess in our light curves can indeed be associated to the presence of CSM resulting from the latest eruption. The next predicted epoch of ejecta-CSM interaction was $\sim$\,+14d. Supporting our prediction, a radio detection with the VLA was reported at +13.6d\citemeth{2026GCN.44239....1O}. Also, as expected, only an upper limit was obtained a few days earlier from ATCA at +7d\citemeth{2026GCN.44403....1J}, while another VLA observation at a later epoch (+22d) again resulted in a non-detection\citemeth{2026GCN.44357....1O}. The emergence of a radio detection close to our predicted interaction epoch, bracketed by upper limits on both sides, supports the presence of multiple shell-like CSM structures. There were no reported radio observations close to the next (+18d) predicted epoch of CSM interaction. Despite having optical spectra close to +18d, we hardly see any interaction features. This can be attributed to the lower resolution of the spectrograph. Additionally, we can not rule out the possibility of the older mass loss episodes resulting with different wind velocities than the simplified assumption of 1000\,km\,s$^{-1}$, hence the locations of CSM shells and the epochs of interaction might be uncertain corresponding to those episodes. 
Even if interaction happens at these radii, the signatures could be subtle, BL-Ic's are not known to show strong interaction features.

\section{Comparison with other X-ray outburst stripped-envelope supernovae}
\label{sec_med:comparison}

Figure\,\ref{fig:lc_comparison} shows the rest-frame absolute light curves of broad-lined Type Ic SNe that were discovered sufficiently early to probe their initial emission, including SNe~1998bw\citemeth{2011AJ....141..163C}, 2006aj\citemeth{2014ApJS..213...19B}, 2018gep\citemeth{2019ApJ...887..169H}, 2020bvc\citemeth{2020ApJ...902...86H}, 2020lao\citemeth{2026A&A...708A.305S}, and the recent EP-discovered events (also Ic-BL) SNe~2024gsa\citemeth{2025ApJ...982L..47V}, 2025kg\citemeth{2025ApJ...988L..13R} and 2025wkm\citemeth{2025arXiv251210239S}. We also include SN~2008D\citemeth{2008Natur.453..469S}, although it was a Type~Ib rather than an Ic-BL, because it remains one of the best-studied stripped-envelope SNe with an X-ray shock-breakout signal and exceptionally early follow-up data. We note that SN~2008D is known to suffer substantial host-galaxy extinction, with $A_{V}\approx2$\,mag\citemeth{2008Natur.453..469S,2008Sci...321.1185M}. No correction for host-galaxy extinction has been applied to any of the SNe except for SN 1998bw\citemeth{2011AJ....141..163C} as given in this plot.

Panel~(a) shows that the $g$-band luminosity evolution of SN~2026gzf is broadly consistent with that of luminous Ic-BL events, with a peak luminosity close to those of SN~2006aj and SN~2020bvc, but with unusually luminous measurements already present during the first day. Panel~(b) shows the same behaviour in the $r$ band: the later rise and peak are similar to those of the comparison Ic-BL sample, whereas the earliest data points lie above the smoother monotonic rise expected for an ordinary radioactive-powered light curve. Panel~(c) shows that SN~2026gzf is also very blue ($g-r\sim-0.5$) at the earliest epochs and then reddens steadily with time, broadly following the colour evolution seen in SN~2006aj and SN~2020bvc.
Taken together, the three panels show that although the later photometric evolution of SN~2026gzf is broadly typical of luminous Ic-BL SNe, its earliest emission includes a distinct bright, blue first-day excess suggestive of interaction with nearby CSM.

\begin{figure}
    \centering
    \includegraphics[width=.8\linewidth]{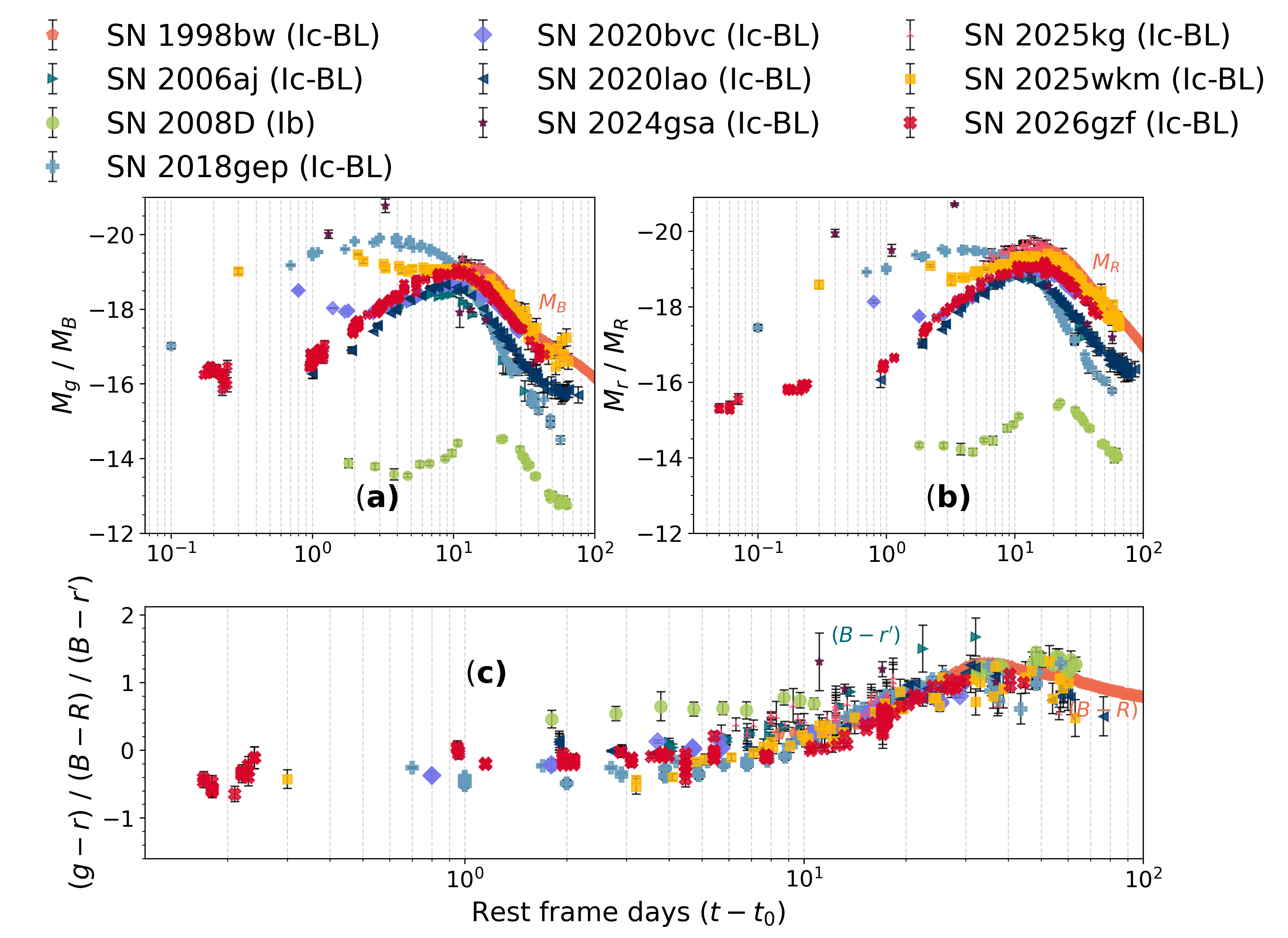}
    \caption{
        \textbf{Lightcurve evolution of SN~2026gzf and comparison with other SNe.}
        The upper panels, (a) and (b), compare the rest-frame red- and blue-band light curves of SN~2026gzf with those of the Ic-BL SNe~1998bw, 2006aj, 2018gep, 2020bvc and 2020lao, together with the recent EP-discovered events SNe~2024gsa, 2025kg and 2025wkm. SN~2008D is also shown, although it was a Type~Ib SN. All light curves are plotted relative to $T_{\rm burst}$. The lower panel, (c), shows the corresponding colour evolution. SN~2026gzf rises rapidly at early times and exhibits unusually blue colours, consistent with excess early emission shaped by interaction with nearby CSM.   
    }
    \label{fig:lc_comparison}
\end{figure}

\section{Comparison of the precursor activity of SN~2026gzf with other transients}
\label{sec_med:comparison_precursor}

We compare the long-term pre-explosion variability of SN~2026gzf with a small, non-exhaustive sample of transients for which precursor activity has been reported, including SNe~1961V\citemeth{2011ApJ...737...76K},
2006jc\citemeth{2007Natur.447..829P},
2009ip\citemeth{2013ApJ...767....1P}, 2018gep\citemeth{2019ApJ...887..169H}, 2023fyq\citemeth{2024A&A...684L..18B}, and UGC 2773-OT\citemeth{2010AJ....139.1451S}, see Fig.\,\ref{fig:precursor_comparison}.
In terms of absolute magnitude, the precursor of SN~2026gzf is among the most luminous in the sample, reaching $M\approx -14$\,mag. The only clearly comparable case in this figure is SN~1961V, whose precursor was also extremely bright and likewise persisted for more than a decade before the main event; it has been interpreted as a pulsational pair-instability event from a very massive progenitor of roughly 100--115\,$M_\odot$\citemeth{2022ApJ...938...57W}. 
Although SN~2026gzf is unlikely to arise from such an extreme progenitor channel, this comparison emphasizes that its long-term precursor activity was unusually luminous for a stripped-envelope SN progenitor.

Moreover, the precursor of SN~2026gzf also shows pronounced short-timescale variability superposed on the longer-term evolution. Similar erratic behavior was seen in SN~2009ip, although that object was a hydrogen-rich Type~II event rather than a stripped-envelope SN. The analogy is therefore not one of progenitor class, but of phenomenology: both objects show evidence for violent variability shortly before explosion. This suggests that such eruptive behavior may not be limited to extended hydrogen-rich progenitors, but could also occur in compact stripped progenitors such as Wolf–Rayet-like stars.

\begin{figure}
    \centering
    \includegraphics[width=\linewidth]{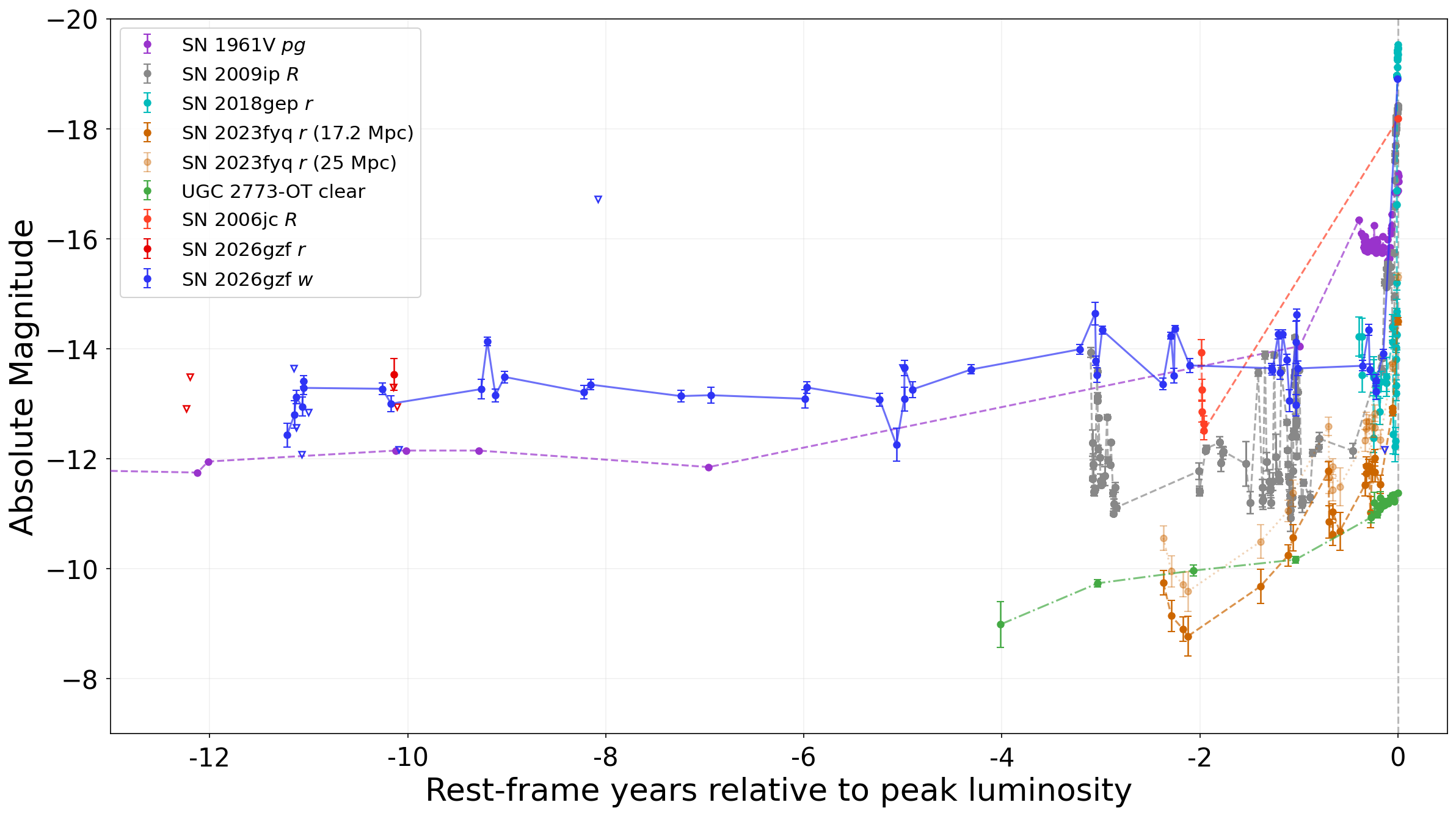}
    \caption{\textbf{Comparison of the long-term precursor activity of SN~2026gzf with other transients.} Absolute magnitude is shown as a function of rest-frame time in years relative to the peak luminosity of the main eruption or SN light curve of each object. Time 0 marks the peak, so all points shown correspond to precursor activity before the main event. The blue and red points show the archival Pan-STARRS $w$- and $r$-band measurements of SN~2026gzf. For comparison, we also show the reported precursor variability of SNe~1961V, 2006jc, 2009ip, 2018gep, 2023fyq, and UGC 2773-OT. SN~2023fyq is shown for two assumed distances because the distance to its Virgo-cluster host remains uncertain; the two tracks illustrate how this uncertainty affects the inferred absolute magnitude of the precursor. Among these objects, the precursor of SN~2026gzf is one of the most luminous, reaching $M\approx -14$\,mag, and also exhibits substantial variability on both year-long and shorter timescales.
    }
    \label{fig:precursor_comparison}
\end{figure}

\section{Host galaxy environments}
\label{sec_med:host}

SN~2026gzf lies in a disturbed system (Fig.,\ref{fig:discovery_lc}, right panel), consisting of the larger galaxy SDSS J095942.99+002504.0 and a nearby blue compact source, SDSS J095942.88+002506.2. The SN is spatially coincident with the latter, indicating that this blue compact source is its most likely immediate host. The larger system has a spectroscopic redshift of $z=0.0345$, independently measured by the 2dF Galaxy Redshift Survey (2dFGRS; TGN352Z077) and DESI Legacy Survey spectroscopy (target ID: 39627799320858502). With an absolute magnitude of about $M_{r}=-16.6$\,mag, the blue compact source could plausibly be a compact dwarf galaxy\citemeth{2015MNRAS.452.1567C} or a particularly luminous star-forming region, possibly associated with or orbiting the larger disturbed galaxy. Because young clusters in interacting galaxies can be very luminous\citemeth{2014AJ....147...60S}, we do not attempt to distinguish among these possibilities given the available spatial resolution and signal-to-noise ratio, and therefore adopt the neutral term ``blue compact source'' throughout this paper.

To characterize the host galaxy and the local environment of SN~2026gzf, we analysed WiFeS integral-field spectroscopy. Due to poor S/N, we stacked two of our WiFeS observations that were taken during dark-time (on 2026-04-08 and 2026-04-16) to improve our data quality. We defined a flux-limited aperture (see Figure \ref{fig:IFU}) to extract a global host spectrum from these stacked IFU cubes, measuring a Balmer decrement of $3.67\pm0.35$, corresponding to $E(B-V)=0.25\pm0.09$\,mag assuming a CCM extinction law with $R_V=3.1$. 
The dust-corrected H$\alpha$ luminosity implies a star-formation rate of $0.37\pm0.08\,M_{\odot}\,\mathrm{yr}^{-1}$, and the gas-phase metallicity derived from the O3N2 diagnostic \citemeth{2013A&A...559A.114M} is $12+\log(\mathrm{O/H})=8.00\pm0.03$. We also extracted a spectrum from a 3x3-spaxel (1'' per spaxel) aperture corresponding with the blue compact source at the SN position. This spectrum is contaminated by SN light, but provides a local estimate of the surrounding environment, with a Balmer decrement of $3.42\pm0.40$, $E(B-V)=0.18\pm0.11$ mag, a dust-corrected star-formation rate of $0.17\pm0.04\,M_{\odot}\,\mathrm{yr}^{-1}$, and $12+\log(\mathrm{O/H})=8.05\pm0.04$. When deriving maps of the host galaxy from the stacked WiFeS datacubes, there remain significant spaxel-to-spaxel fluctuations. We therefore produce maps by extracting spectra from a 3x3 sliding aperture centred on each spaxel, as seen in Figure \ref{fig:IFU}. Given the limited data quality, properties derived for a given spaxel should be treated as indicators of a local region surrounding that spaxel, rather than properties directly within the sightline of the spaxel. 

Although our dust map suggests elevated host extinction near SN~2026gzf, several independent indicators point to little extinction along the direct SN sightline. In particular, the Balmer decrement measured from the LOT spectrum is close to the Case B recombination value of 2.86, suggesting minimal local extinction. This is further supported by the absence of red colours in the SN peak photometry and the lack of Na~I~D absorption in the SN spectra, consistent with the SN lying on the near side of the system along our line of sight. Independent MUSE observations obtained by A.~Martin-Carrillo et al. (in prep.) likewise indicate little or no dust extinction at the SN position. The discrepancy with our estimate may arise because the blue compact source is not fully resolved from the larger galaxy in our data.

\begin{figure}
    \centering
    \begin{minipage}[c]{0.3\linewidth}
        \centering
        \includegraphics[width=\linewidth]{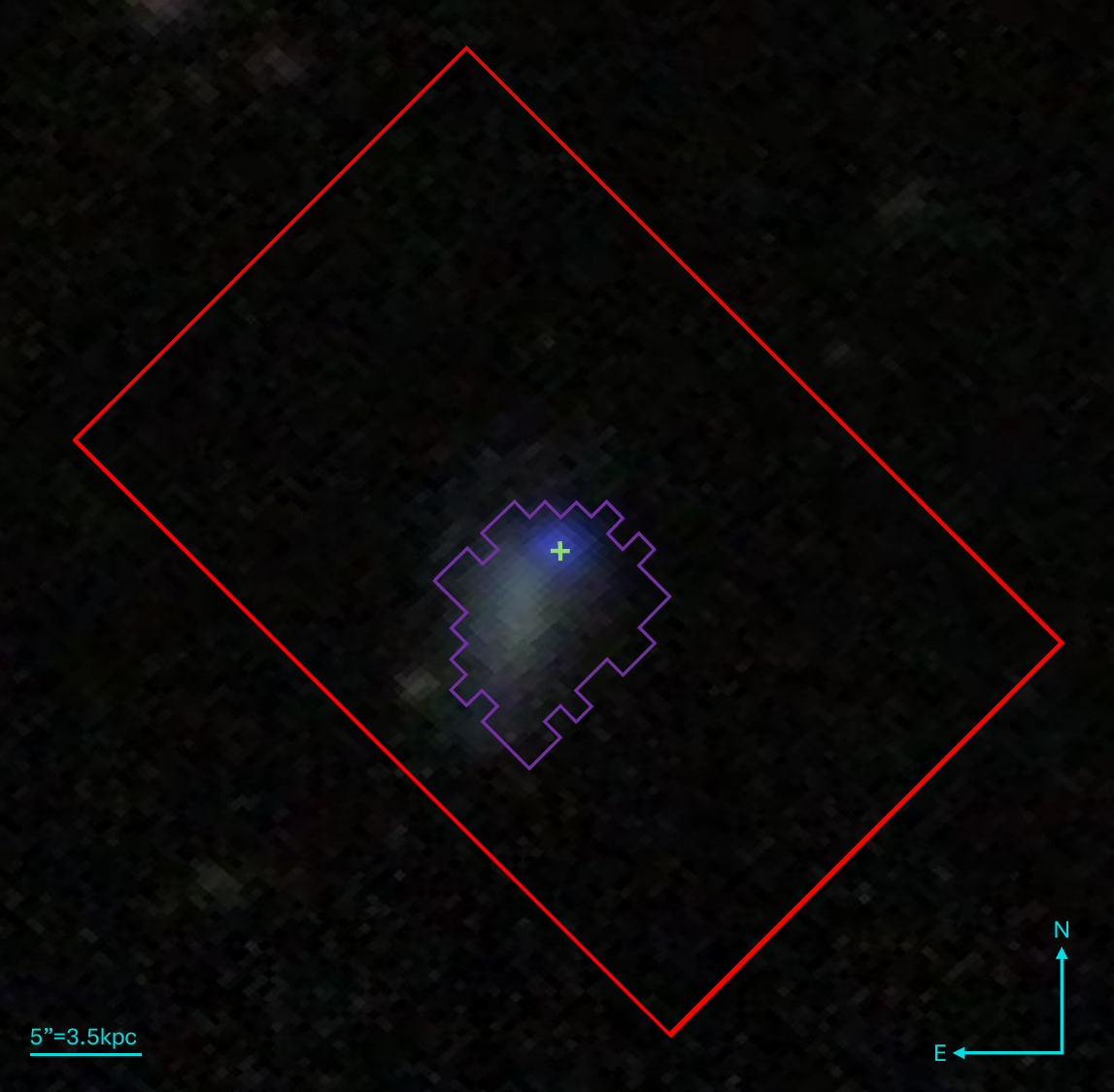}
    \end{minipage}
    \hfill
    \begin{minipage}[c]{0.69\linewidth}
        \centering
        \includegraphics[width=\linewidth]{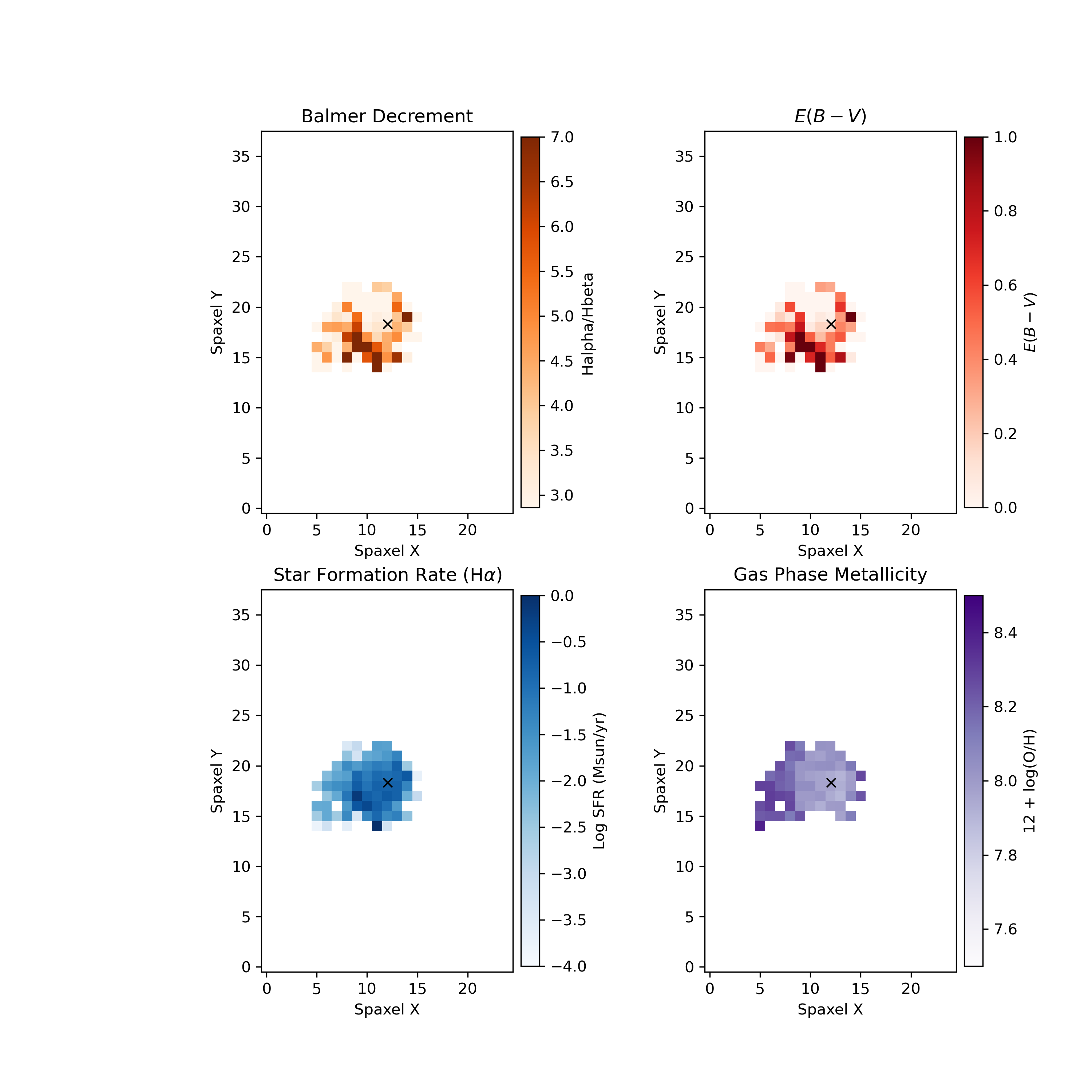}
    \end{minipage}
    
    \caption{\textbf{Host-galaxy colour image and WiFeS IFU maps of the host galaxy of SN~2026gzf.} 
    \textit{Left:} SDSS colour image of the host galaxy of SN~2026gzf, with the red rectangle representing the WiFeS FoV for our observations, and the position of the SN marked as a green cross. The region in purple contains sufficient WiFeS S/N per spaxel for host-property estimation, and this region was used to extract global properties of the host galaxy. 
    \textit{Right:} WiFeS IFU maps of the host galaxy of SN~2026gzf, constructed using a sliding $3\times3$ aperture on stacked datacubes from our dark-time observations (2026-04-08 and 2026-04-16). When rotated by the 45$^\circ$ position angle of the WiFeS observations, these maps correspond to the purple region in the left-hand panel. Shown are the Balmer decrement (upper left), $E(B-V)$ (upper right), H$\alpha$-based star-formation rate (lower left), and gas-phase metallicity (lower right). The SN position is marked with a black ``$\times$''. Owing to the limited signal-to-noise ratio and spatial resolution, the maps provide only smoothed estimates of the local host properties, and the compact blue source at the SN position is not cleanly separable from the surrounding host emission.}
    \label{fig:IFU}
\end{figure}

We modelled the spectral energy distribution for both larger galaxy (SDSS J095942.99+002504.0) and the blue compact source (SDSS J095942.88+002506.2) considered as the host of SN~2026gzf. We characterized the host galaxy using archival ultraviolet-to-infrared imaging from the Galaxy Evolution Explorer (GALEX), Pan-STARRS and unWISE. Using \textsc{hostphot}\citemeth{2022JOSS....7.4508M}, matched-aperture photometry was constructed across all bands using an aperture defined from the Legacy Survey $r$-band image. Specifically, we adopted an aperture size of $11.729'' \times 7.022''$ for the larger galaxy and $1.015'' \times 0.999''$ for the blue compact source.
The SED fitting was performed with \textsc{prospector}\citemeth{2021ApJS..254...22J}, with posterior sampling carried out using \textsc{dynesty}. We adopted \texttt{Prospector-$\alpha$} model and includes nebular line and continuum emission, as well as dust emission. 
We assume a uniform in log prior on the total stellar mass  (\texttt{total\_mass}), covering $10^8 < M/M_{\odot} < 10^{12}$, and model the stellar metallicity (\texttt{logzsol}) with a top-hat prior ranging from $-2 < \log(Z/Z_{\odot}) < 0.19$. The star-formation history (SFH) is modeled non-parametrically via the \texttt{z\_fraction} prescription, complemented by a Beta continuity prior to ensure smooth transitions in the star-formation rate across cosmic time. 
For dust attenuation, the diffuse dust optical depth (\texttt{dust2}, $\hat{\tau}_{2}$) is assigned a uniform prior in the range $[0, 4.0]$. Additionally, we account for IR dust emission using the \citemeth{2007ApJ...657...810} templates, governed by the minimum radiation field intensity ($0.1 < U_{\min} < 25$), the PAH mass fraction ($0.5 < Q_{\text{PAH}} < 7.0$), and the mass fraction of dust exposed to a high-intensity radiation field (\texttt{dust\_gamma}, $\gamma$), for which we assume a prior range of $0.001 < \gamma < 0.15$.
Figures\,\ref{fig:sed_local}, \ref{fig:sed_global} show the host-galaxy broadband photometry and the corresponding best-fitting \textsc{prospector} model. From the posterior medians and $1\sigma$ uncertainties, we infer $\log_{10}(M_\ast/M_\odot)=9.04^{+0.09}_{-0.08}$ for the larger galaxy and $\log_{10}(M_\ast/M_\odot)=8.09^{+0.23}_{-0.07}$ for the blue compact source. Its stellar mass is thus about an order of magnitude lower than that of the larger galaxy, consistent with a dwarf-companion interpretation. By combining stellar mass with the star-formation rate derived above, we can calculate the specific star-formation rate. This yields $\sim1.37$\,Gyr$^{-1}$ for the larger galaxy and $\sim0.34$\,Gyr$^{-1}$ for the blue compact source. Additionally, their $A_V$ values calculated from \texttt{dust2} are $0.076$ and $0.043$\,mag, respectively.

\clearpage

\section*{Extended Data}

\begin{longtable}{cccccc}
\caption{
    Log of optical photometric observations. This table presents a representative sample of the dataset, displaying only the initial observational epochs from various instruments to illustrate the data format and content.
}\\
\hline
MJD & $m$ & $\sigma_m$ & Filter & Telescope & Exptime (s) \\
\hline
\endfirsthead
\hline
MJD & $m$ & $\sigma_m$ & Filter & Telescope & Exptime (s) \\
\hline
\endhead
\hline
\endfoot
61120.568 & 20.74 & 0.06 & $r$ & LOT & 300.0 \\
61120.568 & $>$19.8 &  & $i$ & SLT & 300.0 \\
61121.316 & 19.70 & 0.17 & $g$ & P60 & 30.0 \\ 
61123.052 & 18.34 & 0.01 & $r$ & SWO & 320.0 \\
61124.122 & 17.82 & 0.07 & $g$ & Citra & 60.0 \\ 
61124.377 & 17.98 & 0.06 & $o$ & ATLAS & 30.0 \\
61131.349 & 16.96 & 0.03 & $g$ & PS2 & 60.0 \\ 
61136.021 & 17.01 & 0.00 & $r$ & LSST & 30.0
\label{tab:phot_log}
\end{longtable}

\begin{sidewaystable}
    \renewcommand*{\arraystretch}{1.1}
    \centering
    \footnotesize
    \caption{
        Spectroscopic observing log of SN~2026gzf. $T_{0}$ is defined as MJD 61120.516 (the EP-WXT trigger time).
    }
    \begin{tabular}{ccccccccccc}
        \toprule
        Date & MJD & $T_{\rm start} - T_{0}$ & Telescope & Instrument & Grating & Exp. time  & Slit  & Resolution & Wavelength Range \\
             &     & (days)                  &           &            &         & (seconds) & ('') &       & (\AA)            \\
        \midrule
        2026-03-30 & 61129.897 & 9.38 & NOT & ALFOSC & Grism 4 & 1600 & 1.3 & 710 & 3001--9637 \\

        2026-03-31 & 61130.440 & 9.92  & ANU 2.3 m & WiFeS & B3000/R3000 & $900\times3$ & IFU & 3000/3000 & 3500--5700/5400--9500 \\ 
        
        2026-04-01 & 61131.530 & 11.02 & LOT & UVEX & 300 & $600\times12$ & 1.29 & 2.49 & 4200--7950 \\      

        2026-04-05 & 61135.240 & 14.72 & Keck II & NIRES & - & $120\times4$ & 0.55 & 2700 & 9648–-24666 \\ 

        2026-04-07 & 61137.922 & 17.41 & NOT & ALFOSC & Grism 4 & 1800 & 1.3 & 710 & 3001--9637 \\
        
        2026-04-08  &  61138.387 &  17.87  & ANU 2.3 m & WiFeS & B3000/R3000 & $ 900\times3$ & IFU & 3000/3000 & 3500--5700/5400-9500 \\ 
        
        2026-04-16  &  61146.543 &  26.03  & ANU 2.3 m & WiFeS & B3000/R3000 & $ 900\times3$ & IFU & 3000/3000 & 3500--5700/5400-9500 \\ 
        
        2026-04-23 & 61153.369 & 32.85 & NASA IRTF & SpeX  & - & $120\times12$ & 0.8 & $2000$ & 6905-25450 \\
         
        2026-04-28 & 61158.341 & 37.83 & Keck II & NIRES & - & $120\times8$ & 0.55 & 2700 & 9648–-24666 \\ 
        
          2026-04-28  &  61158.471  &  37.96  & ANU 2.3 m & WiFeS & B3000/R3000 & $ 900\times1$ & IFU & 3000/3000 & 3500--5700/5400-9500 \\ 
          
           2026-04-29  &  61159.466 &  38.95  & ANU 2.3 m & WiFeS & B3000/R3000 & $ 2700\times3$ & IFU & 3000/3000 & 3500--5700/5400-9500 \\ 

           2026-05-03 & 61163.953 & 43.44 & NOT & ALFOSC & Grism 4 & 2042 & 1.0 & 710 & 3001--9637 \\
           
        \bottomrule
    \end{tabular}
    \label{tab:spec_log}
\end{sidewaystable}

\begin{table}
    \caption{
        \textbf{Astrometric measurements at the site of SN~2026gzf.}
        The upper section lists the archival direct-image centroids of the blue compact source used to define the adopted pre-explosion position. The lower section lists representative later SN positions shown for comparison only; these are not used in the weighted mean.
    }
    \begin{tabular}{lcccc}
        \toprule
        Dataset & RA (deg) & Dec (deg) & $\sigma_{\rm RA,sky}$ (arcsec) & $\sigma_{\rm Dec}$ (arcsec) \\
        \midrule
        DECam $r$         & 149.9286600 & 0.4184164 & 0.030 & 0.025 \\
        Subaru $r$        & 149.9286680 & 0.4184338 & 0.118 & 0.119 \\
        SDSS $r$          & 149.9286642 & 0.4184014 & 0.266 & 0.266 \\
        Pan-STARRS $w$    & 149.9286802 & 0.4184167 & 0.067 & 0.067 \\
        \midrule
        Adopted position  & 149.9286637 & 0.4184170 & 0.027 & 0.023 \\
        \midrule
        LSST SN           & 149.9286667 & 0.4184342 & - & -  \\
        LOT SN            & 149.9286830 & 0.4184660 & - & -  \\
        Pan-STARRS SN     & 149.9286701 & 0.4184176 & - & -  \\
        \bottomrule
    \end{tabular}
    \label{tab:astrometry}
\end{table}

\begin{table}
    \caption{
        \textbf{Mean pre-explosion photometry measured at the site of SN~2026gzf.}
        For each survey and band, we list the mean flux density and statistical uncertainty, together with the corresponding AB magnitude and propagated uncertainty, computed using an AB zeropoint of 23.9 for flux densities expressed in \(\mu\)Jy. Within each filter group, the ``Average'' row gives the inverse-variance weighted mean flux across surveys and the weighted scatter in flux space. For the $w$ band, only one survey measurement is available, so the listed uncertainty is the measurement uncertainty rather than an inter-survey weighted scatter.}
    \begin{tabular}{lccccc}
        \toprule
        Survey & Band & Mean flux ($\mu$Jy) & Uncertainty ($\mu$Jy) & AB mag & Uncertainty (mag) \\
        \midrule

        PS1      & $w$ & 28.33 & 0.23 & 20.269 & 0.009 \\
        Average  & $w$ & 28.33 & 0.23 & 20.269 & 0.009 \\
        \midrule

        PS1      & $r$ & 26.46 & 0.88 & 20.344 & 0.036 \\
        ZTF      & $r$ & 27.23 & 0.35 & 20.312 & 0.014 \\
        Subaru   & $r$ & 27.18 & 0.80 & 20.314 & 0.032 \\
        DECam    & $r$ & 27.84 & 0.57 & 20.288 & 0.022 \\
        Average  & $r$ & 27.29 & 0.36 & 20.310 & 0.014 \\
        \midrule

        PS1      & $i$ & 11.82 & 0.75 & 21.218 & 0.069 \\
        ZTF      & $i$ & 15.82 & 1.19 & 20.902 & 0.082 \\
        Subaru   & $i$ & 9.02  & 0.22 & 21.512 & 0.026 \\
        DECam    & $i$ & 12.55 & 0.32 & 21.153 & 0.028 \\
        Average  & $i$ & 10.36 & 1.82 & 21.361 & 0.190 \\
        \midrule

        PS1      & $z$ & 14.40 & 2.75 & 21.004 & 0.207 \\
        Subaru   & $z$ & 6.36  & 0.38 & 21.891 & 0.065 \\
        DECam    & $z$ & 12.05 & 0.36 & 21.198 & 0.032 \\
        Average  & $z$ & 9.40  & 2.87 & 21.467 & 0.331 \\
        \midrule

        ZTF      & $g$ & 33.58 & 0.37 & 20.085 & 0.012 \\
        Subaru   & $g$ & 43.32 & 1.85 & 19.808 & 0.046 \\
        DECam    & $g$ & 45.09 & 0.83 & 19.765 & 0.020 \\
        Average  & $g$ & 35.74 & 4.43 & 20.017 & 0.135 \\
        \bottomrule
    \end{tabular}
    \label{tab:preflux}
\end{table}

\begin{table}
    \caption{
        \textbf{Chance-alignment probabilities for the pre-explosion blue compact source under different matching assumptions.}
        The probabilities were computed from the local surface density of PS1 DR2 catalogue sources in a 5-arcmin control field centred on the explosion site, excluding the inner 5-arcsec host region. For each band, the fiducial apparent magnitude was taken from the band-specific Average AB magnitude listed in Table~\ref{tab:preflux}. 
        \(P_{\rm any}\) uses all catalogue sources of comparable apparent brightness, while \(P_{\rm point}\) restricts the comparison sample to unresolved point-like sources selected using PSF--Kron compactness. Monte Carlo random position tests gave consistent results and are not listed separately here.
    }
    \begin{tabular}{lccccc}
        \toprule
        Run & \(r_{\rm match}\) (arcsec) & \(\Delta m\) (mag) & Band & \(P_{\rm any}\) & \(P_{\rm point}\) \\
        \midrule
        A & 0.3 & \(\pm 0.5\) & \(g\) & \(3.70\times10^{-5}\) & \(1.50\times10^{-5}\) \\
        A & 0.3 & \(\pm 0.5\) & \(r\) & \(2.07\times10^{-4}\) & \(8.90\times10^{-5}\) \\
        A & 0.3 & \(\pm 0.5\) & \(i\) & \(1.16\times10^{-3}\) & \(3.42\times10^{-4}\) \\
        A & 0.3 & \(\pm 0.5\) & \(z\) & \(7.70\times10^{-4}\) & \(1.93\times10^{-4}\) \\
        \midrule
        B & 0.5 & \(\pm 0.5\) & \(g\) & \(1.03\times10^{-4}\) & \(4.17\times10^{-5}\) \\
        B & 0.5 & \(\pm 0.5\) & \(r\) & \(5.75\times10^{-4}\) & \(2.47\times10^{-4}\) \\
        B & 0.5 & \(\pm 0.5\) & \(i\) & \(3.21\times10^{-3}\) & \(9.50\times10^{-4}\) \\
        B & 0.5 & \(\pm 0.5\) & \(z\) & \(2.14\times10^{-3}\) & \(5.36\times10^{-4}\) \\
        \midrule
        C & 0.5 & \(\pm 1.0\) & \(g\) & \(2.75\times10^{-4}\) & \(1.22\times10^{-4}\) \\
        C & 0.5 & \(\pm 1.0\) & \(r\) & \(1.59\times10^{-3}\) & \(5.72\times10^{-4}\) \\
        C & 0.5 & \(\pm 1.0\) & \(i\) & \(5.41\times10^{-3}\) & \(1.34\times10^{-3}\) \\
        C & 0.5 & \(\pm 1.0\) & \(z\) & \(3.03\times10^{-3}\) & \(8.08\times10^{-4}\) \\
        \bottomrule
    \end{tabular}
    \label{tab:chancealign}
\end{table}

\clearpage

\bibliographystylemeth{naturemag}
\bibliographymeth{astroph_references}
\end{methods}

\clearpage

\begin{supplement}

\begin{figure}
  \centering
  \includegraphics[width=0.85\columnwidth]{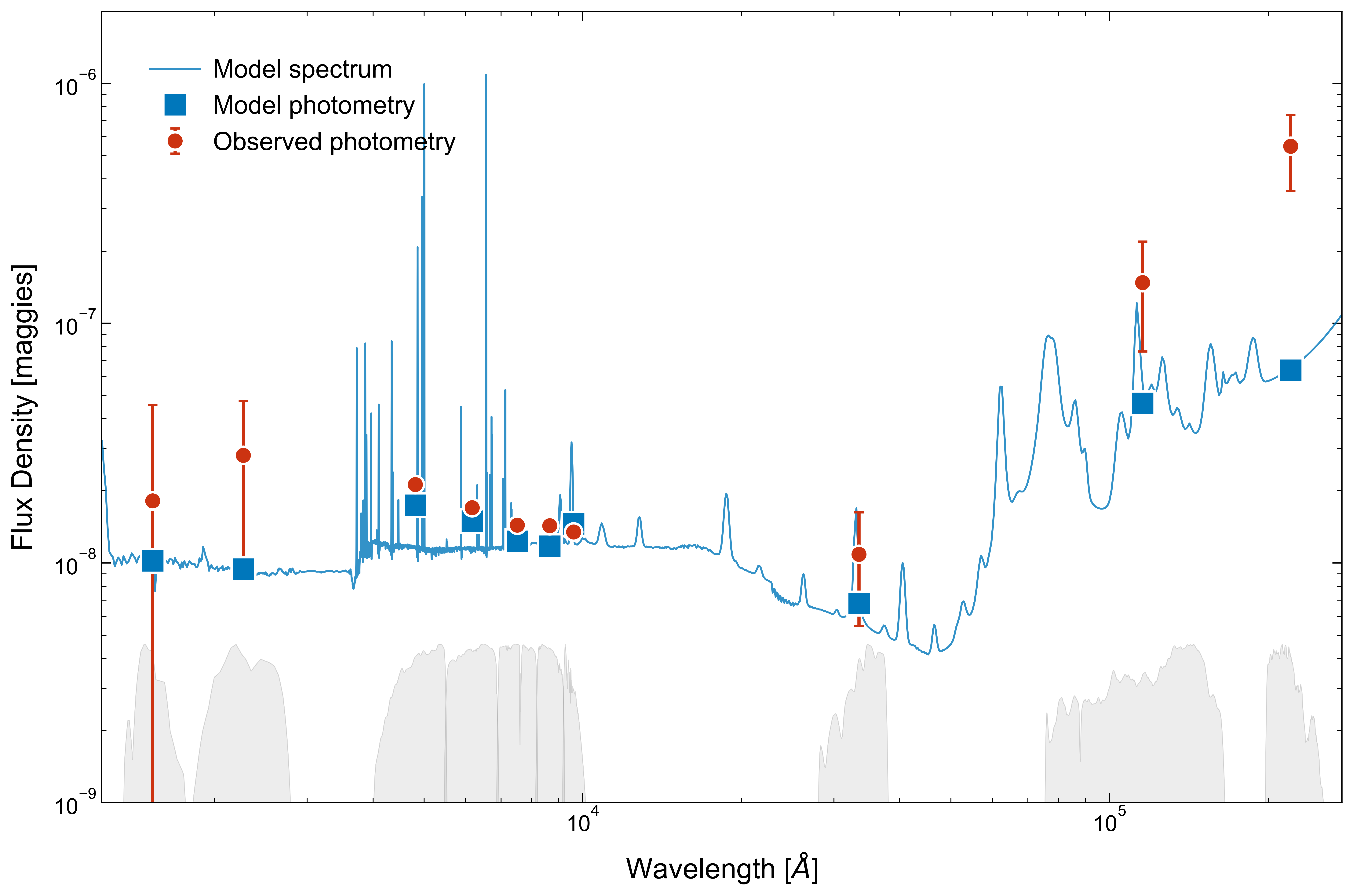}
  \caption{\textbf{Best-fit SED focusing specifically on the smaller galaxy hosting SN~2026gzf.} By utilizing photometry restricted to the smaller galaxy, the model characterizes the stellar population and dust properties of the host component while minimizing light contamination from the larger primary galaxy. The blue solid line represents the best-fit model spectrum, while red circles with error bars denote the observed multi-band photometry. Blue squares indicate the model-predicted photometry integrated over the respective filter bands. The gray shaded regions at the bottom illustrate the transmission curves of the filters used in the fitting process. The fit covers a broad wavelength range from the UV to the mid-IR to constrain the integrated physical properties of the combined system.}
  \label{fig:sed_local}
\end{figure}

\begin{figure}
  \centering
  \includegraphics[width=0.85\columnwidth]{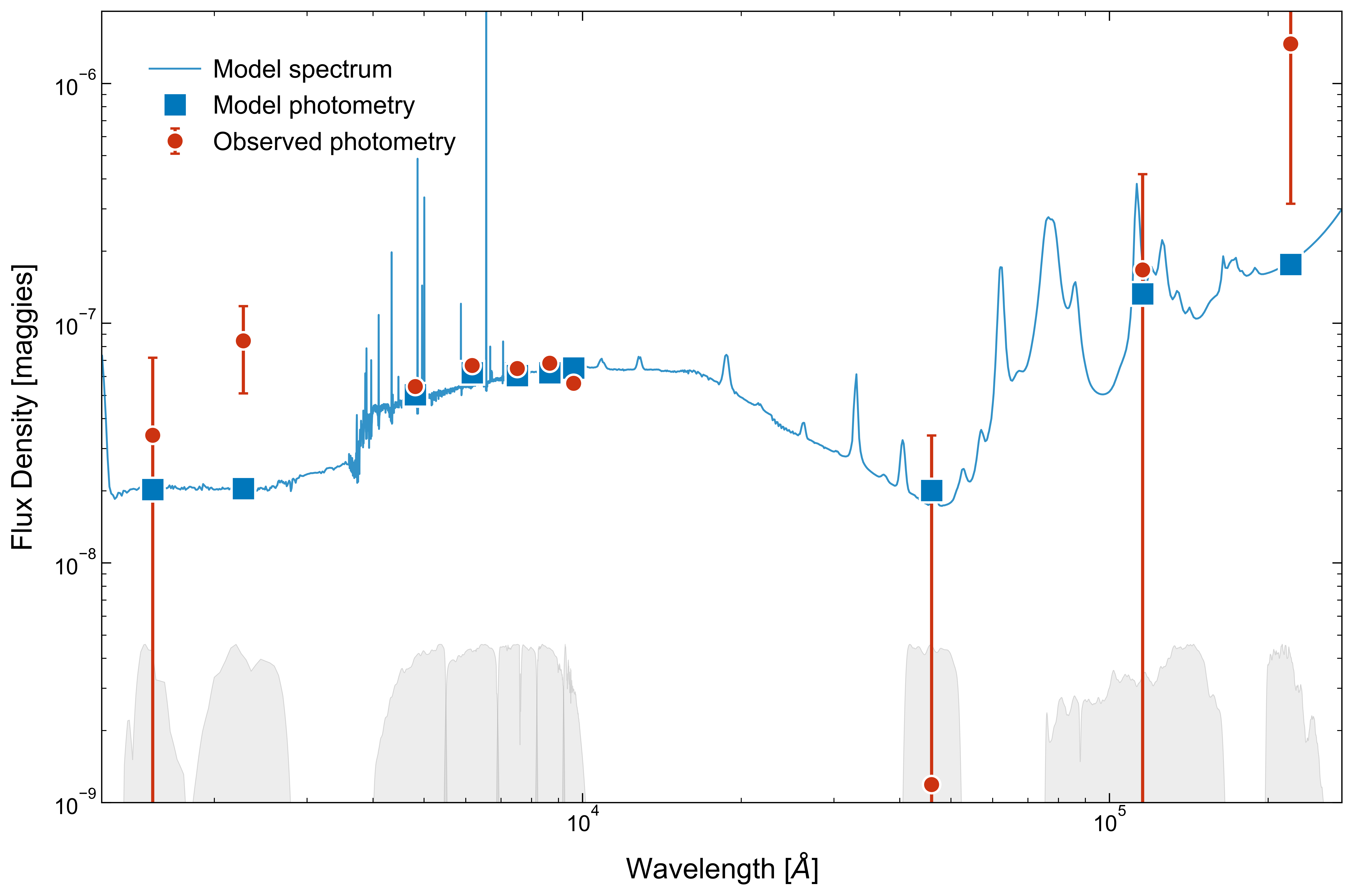}
  \caption{\textbf{Best-fit SED for the entire galaxy system.} This system consists of a larger primary galaxy and a smaller companion hosting SN~2026gzf. Symbols and colour schemes are identical to those in Figure\,\ref{fig:sed_local}.}
  \label{fig:sed_global}
\end{figure}

\clearpage

\end{supplement}

\clearpage

\begin{addendum}

\item 
We thank the Lulin Observatory observing assistants Chi-Sheng Lin, Hsiang-Yao Hsiao, and Wei-Jie Hou for carrying out the observations.
T.-W.C. acknowledges Stefan Taubenberger and Hsing-Wen Lin for useful discussions on spectroscopic interpretation and follow-up strategy. 
NCU GREAT team acknowledges the financial support from the Yushan Fellow Program by the Ministry of Education, Taiwan (MOE-111-YSFMS-0008-001-P1) and the National Science and Technology Council, Taiwan (NSTC grant 114-2112-M-008-021-MY3). 
HNAS team acknowledges the funding from the National Natural Science Foundation of China under grant No. 12303046, No. 12433001, the Startup Research Fund of Henan Academy of Sciences No. 242041217, and the Joint Fund of Henan Province Science and Technology R\&D Program No. 235200810057.
A.M. is supported by the ARC Discovery Early Research Award (DE230100055).
C.-J.L. is supported by the NSTC grant 114-2112-M-004-001-MY2 from the National Science and Technology Council of Taiwan.
C.M.P. acknowledges support from the National Science Foundation Graduate Research Fellowship Program under Grant No. 2236415. Any opinions, findings, and conclusions or recommendations expressed in this material are those of the author(s) and do not necessarily reflect the views of the National Science Foundation.
D.O.J. acknowledges support from NSF grants AST-2407632, AST-2429450, and AST-2510993, NASA grants 80NSSC24M0023 and 80NSSC24K0353, and HST/JWST grants HST-GO-17128.028 and JWST-GO-05324.031, awarded by the Space Telescope Science Institute (STScI), which is operated by the Association of Universities for Research in Astronomy, Inc., for NASA, under contract NAS5-26555. This work is also funded in part by the Gordon and Betty Moore Foundation through Grant GBMF13900 to D.O.J.
K.W.S. acknowledges funding from the Royal Society.
L.R. acknowledges support from the Australian Research Council (ARC) Centre of Excellence for Gravitational Wave Discovery (OzGrav), through project number CE23010001.
M.D.S. is funded by the Independent Research Fund Denmark (IRFD, grant number 10.46540/2032-00022B).
M.N. is supported by the European Research Council (ERC) under the European Union’s Horizon 2020 research and innovation program (grant agreement No. 948381).
T.L.K. acknowledges support via a Warwick Astrophysics prize post-doctoral fellowship, made possible thanks to a generous philanthropic donation.
T.M.R. is part of the Cosmic Dawn Center (DAWN), which is funded by the Danish National Research Foundation under grant DNRF140. T.M.R. acknowledges support from the Research Council of Finland project 350458.
W.B.H. acknowledges support from the National Science Foundation Graduate Research Fellowship Program under Grant Nos. 1842402 and 2236415. Any opinions, findings, conclusions, or recommendations expressed in this material are those of the author(s) and do not necessarily reflect the views of the National Science Foundation.

Numerical computations were in part carried out on PC cluster at the Center for Computational Astrophysics, National Astronomical Observatory of Japan.
This publication has made use of data collected at Lulin Observatory, partly supported by the TAOvA with the NSTC grant 114-2740-M-008-002. 
Based in part on data acquired at the ANU 2.3-metre telescope. The automation of the telescope was made possible through an initial grant provided by the Centre of Gravitational Astrophysics and the Research School of Astronomy and Astrophysics at the Australian National University and through a grant provided by the Australian Research Council through LE230100063. The Lens proposal system is maintained by the AAO Research Data \& Software team as part of the Data Central Science Platform. 
Some of the data presented herein were obtained at Keck Observatory, which is a private 501(c)3 non-profit organization operated as a scientific partnership among the California Institute of Technology, the University of California, and the National Aeronautics and Space Administration. The Observatory was made possible by the generous financial support of the W. M. Keck Foundation.
The authors wish to recognize and acknowledge the very significant cultural role and reverence that the summit of Maunakea has always had within the Native Hawaiian community. We are most fortunate to have the opportunity to conduct observations from this mountain.
Astronomer observing with the Infrared Telescope Facility, which is operated by the University of Hawaii under contract 80HQTR24DA010 with the National Aeronautics and Space Administration.
The DESI Legacy Imaging Surveys consist of three individual and complementary projects: the Dark Energy Camera Legacy Survey (DECaLS), the Beijing-Arizona Sky Survey (BASS), and the Mayall z-band Legacy Survey (MzLS). DECaLS, BASS and MzLS together include data obtained, respectively, at the Blanco telescope, Cerro Tololo Inter-American Observatory, NSF's NOIRLab; the Bok telescope, Steward Observatory, University of Arizona; and the Mayall telescope, Kitt Peak National Observatory, NOIRLab. NOIRLab is operated by the Association of Universities for Research in Astronomy (AURA) under a cooperative agreement with the National Science Foundation. Pipeline processing and analyses of the data were supported by NOIRLab and the Lawrence Berkeley National Laboratory (LBNL). Legacy Surveys also uses data products from the Near-Earth Object Wide-field Infrared Survey Explorer (NEOWISE), a project of the Jet Propulsion Laboratory/California Institute of Technology, funded by the National Aeronautics and Space Administration. Legacy Surveys was supported by: the Director, Office of Science, Office of High Energy Physics of the U.S. Department of Energy; the National Energy Research Scientific Computing Center, a DOE Office of Science User Facility; the U.S. National Science Foundation, Division of Astronomical Sciences; the National Astronomical Observatories of China, the Chinese Academy of Sciences and the Chinese National Natural Science Foundation. LBNL is managed by the Regents of the University of California under contract to the U.S. Department of Energy. The complete acknowledgments can be found at https://www.legacysurvey.org/acknowledgment/.

\item[Contributions] T.-W.C. led the project, coordinated the follow-up campaign, and is the primary author of the manuscript. A.A. led the optical data reduction and photometric analysis, and contributed to the Lulin/Kinder discovery and follow-up observations. S.Y. contributed to the data analysis, astrometric and source-association analysis, and spectroscopic interpretation. S.J.S. contributed to the interpretation of the source association and precursor activity. T.J.M. led the hydrodynamical light-curve modelling and circumstellar-interaction interpretation. S.J.B. contributed to the precursor activity analysis. M.D.S. contributed to the POISE photometric, NUTS and HISS spectroscopic data and stripped-envelope supernova context. B.M. and B.P.S. contributed to the WiFeS observations and host-environment analysis. M.N. performed the semi-analytic light curve modelling and contributed to the interpretation and writing of the manuscript. J.H.G. contributed to Pan-STARRS follow-up and spectroscopic classification and interpretation. A.K.H.K. contributed to the Lulin follow-up and shock-breakout interpretation. A.D. led the TARDIS spectral modelling and light curve comparison. C.-H.L. performed the prospector fitting. Y.-H.L. contributed to precursor comparison. 
Y.-C.C., C.-H.L., Y.-H.L., Z.-Y.C., K.N.T.H., M.-H.L., C.-C.N., A.S.K., T.M., S.S., D.-C.Q., Z.-N.W. and Z.Z. contributed to the Lulin/Kinder observations, data processing and early counterpart identification. 
K.A., A. M., C.L. L.R. and M.S. contributed to the WiFeS observations.
M.E.H., K.W.S., K.C.C., T.d.B., C.-C.L., T.B.L., E.A.M., G.S.H.P., and R.W. contributed to Pan-STARRS observations, archival imaging, calibration, data products and survey infrastructure. 
E.K., T.L.K. and T.M.R. contributed to NOT spectroscopic observations. 
H.K. contributed to spectroscopic interpretation. 
C.-J.L. and J.T. contributed to host-environment interpretation. 
C.-H.L. contributed Citra observations and data products.
C.A., W.B.H., D.J., K.M., and C.M.P. contributed near-infrared spectroscopic observations and reductions. 
C.R.B., E.Y.H., W.B.H., N.M. and H.X. contributed to the POISE photometric observations and data products.
All authors discussed the results and contributed to editing the manuscript.

\item[Competing Interests] The authors declare no competing interests.

\item[Correspondence] Correspondence and requests for materials should be addressed to Ting-Wan~Chen~(email: twchen@astro.ncu.edu.tw) or Amar~Aryan~(email: amar@astro.ncu.edu.tw) or
Sheng~Yang~(email: sheng.yang@hnas.ac.cn).

\item[Data Availability] The optical spectra of SN~2026gzf that support the findings of this study have been made available via the WISeREP archive (\url{https://www.wiserep.org/object/14508}). The photometry is listed in the Supplementary Information. Photometry from the ATLAS were obtained from a public source (\url{https://fallingstar-data.com/forcedphot/}), ZTF and LSST points are from the Lasair broker (\url{https://lasair-ztf.lsst.ac.uk/objects/ZTF26aaonmha/} and \url{https://lasair.lsst.ac.uk/}).

\item[Code Availability] Upon request, the corresponding author will provide code used to produce the figures. 

\end{addendum}

\end{document}